\documentclass[11pt, a4paper]{article}
\pdfoutput=1

\usepackage{comment}
\usepackage{jcappub}
\usepackage{amsmath}
\usepackage{amsfonts}
\usepackage{amssymb}
\usepackage{graphicx}
\usepackage{mathrsfs}
\usepackage{latexsym}
\usepackage[usenames,dvipsnames]{xcolor}
\usepackage{slashed, cancel}
\usepackage{hyperref}
\usepackage{afterpage}
\usepackage{soul} 
\usepackage[font=footnotesize,labelfont=bf]{caption} 
\usepackage{enumitem}

\definecolor{rossos}{rgb}{0.8,0.2,0.3}
\definecolor{bluscuro}{rgb}{0.15, 0.2, .85}
\definecolor{bluchiaro}{cmyk}{1,.3,0.,0.1}
\hypersetup{colorlinks, citecolor=bluscuro, linkcolor=bluscuro, urlcolor=bluscuro}

\makeatother   
\newcommand\vv[1]{\vec{#1}}
\newcommand{\GeV}{{\rm \,GeV}}
\newcommand{\MeV}{{\rm \,MeV}}

\DeclareMathOperator\erf{erf}

\def\be   {\begin{equation}}   \def\ee   {\end{equation}}
\def\ba   {\begin{array}}      \def\ea   {\end{array}}
\def\bea  {\begin{eqnarray}}   \def\eea  {\end{eqnarray}}
\def\bean {\begin{eqnarray*}}  \def\eean {\end{eqnarray*}}
\def\nn{\nonumber}

\begin{document}

\title{Evaporation and scattering of momentum- and velocity-dependent dark matter in the Sun}

\author[1]{Giorgio Busoni,}
 \emailAdd{giorgio.busoni@unimelb.edu.au}

\author[2]{Andrea De Simone,}
 \emailAdd{andrea.desimone@sissa.it}

\author[3]{Pat Scott}
 \emailAdd{p.scott@imperial.ac.uk}

\author[3]{and Aaron C. Vincent}
 \emailAdd{aaron.vincent@imperial.ac.uk}

\affiliation[1]{ARC Centre of Excellence for Particle Physics at the Terascale \\
School of Physics, The University of Melbourne, Victoria 3010, Australia}
\affiliation[2]{SISSA and INFN, Sezione di Trieste, Via Bonomea 265, 34136 Trieste, Italy}
\affiliation[3]{Department of Physics, Imperial College London, Blackett Laboratory, Prince Consort Road, London SW7 2AZ, UK}

\abstract
{
  Dark matter with momentum- or velocity-dependent interactions with nuclei has shown significant promise for explaining the so-called Solar Abundance Problem, a longstanding discrepancy between solar spectroscopy and helioseismology.  The best-fit models are all rather light, typically with masses in the range of 3--5\,GeV.  This is exactly the mass range where dark matter evaporation from the Sun can be important, but to date no detailed calculation of the evaporation of such models has been performed.  Here we carry out this calculation, for the first time including arbitrary velocity- and momentum-dependent interactions, thermal effects, and a completely general treatment valid from the optically thin limit all the way through to the optically thick regime.  We find that depending on the dark matter mass, interaction strength and type, the mass below which evaporation is relevant can vary from 1 to 4\,GeV.  This has the effect of weakening some of the better-fitting solutions to the Solar Abundance Problem, but also improving a number of others.  As a by-product, we also provide an improved derivation of the capture rate that takes into account thermal and optical depth effects, allowing  the standard result to be smoothly matched to the well-known saturation limit.
}

\maketitle

\section{Introduction}

The search for the particle nature of dark matter (DM) has been a major focus of both the particle and astrophysics communities for several decades now. In the vast majority of models considered in the literature, a non-gravitational link between the dark sector and the standard model (SM) implies some amount of elastic scattering with nucleons, either directly with quarks, or through loop effects. This observation has been the motivation behind a concerted effort in large-scale direct detection experiments such as LUX \cite{LUXRun2}, SuperCDMS \cite{CDMSLite}, PICO \cite{PICO15,PICO60}, Panda-X \cite{PandaX2016} and XENON1T \cite{Aprile:2012zx}, which aim to measure the recoil energies of heavy nuclei following collisions with DM particles from the Galactic halo. Most of these experiments were designed with the canonical weakly-interacting massive particle (WIMP) in mind, so the target nuclear masses and threshold energies are optimally suited to detect particles with masses in the 100--1000\,GeV range. However, GeV-scale dark matter models are not at all disfavoured, and in some cases even preferred. For example, asymmetric (A)DM, a theoretical framework motivated by the baryon asymmetry, naturally predicts a mass of around $m_\chi/m_p \sim \Omega_{DM}/\Omega_b \sim 5$ \cite{Zurek14}.

It has been known for quite some time \cite{Gould87a,Taoso10,FrandsenSarkar} that the Sun is a very good laboratory in which to search for such low-mass particles, as mass-matching of 1--4\,GeV DM candidates with hydrogen and helium results in highly efficient momentum transfer in scattering events. If particles from the DM halo scatter in the Sun to energies below the local escape velocity, they become gravitationally bound, eventually returning to rescatter and rapidly settle into a stable configuration near the centre \cite{Widmark:2017yvd}. If the particle comes from the low-velocity tail of the Galactic DM distribution, then the momentum transfer needed to capture a DM particle can be quite small.  This makes solar capture highly complementary to direct detection experiments, which instead probe the high-velocity part of the distribution.  This aspect also endows capture and direct detection with quite different sensitivities to interactions that scale with the momentum exchanged in collisions ($q_{tr}$) or the DM-nucleon relative velocity ($v_r$).

DM capture by the Sun has two observable effects: 1) the production of observable GeV-energy neutrinos from self-annihilation into SM particles \cite{Bergstrom98b,Barger02,Blennow08,Wikstrom09,IC22Methods,Silverwood12,SuperK15,IC79,IC79_SUSY}; and  2) the transfer of heat from the solar core to outer regions, due to the large mean free path of weakly interacting particles in the Sun \cite{Spergel85,Nauenberg87,GouldRaffelt90a,GouldRaffelt90b,Vincent13}. Despite contributing a very small fraction of the Solar mass (at most a part in $10^{10}$), the additional heat transport can have dramatic effects on standard (MeV-energy) solar neutrino production, as well as the solar structure itself, which can be probed with helioseismology. If the particles interact weakly enough to carry energy on macroscopic scales, yet have a large enough cross-section to efficiently transfer energy, then these effects can be dramatic. This ``sweet spot'' is called the Knudsen transition, where local thermal equilibrium (LTE) gives way to non-local, Knudsen transport. It has recently been learnt \cite{Vincent13,Vincent14,Vincent15,Vincent16, Geytenbeek16} that for certain types of interaction, DM capture and energy transport can actually improve the fit of the Standard Solar Model (SSM) to helioseismological data, potentially solving the decade-old Solar Abundance Problem.  This problem represents a $> 6\sigma$ mismatch between SSM predictions and the sound speed profile, the location of the base of the convection zone, the surface helium abundance and the structure of the core \cite{AspIV,Bahcall05,Serenelli09,Scott09Ni,AGSS,Serenelli11,Villante14,Serenelli16}. Such a solution requires going beyond the classic spin-independent (SI) and spin-dependent (SD) ``billiard ball'' interaction models, to a more general parameterisation that allows for the scattering cross-section to depend on both the relative velocity of a collision $v_r$ and the transferred momentum $q_{tr}$. Although the best solutions in many models are excluded by recent direct detection experiments \cite{CRESST_momdep,CDMSLite}, a handful are still allowed \cite{Vincent16}.

Crucially, the strongest effects on solar physics come from light DM particles, with masses below $\sim 5$\,GeV. At such low masses, there is a non-negligible chance that a scattering event will lead to ejection of the DM particle from the Sun, i.e.\ evaporation. The original formalism to calculate evaporation rates was developed in Ref.\ \cite{Gould87a}, where it was shown that evaporation prohibits accumulation of DM of mass less than 4\,GeV in the Sun, if its scattering cross-section is constant with respect to $v_r$ and $q_{tr}$. The corresponding evaporation mass for velocity- and momentum-dependent DM is not known.  Given the very different phenomenology of such models for capture and transport, it is reasonable to expect that the impacts of evaporation will also vary with the interaction.

This work has two main goals: first, to update the capture and evaporation formalism of \cite{Gould87a}, in order to include a more self-consistent description of thermal and rescattering effects governing these processes; and second, to extend the computation of the evaporation rate to the generalised form factor models described in Refs.\ \cite{Vincent15,Vincent16}.  Ref. \cite{Garani:2017jcj} has performed this computation for two of the models considered here, although their focus was mainly on DM-\textit{electron} scattering. Other recent calculations of the evaporation \cite{Busoni13, Liang16} and capture rates \cite{Vincent14,Lopes14,Vincent15} in the Sun have not included the thermal or rescattering (optical depth) effects. The very recent Ref. \cite{Bramante:2017xlb} looked at capture through multiple scattering events, though their focus was on constant cross-sections and much heavier ($\sim$ TeV) dark matter.

In Sec.\ \ref{sec:defs}, we begin by defining and calculating the thermal quantities involved in the full evaluation of the capture and evaporation rates, including the velocity distributions, effective DM temperature, thermally-averaged cross-sections and the optical depth. Sections \ref{sec:capture} and \ref{sec:evaporation} present the improved capture and evaporation rate calculations, respectively. We conclude in Sec.\ \ref{sec:conclusions}, and provide details of the more involved calculations in Appendices \ref{sec:geomlimit}, \ref{sec:reldistr} and \ref{sec:calcdetails}.

\section{Dark matter microphysics, distributions and thermodynamics}
\label{sec:defs}

\subsection{Cross-sections and kinematics}

In this paper we study solar capture and evaporation of DM with nuclear couplings proportional to some power of the momentum exchanged in collisions ($q_{tr}$), or to some power of the DM-nucleus relative velocity $v_r$.  This leads to the following differential cross-sections for scattering of DM and nuclear species $i$:
\bea
\frac{d\sigma_i}{d\cos\theta}(v_r,\theta)=\sigma_{0,i} \left(\frac{v_r}{v_0}\right)^{2n}\label{eq:sigmavrel},\\
\frac{d\sigma_i}{d\cos\theta}(v_r,\theta)=\sigma_{0,i} \left(\frac{q_{tr}}{q_0}\right)^{2n}\label{eq:sigmaqtr}.
\eea
In general we are most interested in the cases where $n=-1$, 1 or 2, as these non-trivial velocity and momentum-dependences have been seen to arise in various concrete particle models for DM.  We will also consider the special case $n=0$ of a constant cross-section.

The momentum transfer can be written as a function of the center of mass scattering angle $\theta_{cm}$\footnote{For the sake of readability, we keep relevant factors of $c$, $\hbar$ and $k_B$ implicit.}
\be
q_{tr}^2=(1-\cos\theta_{cm})v_r^2 \frac{2m_{\chi}^2}{(1+\mu)^{2}} \label{eq:qtrcos},
\ee
where
\be
\mu = \frac{m_\chi}{m_i}.
\ee

The values of the normalisation constants $v_0$ and $q_0$ are arbitrary, and simply set the relative velocity and momentum transfer at which the constant part of the reference cross-section $\sigma_{0,i}$ is defined. We choose
\bea
v_0 &=& 220 \mathrm{km\,s}^{-1},\\
q_0 &=& 40 \MeV.
\eea
These correspond to typical values encountered in Earth-based direct detection experiments. The reference cross-section for each species can be obtained from the reference DM-nucleon cross-section $\sigma_0\equiv\sigma_{0,\mathrm{H}}$. For spin-independent (SI) interactions, this is given by
\bea
\sigma_{0,i} &=& \sigma_i |F_i(q_{tr})|^2,\label{eq:nucleoncrosssection}\\
\sigma_i &\equiv& \sigma_0 A_i^2\left(\frac{m_i}{m_p}\right)^2 \left(\frac{m_{\chi}+m_p}{m_{\chi}+m_i}\right)^2.
\eea
Here $\sigma_i$ is the constant part, and $F_i(q_{tr})$ is the nuclear form factor. For scattering on protons, $|F_H(q_{tr})|^2 = 1$ for all $q_{tr}$.  This remains true for heavier species only at small momentum transfer; at large momentum transfer, this term introduces an additional momentum dependence to the differential cross-section.  To describe this, wherever the momentum transfer can be large, we use the exponential (Helm) form factor \cite{Helm:1956zz}
\be
|F_i(E_R)|^2=\exp(-E_R/E_i),
\ee
with
\bea
E_i &\equiv& 3/(2 m_i \Lambda_i^2),\\
\Lambda_i &\equiv& [ 0.91 \left( {m_i}/{\rm{GeV}} \right)^{1/3} + 0.3] \,\, \rm{fm},
\eea
where $E_R = q_{tr}^2/(2m_i)$ is the nuclear recoil energy.

For spin-dependent (SD) interactions,
\be
\sigma_{0,i} = \sigma_i = \sigma_0 \frac{4(J_i+1)}{3J_i}|\langle S_{\mathrm{p},i}\rangle+\langle S_{\mathrm{n},i}\rangle|^2 \left(\frac{m_i}{m_p}\right)^2 \left(\frac{m_{\chi}+m_p}{m_{\chi}+m_i}\right)^2,
\ee
where $J_i$ is the nuclear spin and $\langle S_{\mathrm{p},i}\rangle$ and $\langle S_{\mathrm{n},i}\rangle$ are the respective expectation values of the spins of the proton and neutron subsystems.

\subsection{Dark matter temperature}
In the non-local Knudsen regime, the dark matter profile is fully specified by a single temperature $T_\chi$, which is a weighted average of the stellar temperature with which the DM is in thermal contact.

The relative speed distribution for two particle species following Maxwell-Boltzmann velocity distributions with temperatures $T_\chi$ and $T(r)$ is
\be
f(v_r)dv_r=\sqrt{\frac{2}{\pi}}v_r^2\left(\frac{T_\chi}{m_{\chi}}+\frac{T(r)}{m_i}\right)^{-\frac32} e^{-\frac{v_r^2}{2\left(\frac{T_\chi}{m_{\chi}}+\frac{T(r)}{m_i}\right)}}dv_r.
\label{eq:distrib2t}
\ee
The $\ell$th moment of this distribution is:
\be
\langle v^\ell \rangle = \frac{2}{\sqrt{\pi}} \Gamma\left(\frac{3+\ell}{2}\right)\left(\frac{2T_\chi}{m_{\chi}}+\frac{2T(r)}{m_i}\right)^{\ell/2}\label{eq:momdistr}.
\ee
Using Eqs.\ \ref{eq:sigmavrel}, \ref{eq:sigmaqtr}, \ref{eq:qtrcos} and \ref{eq:momdistr}, we can also calculate the thermally-averaged cross-sections.  When considering scattering of the bound population of DM with nuclei in the Sun, $ \frac{T(r)}{m_i} \ll 1$, so the integral is dominated by small relative velocities, and the momentum transfer is typically very small.  This means that to a very good approximation $|F_i(q_{tr})|^2 \sim 1$. We then compute the thermally-averaged \textit{total} cross sections (integrating over $\cos \theta$, which yields an overall factor of 2 for the velocity-dependent case with respect to Eq.\ \ref{eq:sigmavrel}):
\bea
\langle\sigma_v\rangle_i(T_1,T_2) &=& \sigma_i \frac{2^{2+n} \Gamma(\frac{3}{2}+n)\left(\frac{T_1}{m_{\chi}}+\frac{T_2}{m_i}\right)^{n}}{\sqrt{\pi}v_0^{2n}}\label{eq:sigmavrelav},\\
\langle\sigma_q\rangle_i(T_1,T_2) &=& \sigma_i \frac{4 \Gamma(\frac{3}{2}+n)\left(\frac{T_1}{m_{\chi}}+\frac{T_2}{m_i}\right)^{n}}{\sqrt{\pi}q_0^{2n}} \frac{m_{\chi}^{2n}}{\mu_+^{2n}} f(n) \label{eq:sigmaqtrav}.
\eea
Here $\mu_\pm \equiv (1\pm\mu)/2$.  For $n \le 0$, $f(n)=1$; for $n > 0$,
\be
\label{eq:angular_average}
f(n)\equiv \frac{1}{2} \int_{-1}^1 d\cos\theta (1-\cos\theta)^n = \frac{2^{n}}{n+1}.
\ee
In the same way, we can obtain the thermally-averaged product of the nuclear scattering cross-section and the DM-nucleus relative velocity $\langle\sigma v\rangle$ for scattering between bound DM and solar nuclei,
\bea
\langle\sigma_v v\rangle_i(T_{1},T_2) &=& \sigma_i \frac{2^{n+5/2} \Gamma \left(n+2\right) \left(\frac{T_{1}}{m_{\chi}}+\frac{T_2}{m_i}\right)^{n+1/2}}{\sqrt{\pi }v_0^{2n}},\label{eq:sigmavrela}\\
\langle\sigma_q v\rangle_i(T_{1},T_2) &=& \sigma_i \frac{2^{5/2} \Gamma \left(n+2\right) \left(\frac{T_{1}}{m_{\chi}}+\frac{T_2}{m_i}\right)^{n+1/2}}{\sqrt{\pi }q_0^{2n}} \frac{m_{\chi}^{2n}}{\mu_+^{2n}} f(n).\label{eq:sigmaqtra}
\eea

\begin{figure}[tp]
    \includegraphics[width=0.5\textwidth]{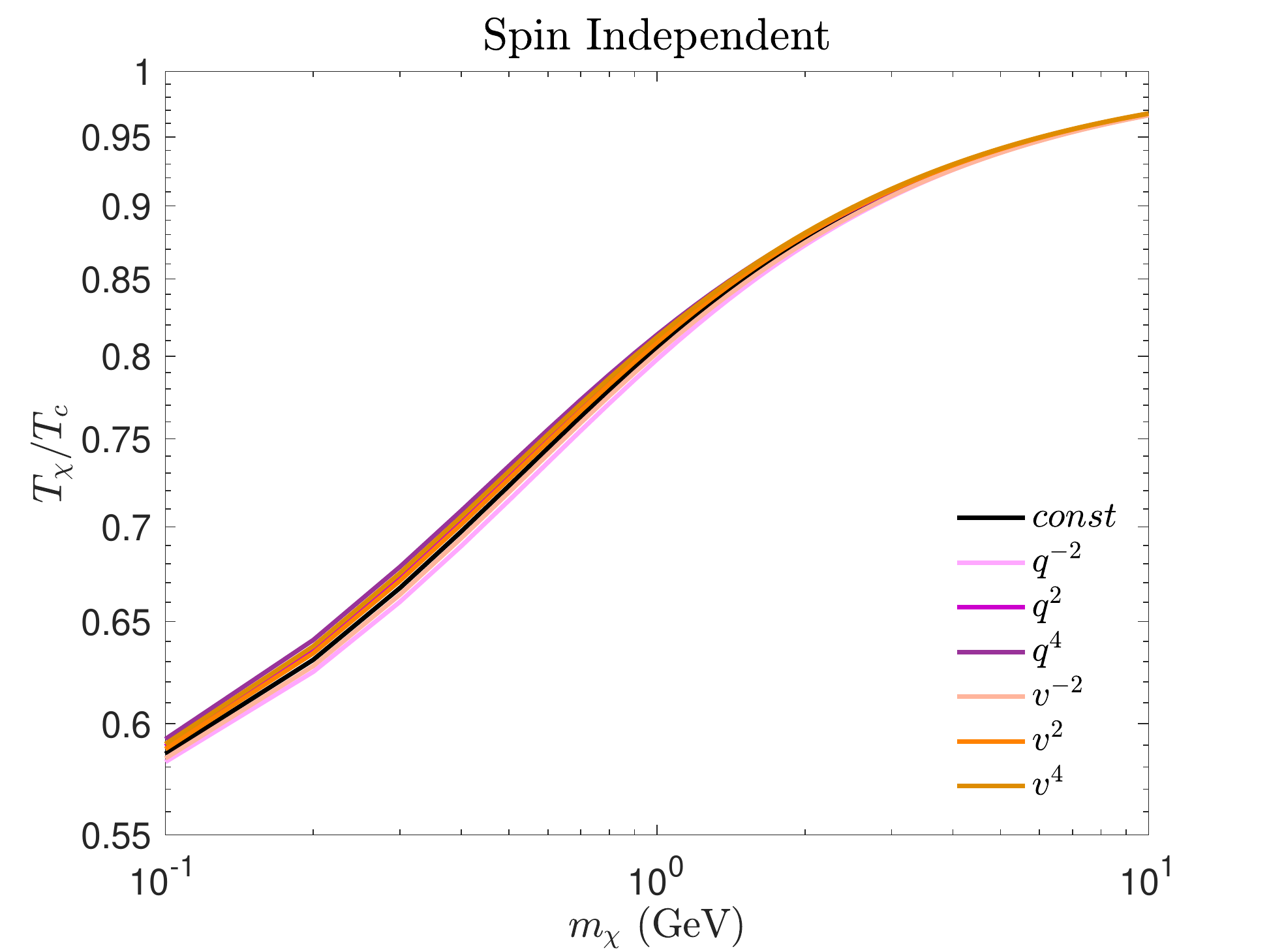}    \includegraphics[width=0.5\textwidth]{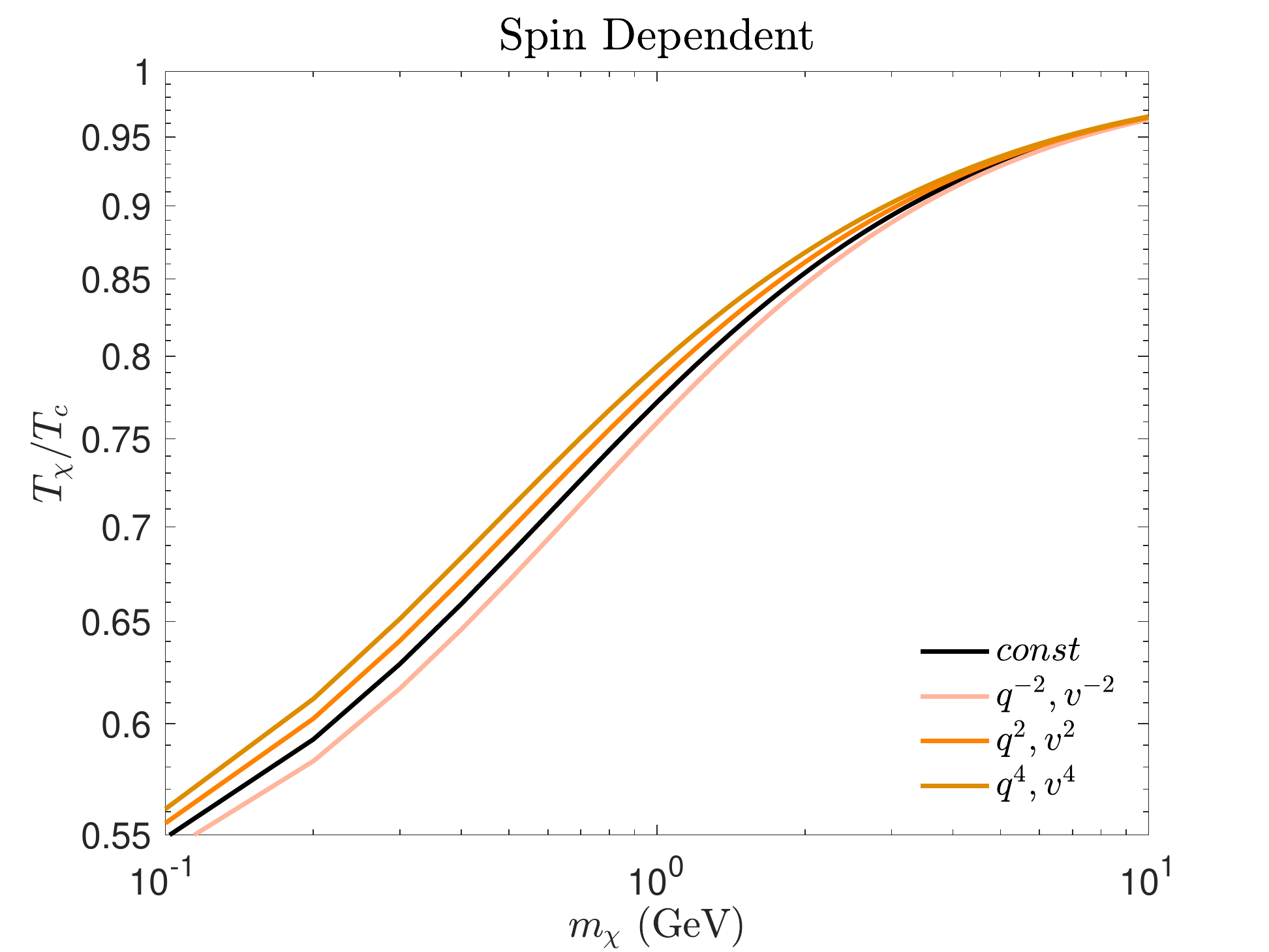}
    \caption{Dark matter temperature $T_\chi$ in units of the central temperature $T_\text{c}$, as a function of the dark matter particle mass and interaction type, using the AGSS09ph solar model \cite{AGSS,Serenelli09}.  \textit{Left}: spin-independent scattering; \textit{Right}: spin-dependent scattering. In the spin-dependent case, both $q^n$ and $v^n$ interactions give identical temperatures.}
    \label{fig:DMtemp}
\end{figure}
Using Eqs.~(\ref{eq:sigmavrela})-(\ref{eq:sigmaqtra}), we can now implicitly define the temperature of the DM isothermal component in terms of the isothermal density profile $n_{\rm iso}$ (defined later in Eq.~\ref{eq:niso}), as
\be
T_\chi = \frac{\int_0^R dr 4\pi r^2 T(r) \sum_i \langle\sigma_{v,q}v\rangle_i\left[T_\chi,T(r)\right]  n_i(r) n_{\rm iso}(r,T_\chi)}{\int_0^R dr 4\pi r^2 \sum_i \langle\sigma_{v,q}v\rangle_i\left[T_\chi,T(r)\right] n_i(r) n_{\rm iso}(r,T_\chi)}.
\ee
This equation does not depend on the value of $\sigma_0$, nor on the overall DM population in the Sun, so we can solve it once and for all for different values of $m_{\chi}$ and $n$. We show the resulting temperature curves for the Sun in Fig.\ \ref{fig:DMtemp}, using the AGSS09ph solar model \cite{AGSS,Serenelli09} for the run of nuclear densities.  As can be seen, the DM temperature does not differ greatly for different interaction types. This definition of the DM temperature is a significant improvement over that of Ref.\ \citep{Spergel:1984re}, as it accounts for scattering of DM on multiple nuclear species, and does not rely on the assumption that the net energy flux across the solar surface due to dark matter scattering is zero.  This second assumption is equivalent to assuming the total evaporation rate is zero (though we note that it is possible to correct the temperature of Ref.\ \citep{Spergel:1984re} for the impacts of evaporation \citep{Garani:2017jcj}).

We end this section on a technical note. By setting $f(n)=1$ for $n=-1$ in the thermal averages, we have implicitly employed the momentum-transfer cross-section $\sigma_{q_{tr}} \equiv (1-\cos\theta) \sigma$ for the $q^{-2}$ case. This is necessary to avoid divergence in the forward-scattering limit, where momentum transfer is negligible. This regularization method avoids the imposition of an arbitrary cutoff or screening length.

\begin{figure}[tp]
    \centering
    \includegraphics[width=0.45\textwidth]{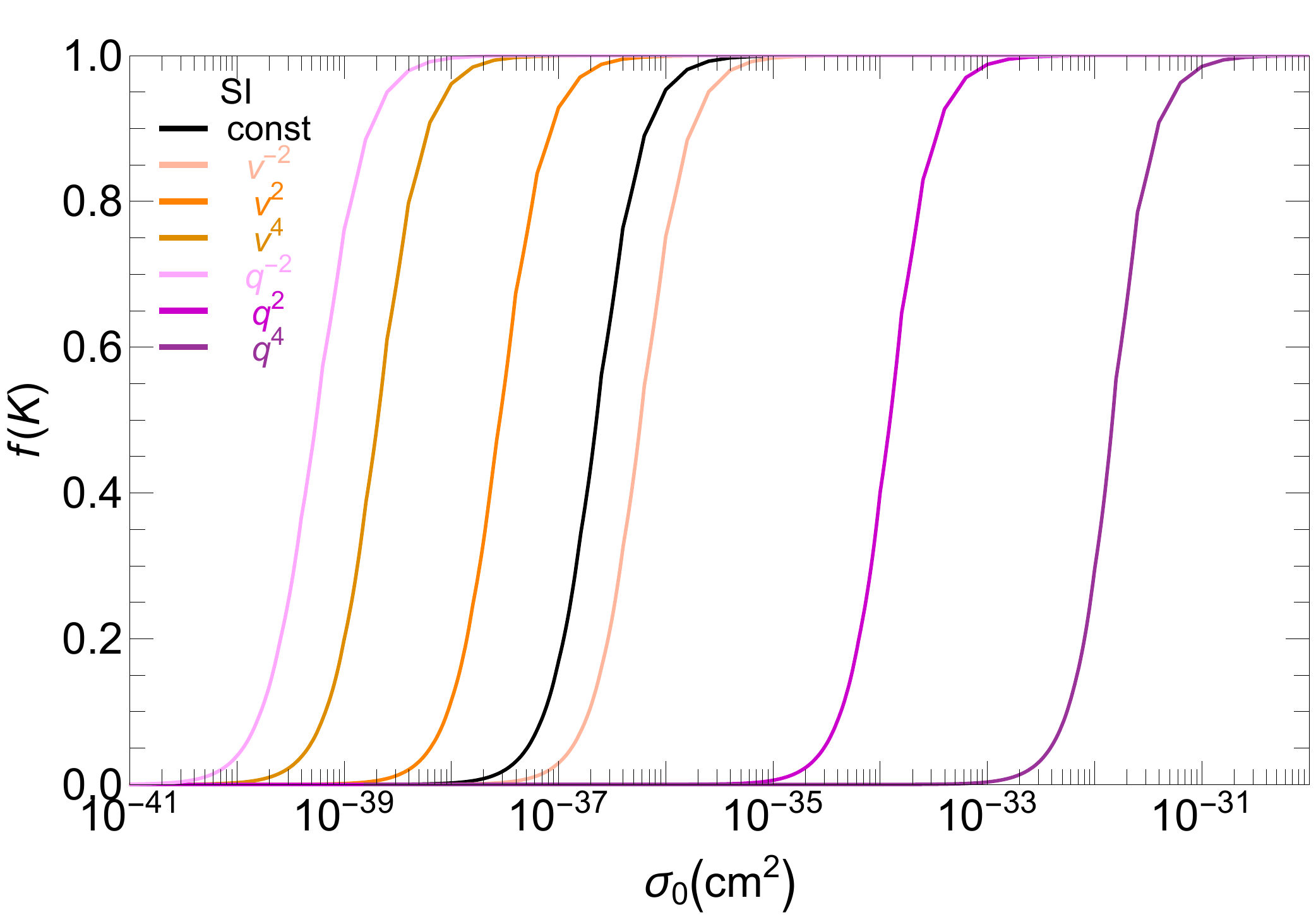}
    \includegraphics[width=0.45\textwidth]{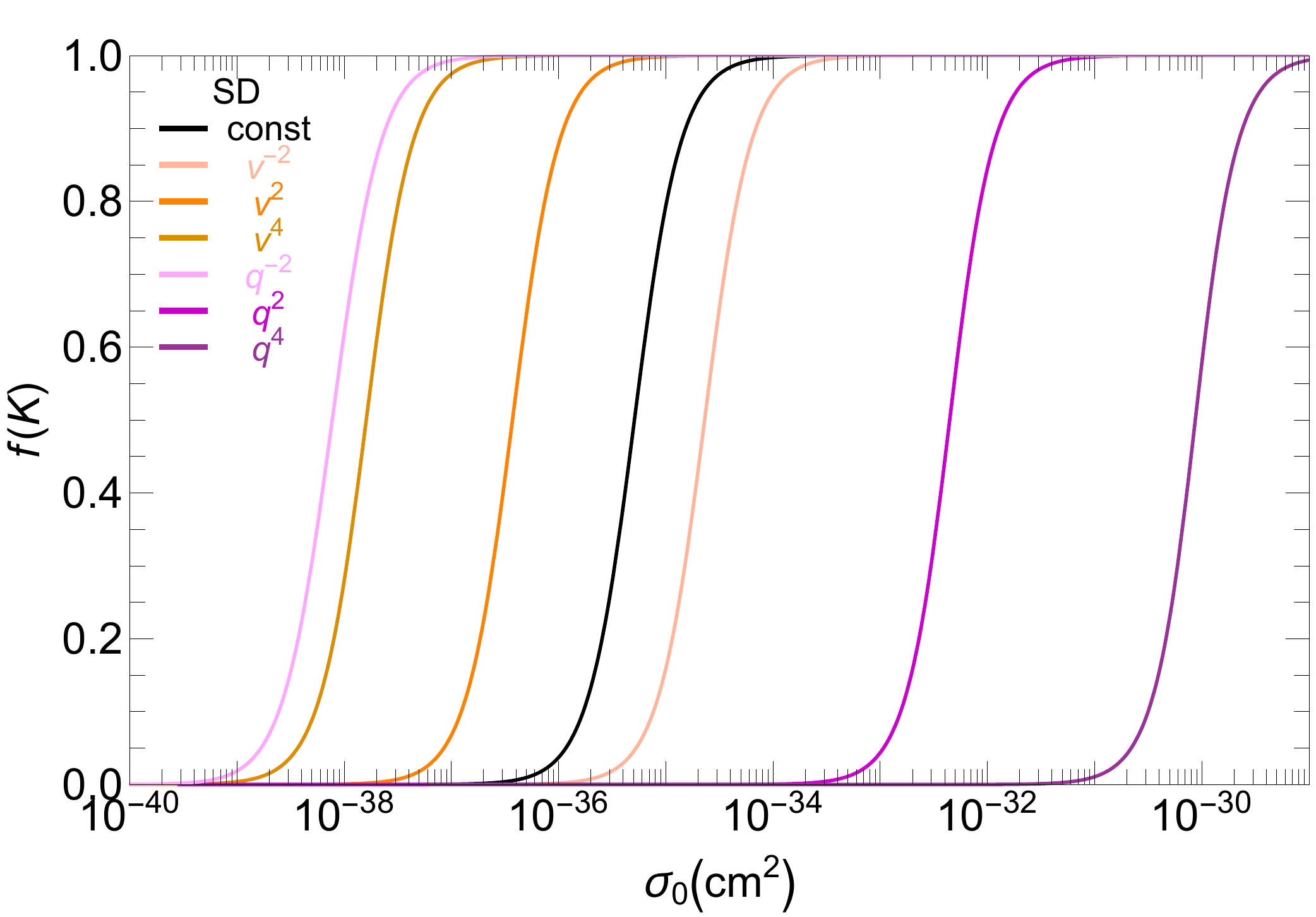}
    \caption{Value of $\mathfrak{f}(K)$ as a function of the cross-section, for $m_\chi=1$\,\GeV and spin-independent (\textit{left}) and spin-dependent (\textit{right}) interactions.}
    \label{fig:kunnumb}
\end{figure}


\subsection{Dark matter velocity distribution}
We approximate the DM population in the Sun by a mixture of an isothermal component (following a Maxwell-Boltzmann velocity distribution), and a component tracking the local temperature of the gas (the LTE component).  Indeed, Gould and Raffelt \cite{GouldRaffelt90a} found that the true distribution of weakly interacting particles can be accurately modelled via interpolating functions that connect the LTE and non-local isothermal component. The overall dark matter velocity distribution inside the Sun relevant for evaporation is then:
\be
f_{evap}(v)=\left[1-\mathfrak{f}(K)\right]n_{\rm iso}(r) e^{-\frac{m_{\chi}v^2}{2T_\chi}} + \mathfrak{f}(K) n_{\rm LTE}(r) e^{-\frac{m_{\chi}v^2}{2T(r)}}\label{eq:dmsundistrib},
\ee
where
\be
\mathfrak{f}(K)\equiv\frac{1}{1+(K/K_0)^2}\label{eq:fk}.
\ee
Here $K_0=0.4$, and $K$ is the Knudsen number, defined as
\be
K\equiv\frac{\lambda}{r_\chi}.
\ee
The scale radius $r_\chi$ of the DM cloud is given by
\be
r_\chi=\sqrt{\frac{3T_c}{2\pi G \rho_c m_{\chi}}},
\ee
and the typical inter-scattering distance $\lambda$ is given by
\bea
\lambda\equiv\left(\sum_i n_i(0) \langle\sigma_{v,q}\rangle_i(T_\chi,T_c)\right)^{-1}.
\eea
The resulting values of $\mathfrak{f}(K)$ are shown in Fig.\ \ref{fig:kunnumb}, for the case of $m_\chi = 1$\,GeV.

The two spatial distributions are given by
\bea
n_{\rm iso}(r)&=&N_{\rm iso} e^{-m_{\chi}\phi(r)/T_\chi}\label{eq:niso},\\
n_{\rm LTE}(r)&=&N_{\rm LTE} \left[\frac{T(r)}{T(0)}\right]^\frac32  e^{-\int_0^r dr' \frac{\alpha(r') dT/dr'(r') + m_{\chi} d\phi/dr'(r') }{T(r')}},
\eea
where $\alpha(r)$ is the thermal diffusivity, which depends on both the DM mass and the $v_r/q_{tr}$-dependence of the differential cross-section.  The quantities $N_{\rm iso}$ and $N_{\rm LTE}$ are normalisation constants, defined such that the integral over each distribution is 1. We take the values of $\alpha$ from the calculations of Ref.\ \citep{Vincent13}.

\begin{figure}[tp]
    \centering
    \includegraphics[width=0.48\textwidth]{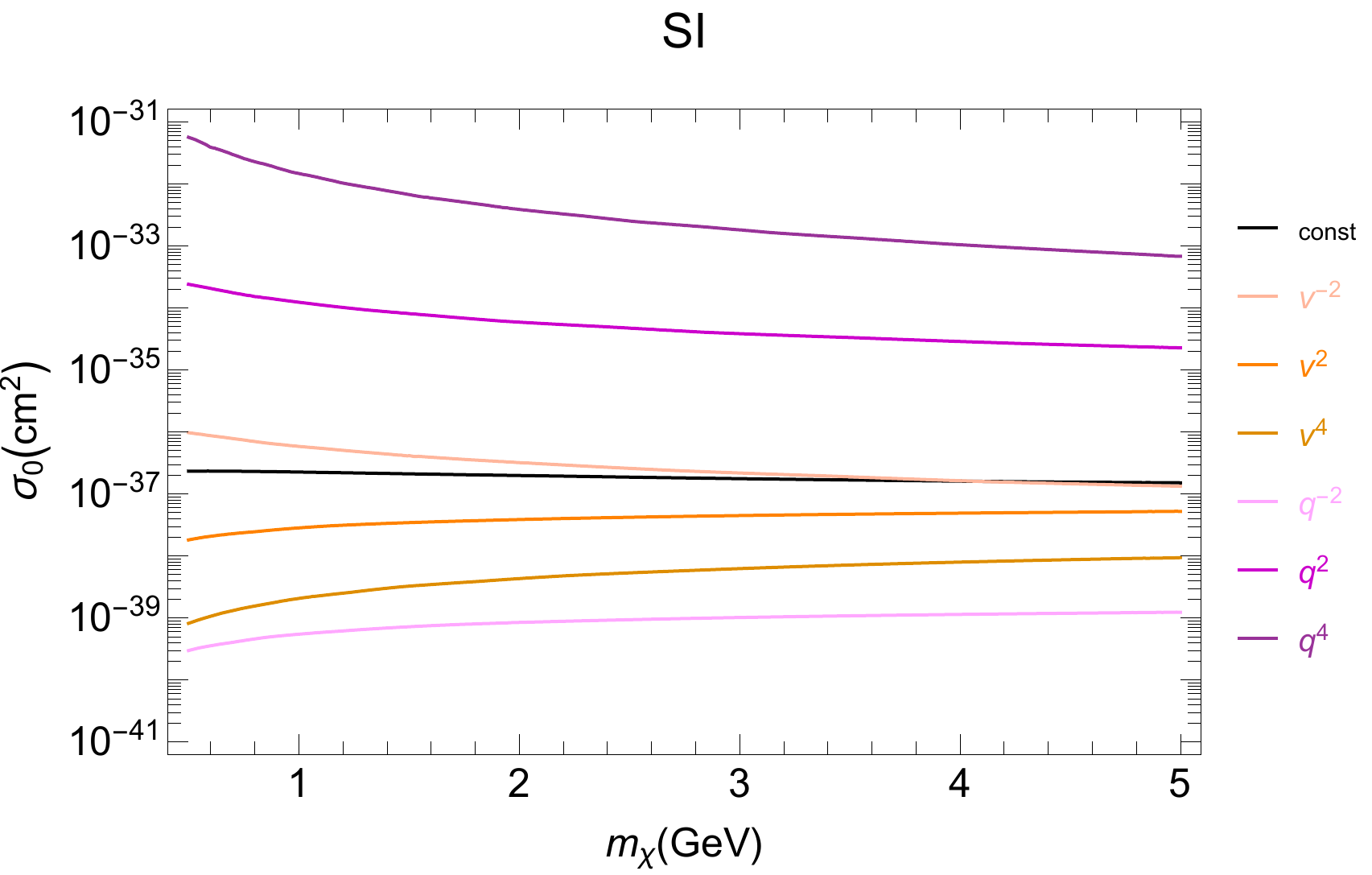}
    \includegraphics[width=0.48\textwidth]{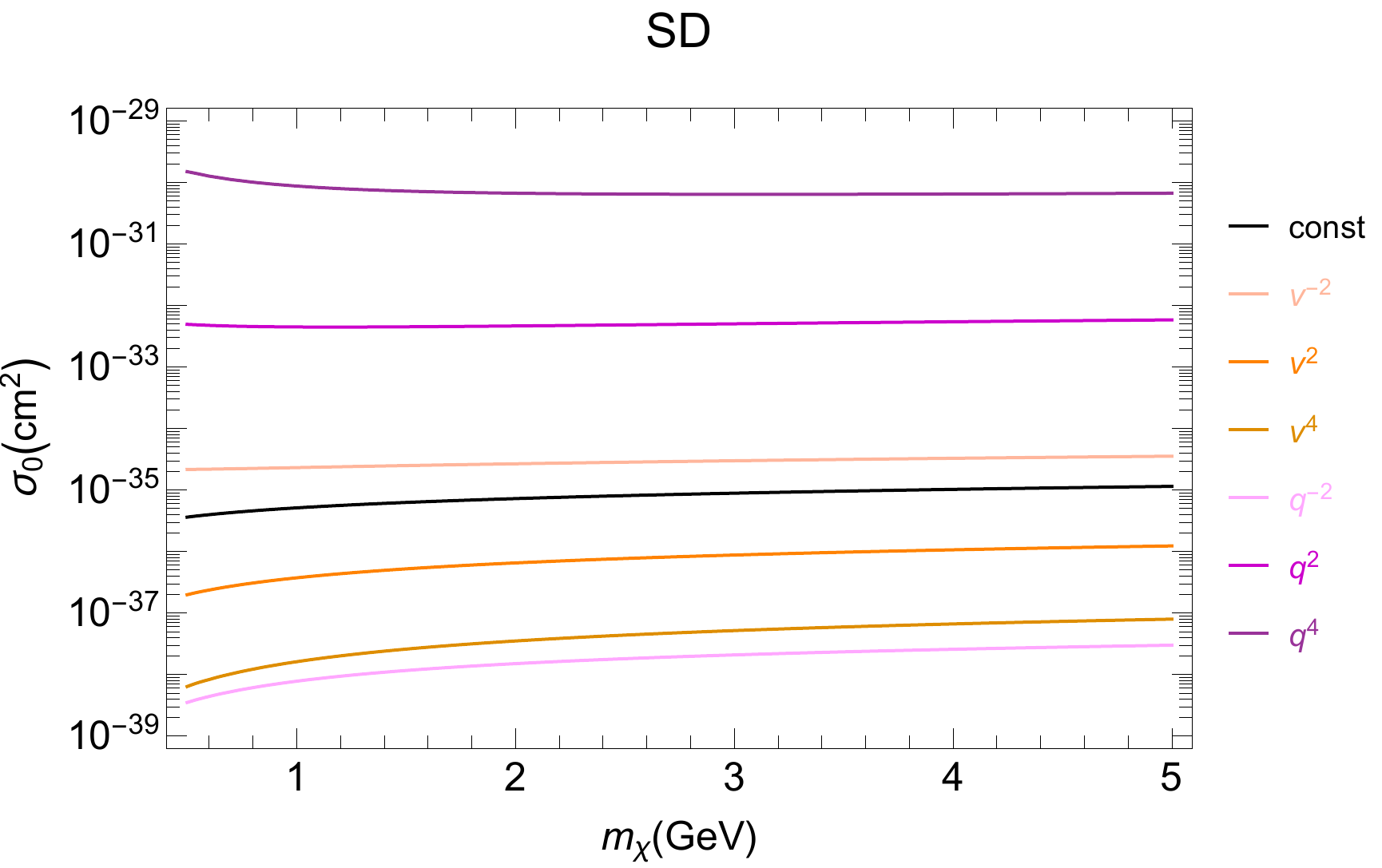}
    \caption{Masses and cross-sections leading to $\mathfrak{f}(K)=0.5$, the point at which heat transport by DM in the Sun transitions between the isothermal and Knudsen regimes.  Here we show curves for different interaction types $\sigma \propto q^n, v^n$, assuming spin-independent scattering (\textit{left}) or spin-dependent scattering (\textit{right}). }
    \label{fig:kuntrans}
\end{figure}

In Fig.\ \ref{fig:kuntrans}, we show the critical value of the cross-section where the transition from the isothermal to the Knudsen regime takes place, as a function of mass.  We define this as the cross-section for which $\mathfrak{f}(K)=0.5$.


\subsection{Optical depth}
\label{sec:optdpt}
\begin{figure}[tp]
    \centering
    \includegraphics[width=0.5\textwidth]{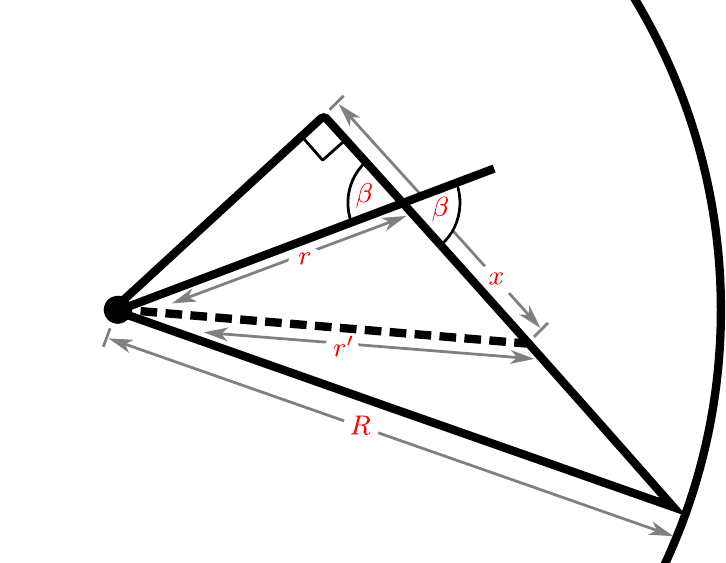}
    \caption{Definitions of the quantities $x$, $z\equiv\cos\beta$ and $r'^2 \equiv x^2+r^2(1-z^2)$ used for calculating the extinction coefficient due to the optical depth seen by a dark matter particle (Eqs.\ \ref{eq:extinction} and \protect\ref{eq:opticdepth}).}
    \label{fig:diagram}
\end{figure}

\begin{figure}[tp]
    \centering
    \includegraphics[width=0.9\textwidth]{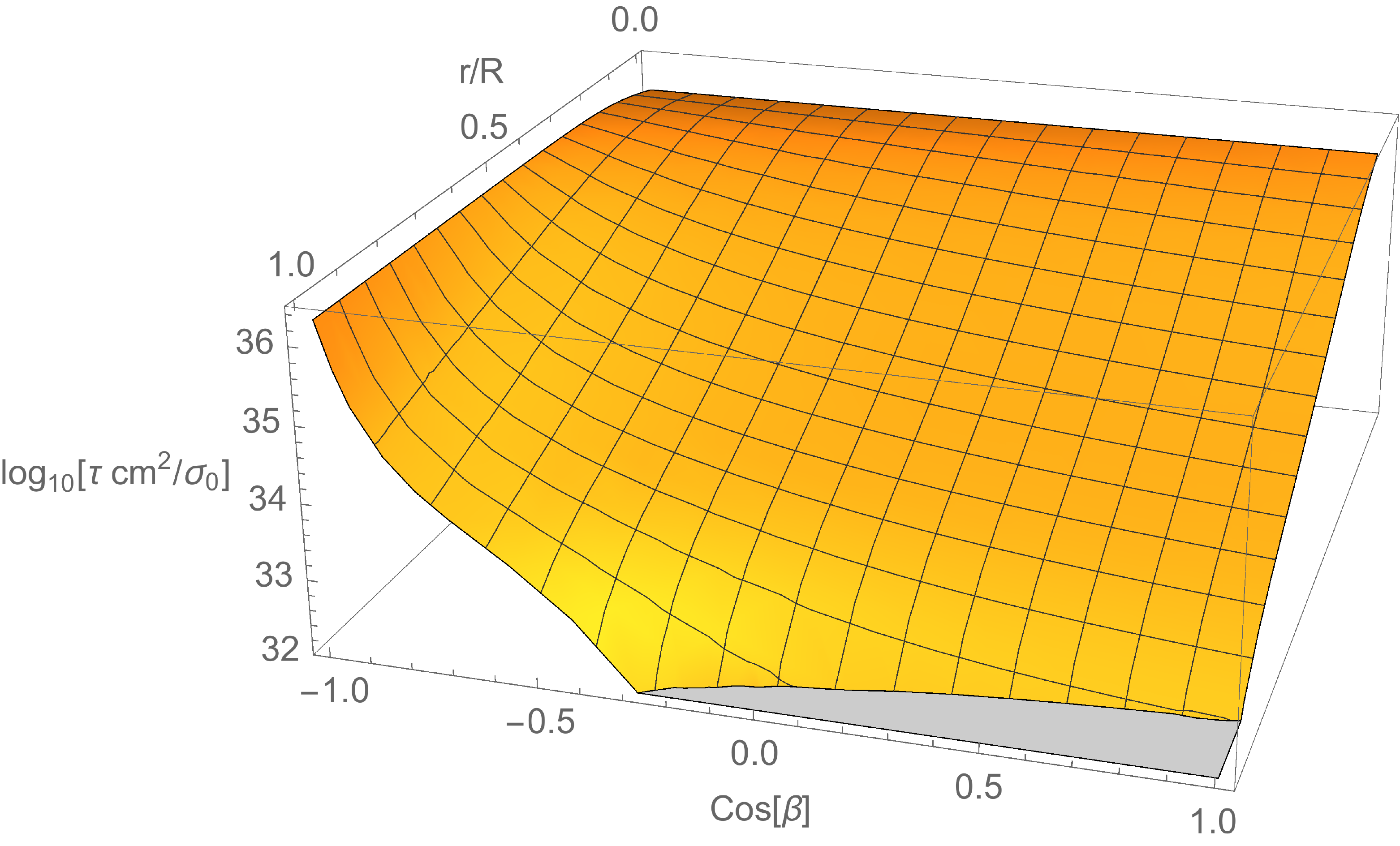}
    \caption{The value of the optical depth $\tau$ as a function of $z=\cos\beta$ and the distance $r$, for $m_\chi=1\GeV$ with constant ($n=0$) spin-dependent interactions.  Other cases have similar behaviours.}
    \label{fig:opttau}
\end{figure}

\begin{figure}[tp]
    \centering
    \includegraphics[width=0.80\textwidth]{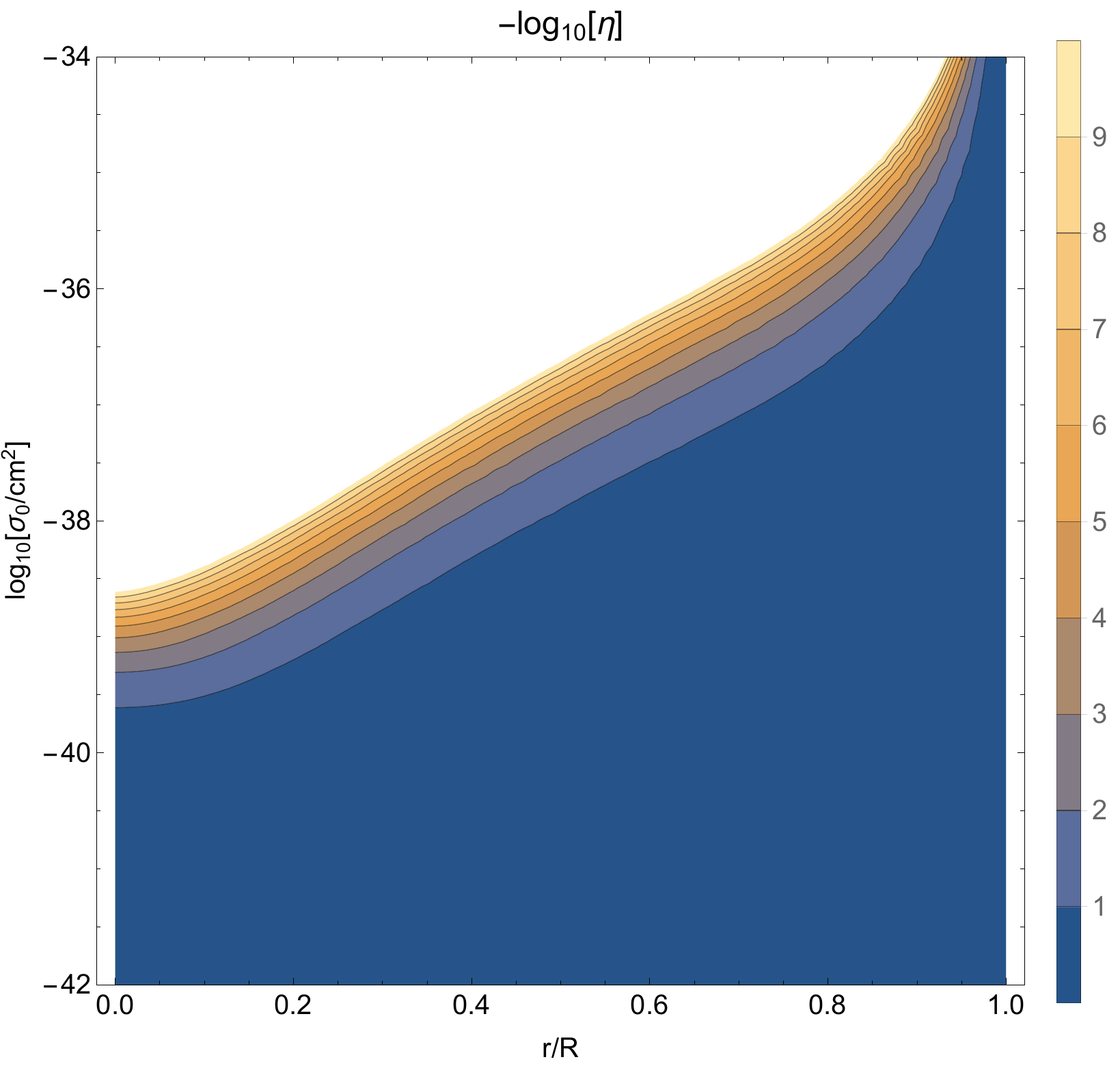}
    \caption{The extinction coefficient $\eta$ as a function of $\sigma_0$ and the distance $r$ from the centre of the Sun, for $m_\chi=1\GeV$, $n=0$, spin-dependent interactions. Other cases have similar behaviours. We have limited the plot range to the interval $[0,10]$.}
    \label{fig:opteta}
\end{figure}

When the DM-nucleon cross-section becomes large, the flux of DM particles traversing the Sun is significantly reduced over its path length.  To account for this effect, we add to all volume integrals an additional extinction factor $\eta(r)$, defined as
\be
\eta(r) =\frac{1}{2}\int_{-1}^1 dz e^{-\tau(r,z)}\label{eq:extinction},
\ee
with the optical depth $\tau$ defined as
\be
\tau(r,z)=\int_{rz}^{\sqrt{R^2-r^2(1-z^2)}} dx \sum_i n_i(r') \langle\sigma_i\rangle_{evap, cap}\label{eq:opticdepth}.
\ee
Here $z \equiv \cos\beta$ and $r'^2 \equiv x^2+r^2(1-z^2)$.  The definitions of the various geometric quantities is shown in more detail in Fig.\ \ref{fig:diagram}. Here we neglect additional (small) corrections expected due to multiple scattering events and departures of the DM velocity distribution from pure isotropy \cite{Gould:1989tu,Garani:2017jcj}. In Figs.\ \ref{fig:opttau} and Fig.\ \ref{fig:opteta} we show the resulting optical depth and suppression factor, for evaporation of an example 1\,\GeV\ DM candidate with a constant spin-dependent cross-section.

Using this average optical depth formalism and taking the limit of very large cross-sections, it is possible to reproduce the geometric limit for the capture cross-section. We give the analytic proof in Appendix \ref{sec:geomlimitproof}, and carry it through to the geometric limit for the actual capture rate in Appendix \ref{sec:geomlimitformulae}, for $n=0$.

In calculating the optical depth (Eq.\ \ref{eq:opticdepth}), the average cross-section must be calculated using the appropriate DM velocity distribution. Therefore, apart from the constant cross-section case, where the optical depth for capture and evaporation coincide, in all other cases this calculation must employ \textit{different} DM velocity distributions for capture and evaporation.

\subsubsection{Optical depth for DM capture}
For capture, the optical depth expresses the probability that an incoming DM particle would not yet have been scattered off its path before reaching an interaction point at some height $r$ in the star.  For this calculation, we should therefore use the velocity distribution of the incoming, unbound DM from Galactic halo.  This results in a relative velocity distribution between unbound halo DM and nuclei in the Sun of (see Appendix \ref{sec:reldistr} for more details)
\be
f_{cap}(u_r)du_r = \frac{u_r}{v_\odot} \sqrt{\frac{3}{2\pi(v_d^2+3T\mu/m_{\chi})}} \left(e^{-\frac{3(u_r-v_\odot)^2}{2(3T\mu/m_{\chi}+v_d^2)}}-e^{-\frac{3(u_r+v_\odot)^2}{2(3T\mu/m_{\chi}+v_d^2)}}\right)du_r\label{cap_distrib}.
\ee
Here $u_r$ is the relative velocity far away from the Sun.  By the time a DM particle arrives to collide with a nucleus in the Sun, $u_r$ will have been boosted by the local escape velocity $v_e$, giving a DM-nucleus relative velocity $w_r(r)=\sqrt{u_r^2+v_e^2(r)}$.  Taking into account this velocity boost from the Sun's potential well, the final distribution of DM-nucleon relative velocities at the interaction point relevant for capture is
\be
f_{cap}(w_r)dw_r = \frac{w_r}{v_\odot} \sqrt{\frac{3}{2\pi(v_d^2+3T\mu/m_{\chi})}} \left(e^{-\frac{3\left(\sqrt{w_r^2-v_e^2}-v_\odot\right)^2}{2(3T\mu/m_{\chi}+v_d^2)}}-e^{-\frac{3\left(\sqrt{w_r^2-v_e^2}+v_\odot\right)^2}{2(3T\mu/m_{\chi}+v_d^2)}}\right)dw_r.
\ee
The moments of this distribution are
\bea
\langle w_r^4    \rangle &=& (v_\odot^2+ v_e^2 + v_d^2+3T\mu/m_{\chi})^2+\tfrac{2}{3}(v_d^2+3T\mu/m_{\chi})(2v_\odot^2+ v_d^2+3T\mu/m_{\chi})\label{eq:vmomentcapn2},\\
\langle w_r^2    \rangle &=& v_\odot^2+ v_e^2 + v_d^2+3T\mu/m_{\chi}\label{eq:vmomentcapn1},\\
\langle w_r^{-2} \rangle &=& AH_{-1}\left(Av_e,Av_\odot\right)/(v_\odot\sqrt{\pi})\label{eq:vmomentcapnminus1},
\eea
where
\bea
A^2 &\equiv& \tfrac32(v_d^2 +3T\mu/m_{\chi})^{-1},\\
H_{-1}(x,y) &\equiv& \int_0^\infty \frac{t}{t^2+x^2}\left(e^{-(t-y)^2}-e^{-(t+y)^2}\right)dt.
\label{eq:Hm1def}
\eea
From this, we can calculate the average cross-sections in the usual way, with
\bea
\langle\sigma_v\rangle_{i,cap} &=& 2\sigma_{0,i}\langle v^{2n}\rangle/v_0^{2n}\label{eq:sigmagen_v},\\
\langle\sigma_q\rangle_{i,cap} &=& 2^{-n} f(n) \langle\sigma_v\rangle_{i,cap} (v_0/q_0)^{2n} (m_{\chi} / \mu_+)^{2n}\label{eq:sigmagen_q}.
\eea

\begin{figure}[tp]
    \centering
    \includegraphics[width=0.45\textwidth]{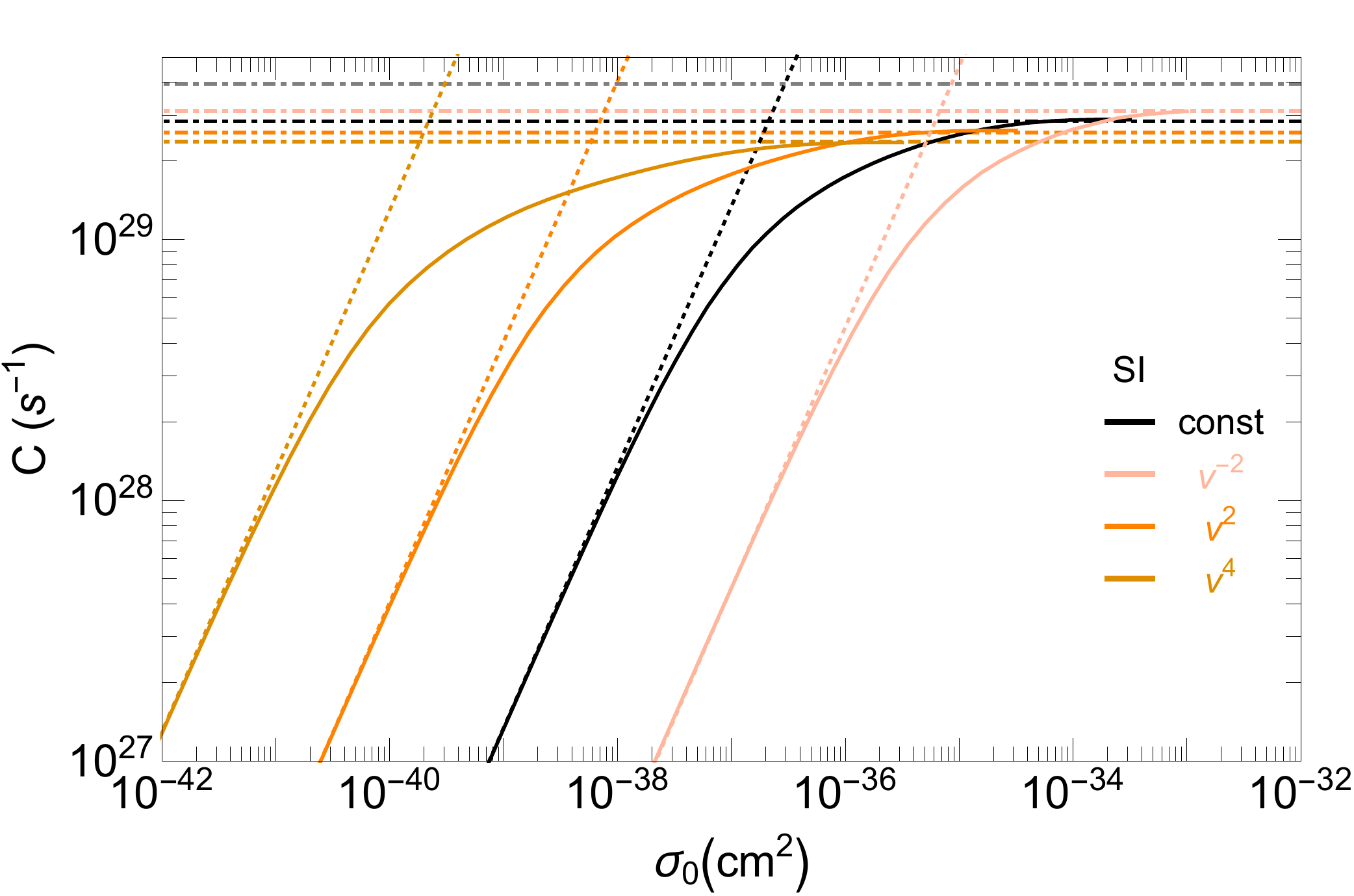}
    \includegraphics[width=0.45\textwidth]{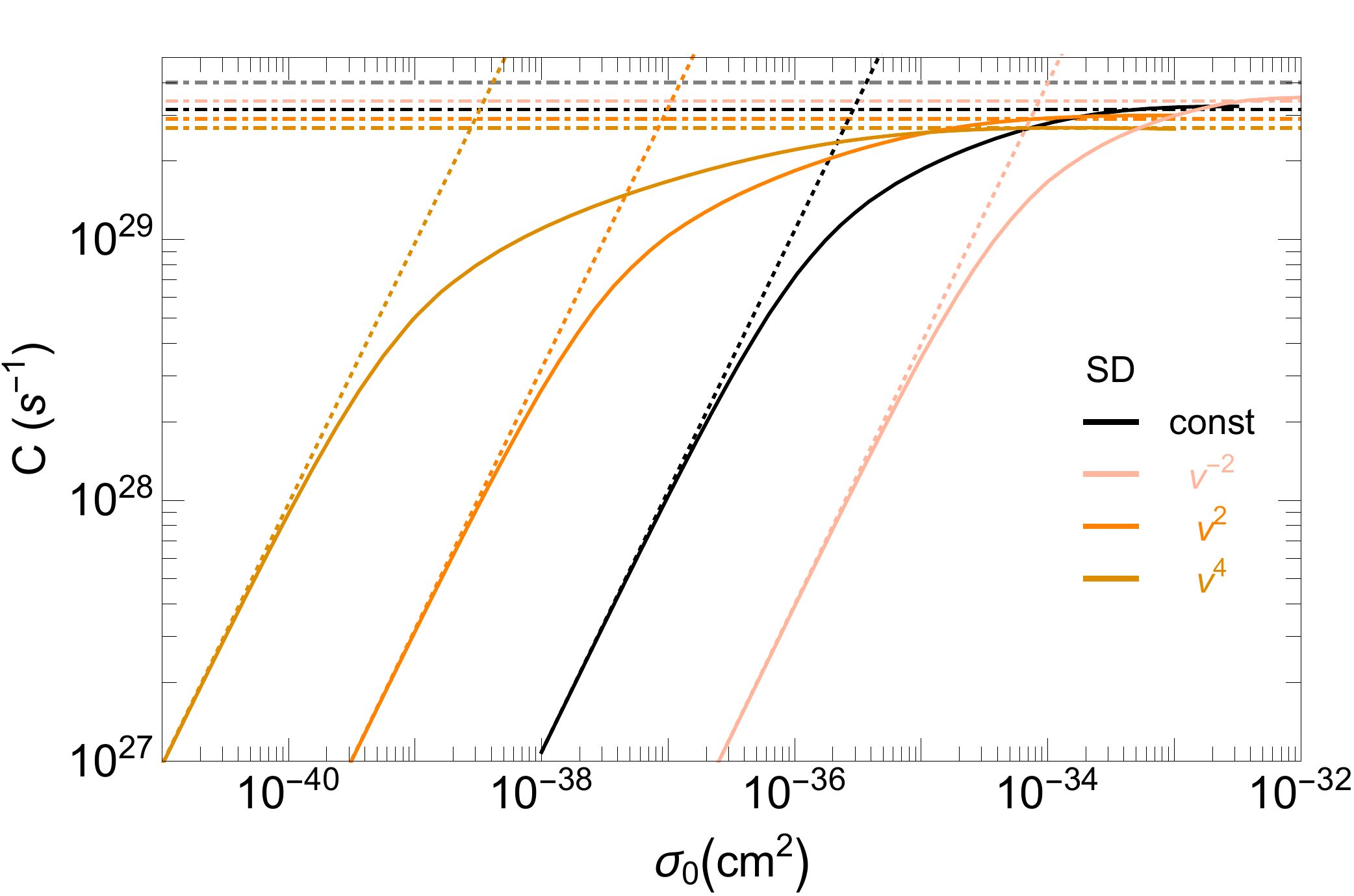}\\
    \includegraphics[width=0.45\textwidth]{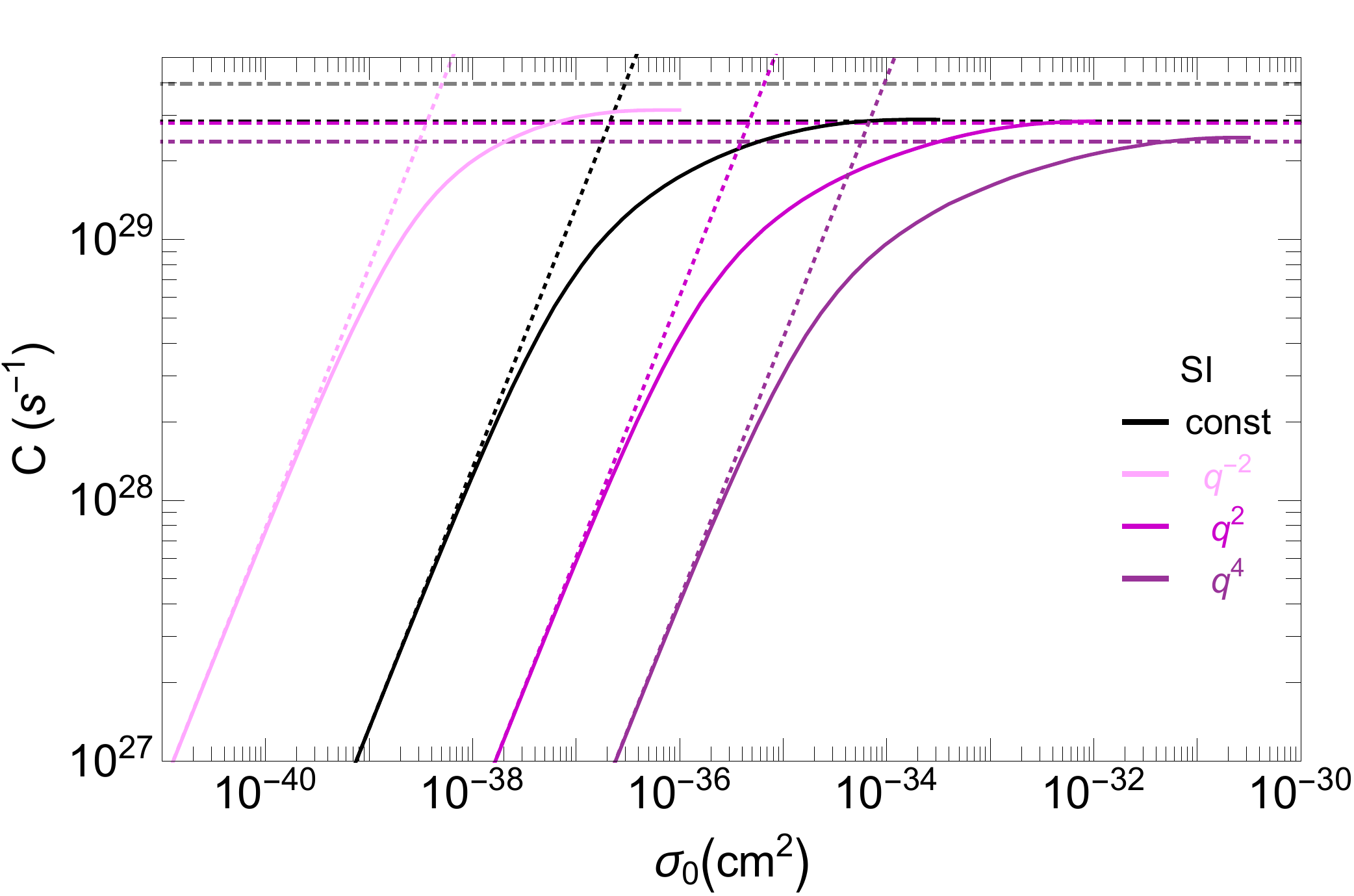}
    \includegraphics[width=0.45\textwidth]{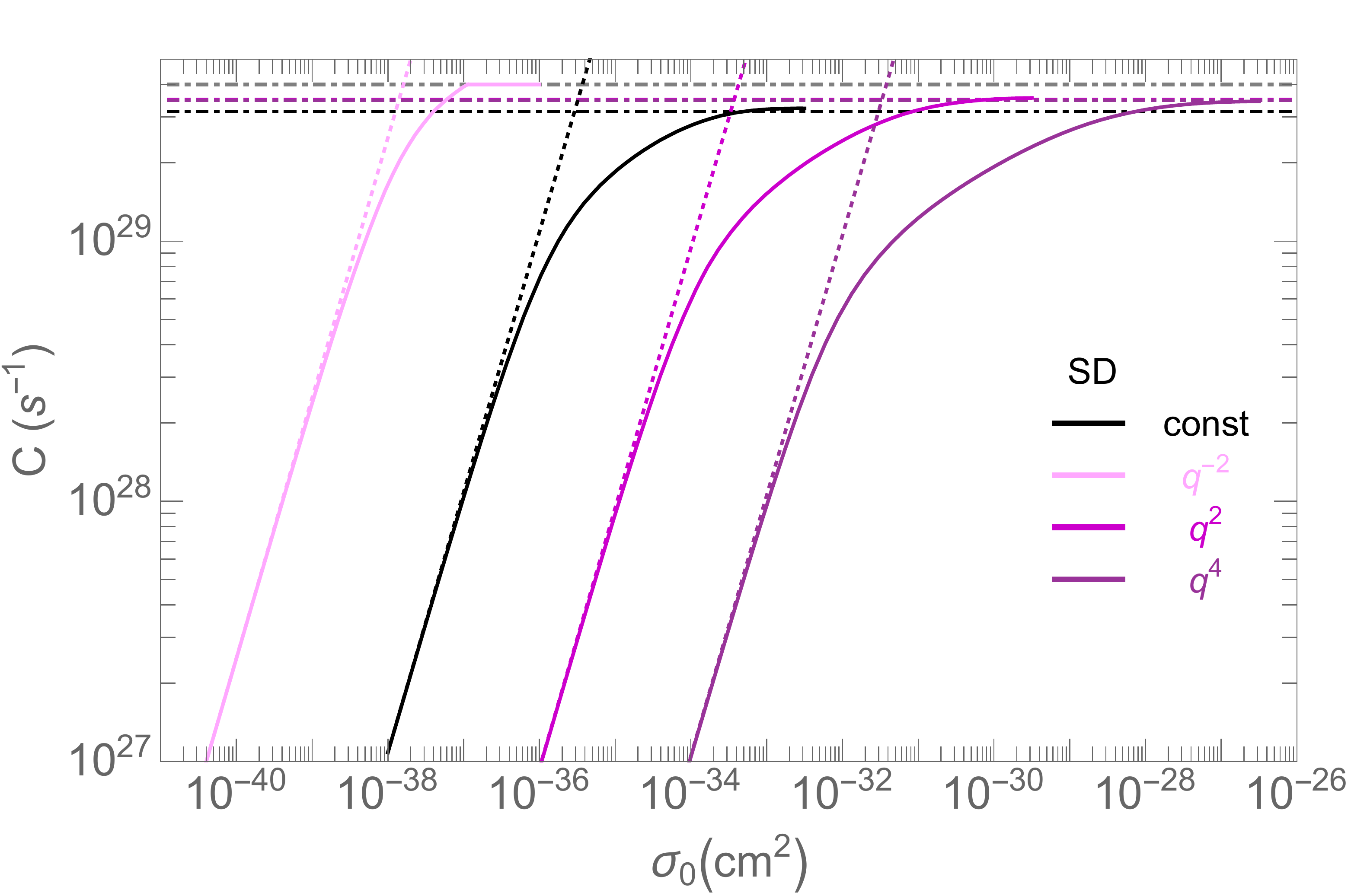}\\
    \caption{Capture rates as a function of the reference scattering cross-section $\sigma_0$, for $n=0,1,2,-1$, $m_\chi=2\GeV$.  The upper panels show $v_r^{2n}$ scattering, the lower ones show $q_{tr}^{2n}$ scattering, the left panels are for SI interactions and the right panels for SD. The solid lines indicate the result using the full treatment including the optical depth factors.  Dashed lines are the standard capture rate, which uses the optically thin approximation, and the dotted lines indicate the respective saturation limits.  The dot-dash grey line indicates the absolute geometric upper limit.  Percent-level discrepancies between the limiting values of some capture curves and their corresponding saturation limits are due to slow numerical convergence, which becomes an issue in the limit where the calculation of the full capture rate is dominated by scattering in an infinitely thin shell near the solar surface.}
    \label{fig:capturethickv}
\end{figure}

\begin{figure}[p]
    \centering
    \includegraphics[width=0.45\textwidth]{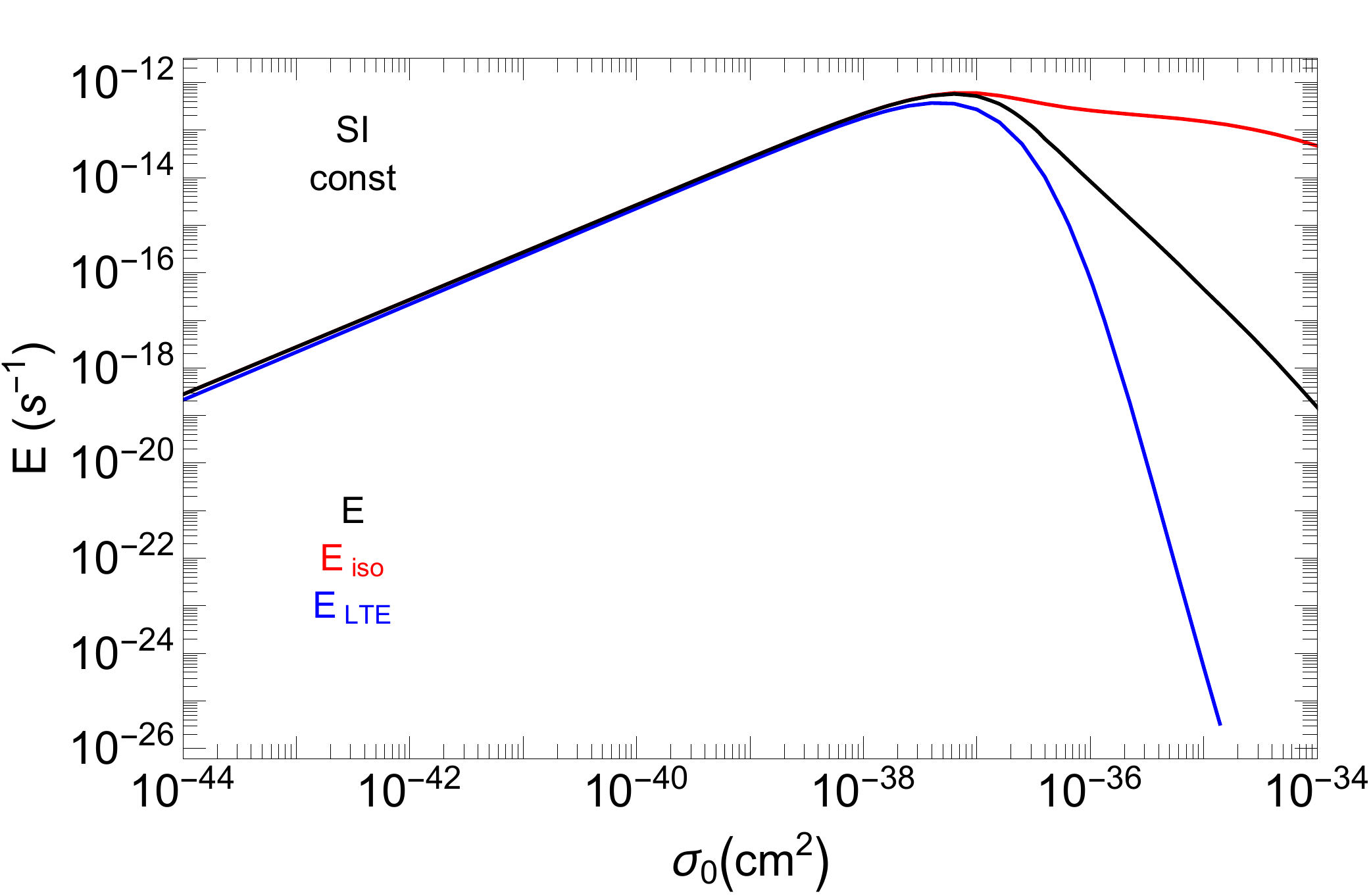}\\
    \includegraphics[width=0.45\textwidth]{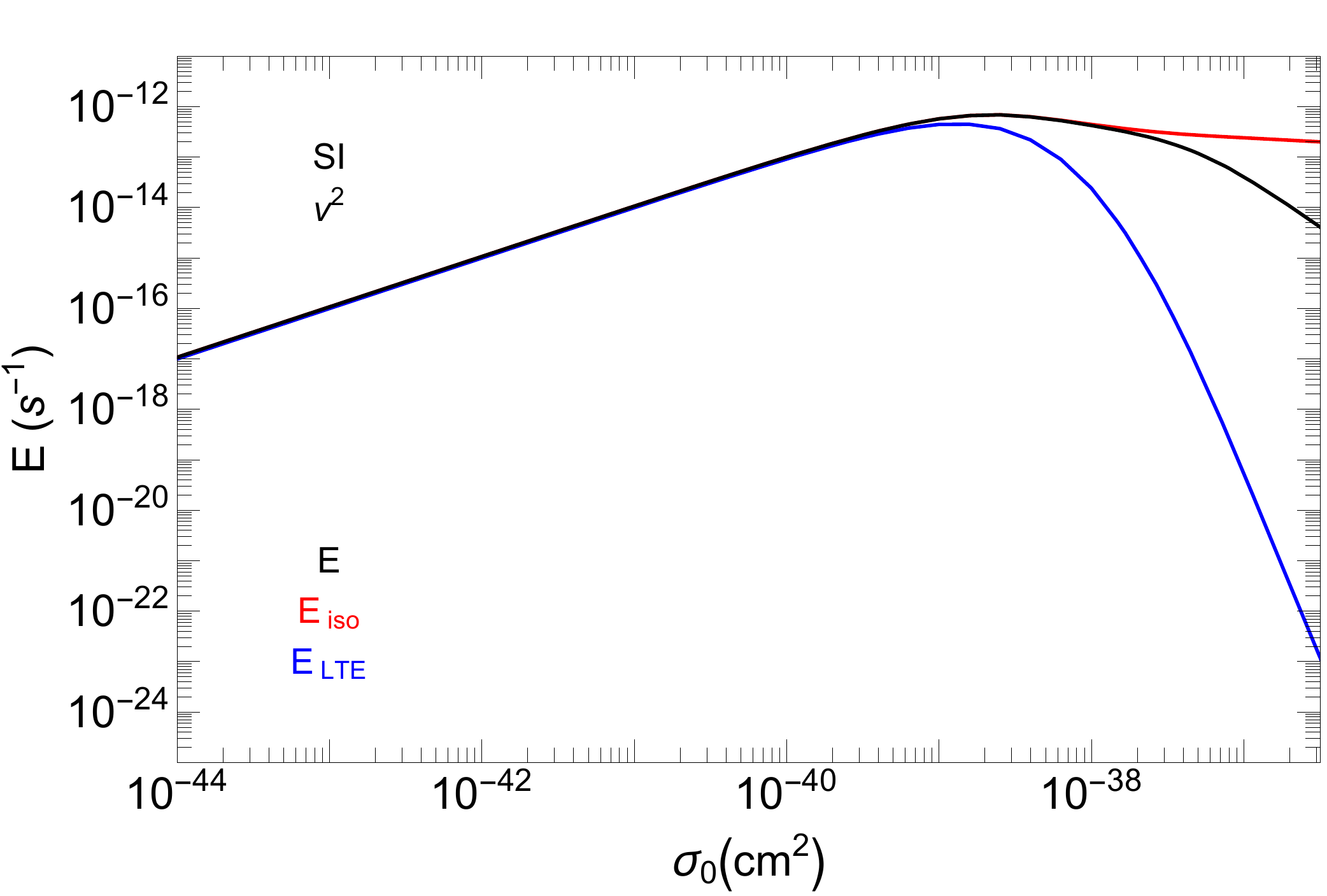}
    \includegraphics[width=0.45\textwidth]{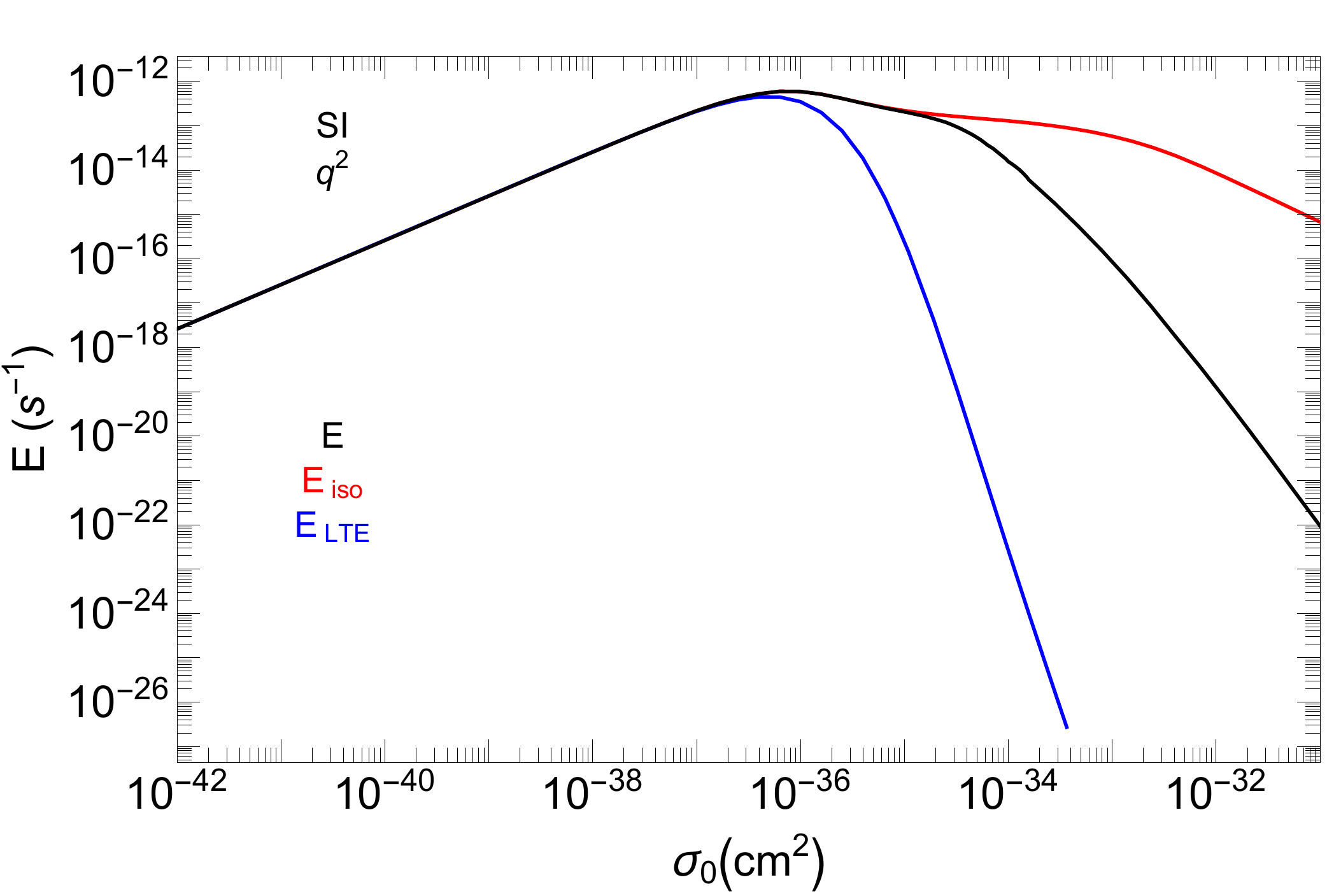}\\
    \includegraphics[width=0.45\textwidth]{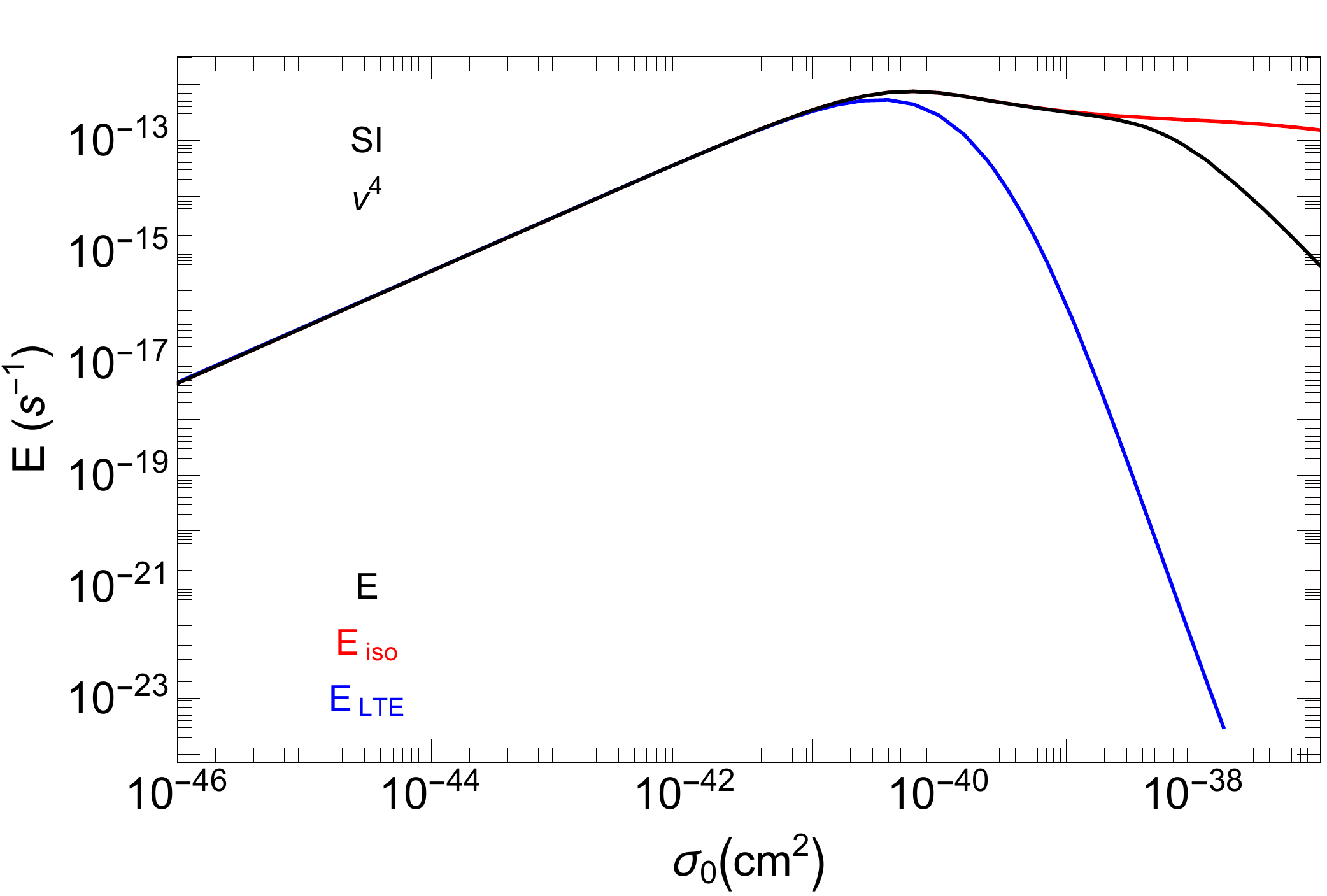}
    \includegraphics[width=0.45\textwidth]{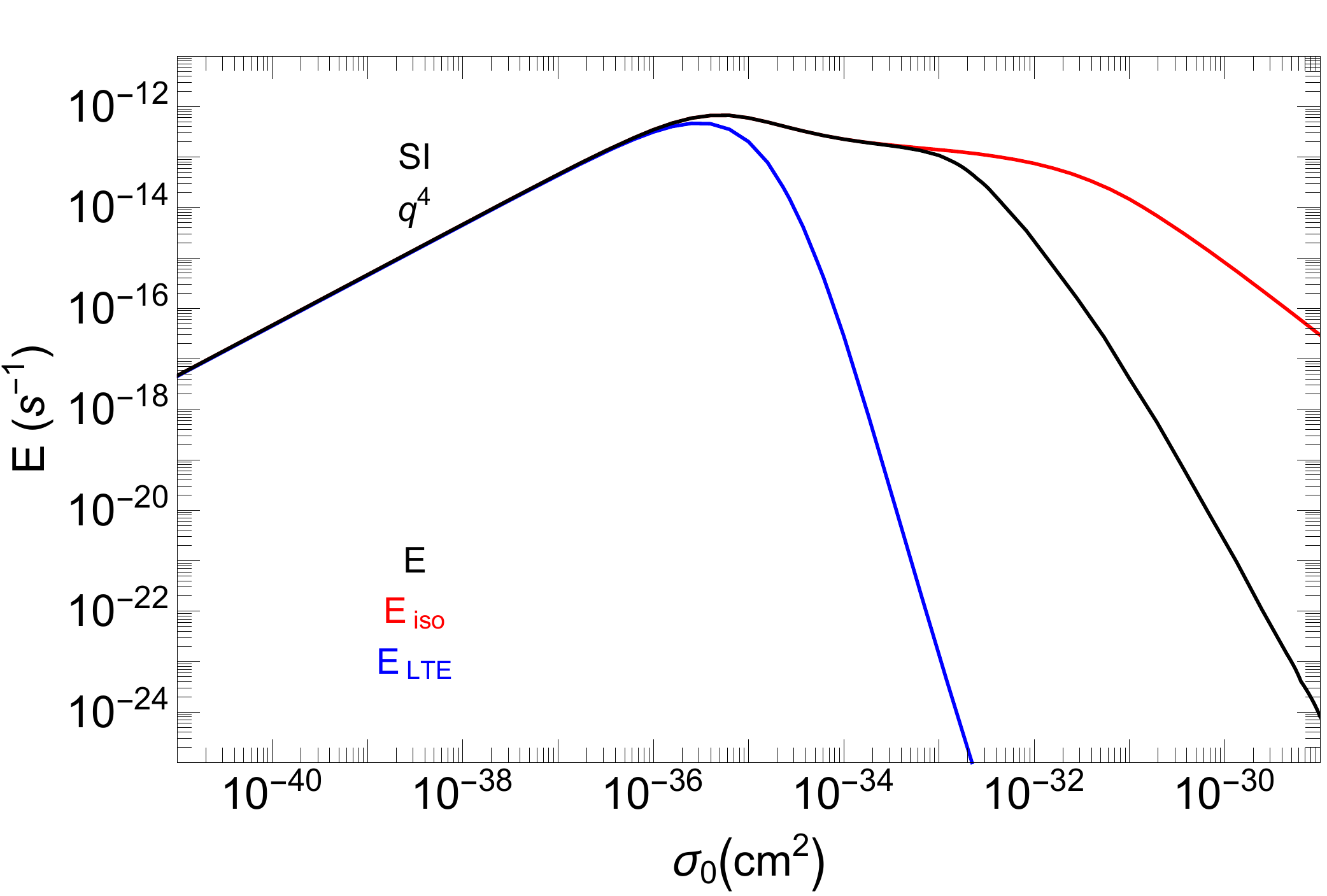}\\
    \includegraphics[width=0.45\textwidth]{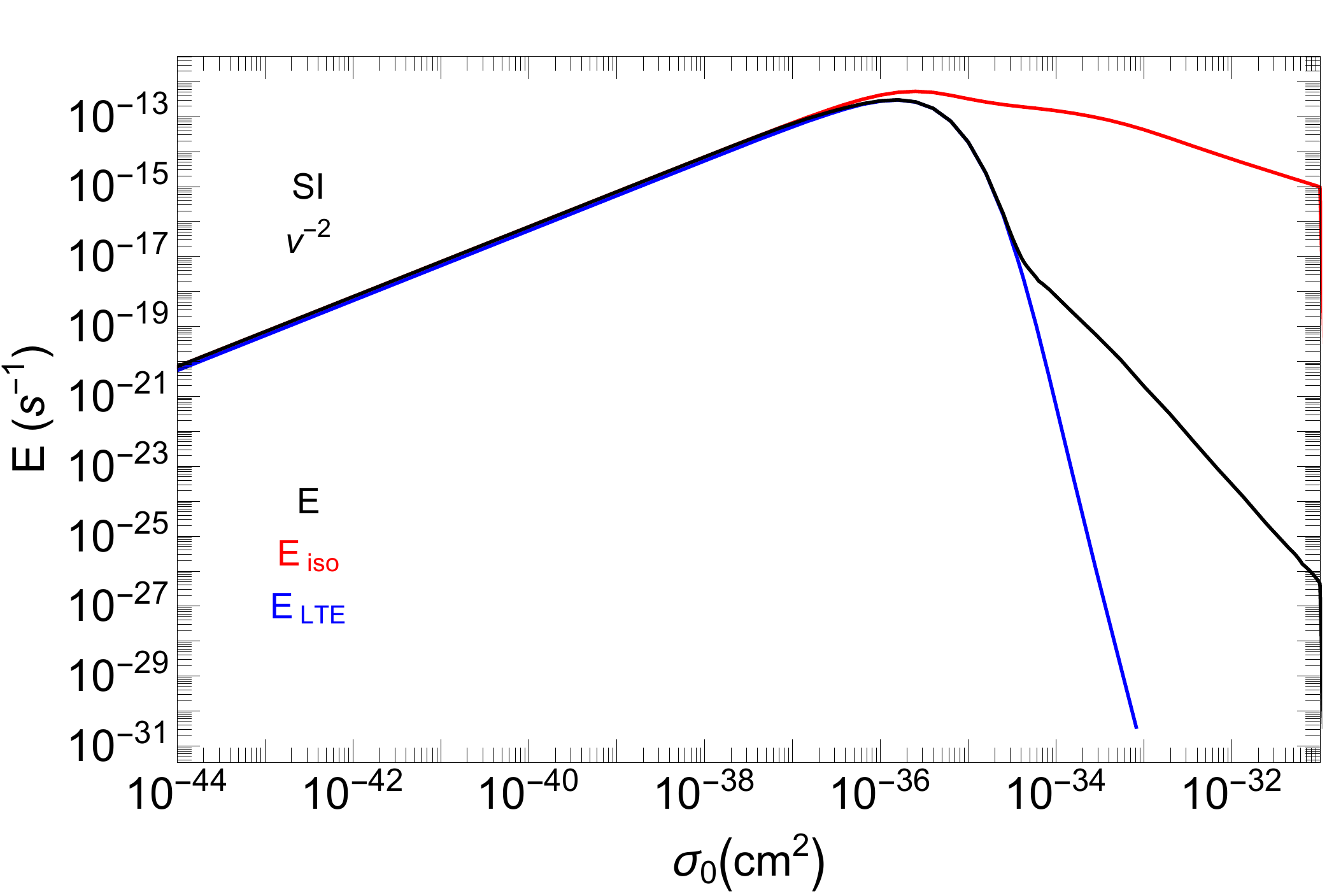}
    \includegraphics[width=0.45\textwidth]{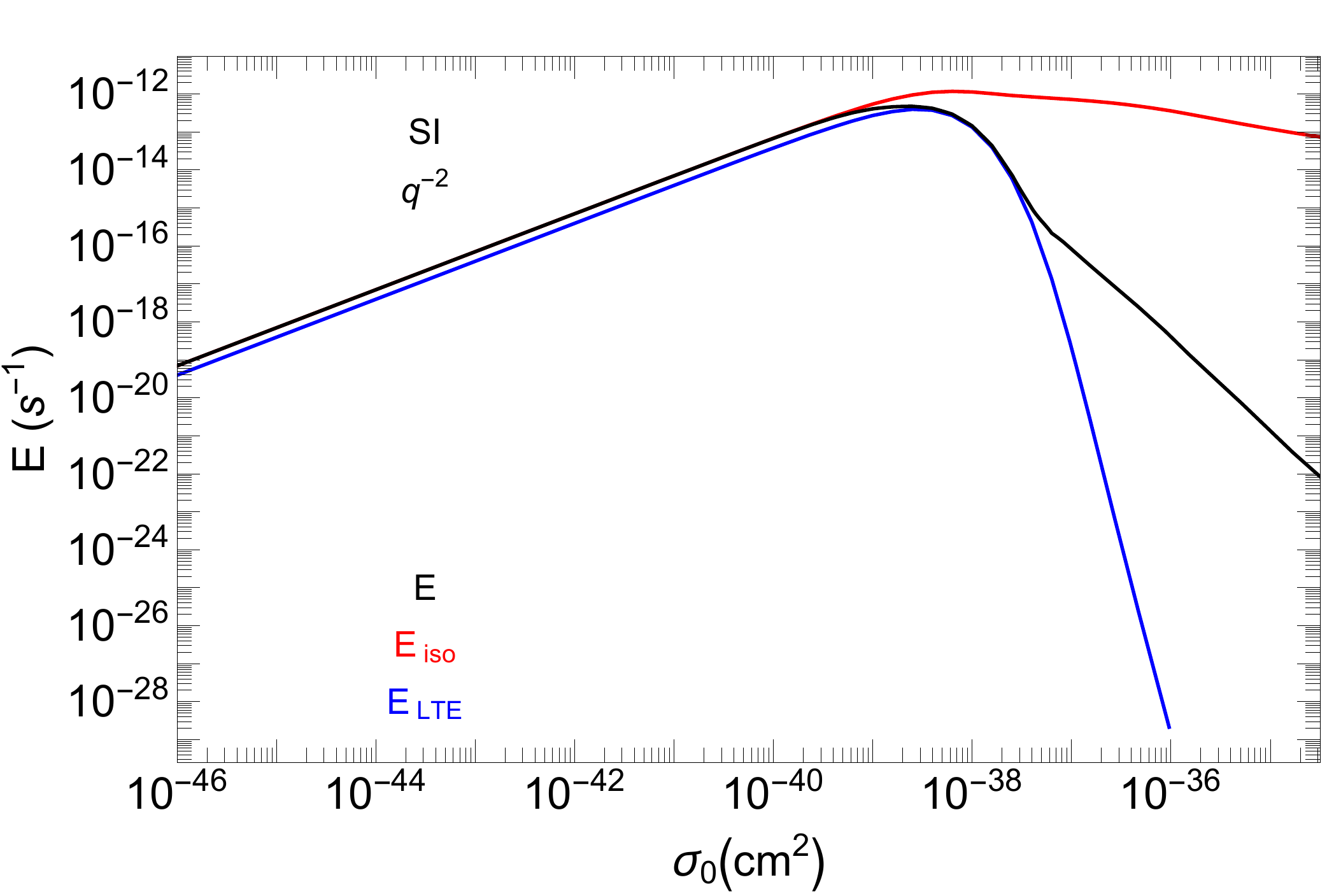}
    \caption{Evaporation rates of a 3\,GeV dark matter particle in the Sun, for different reference scattering cross-sections $\sigma_0$ and spin-independent (SI) dark matter interactions with nuclei.  The upper plot is for an SI cross-section that depends on neither momentum exchange nor relative velocity, left panels refer to $v_r$-dependent scattering, right panels refer to $q_{tr}$-dependent scattering, and the lower three rows are for $n=1,2$ and $-1$, respectively.  Curves show the evaporation rates that would follow if energy transport in the Sun were either purely local ($E_{\rm LTE}$) or purely Knudsen/isothermal ($E_{\rm iso}$), as well as the total weighted combination of the two ($E$). The evaporation rate can be seen to increase with increasing cross-section, until the gas becomes optically thick to scattering, beyond which evaporation is suppressed. }
    \label{fig:evapisolte}
\end{figure}

\begin{figure}[p]
    \centering
    \includegraphics[width=0.45\textwidth]{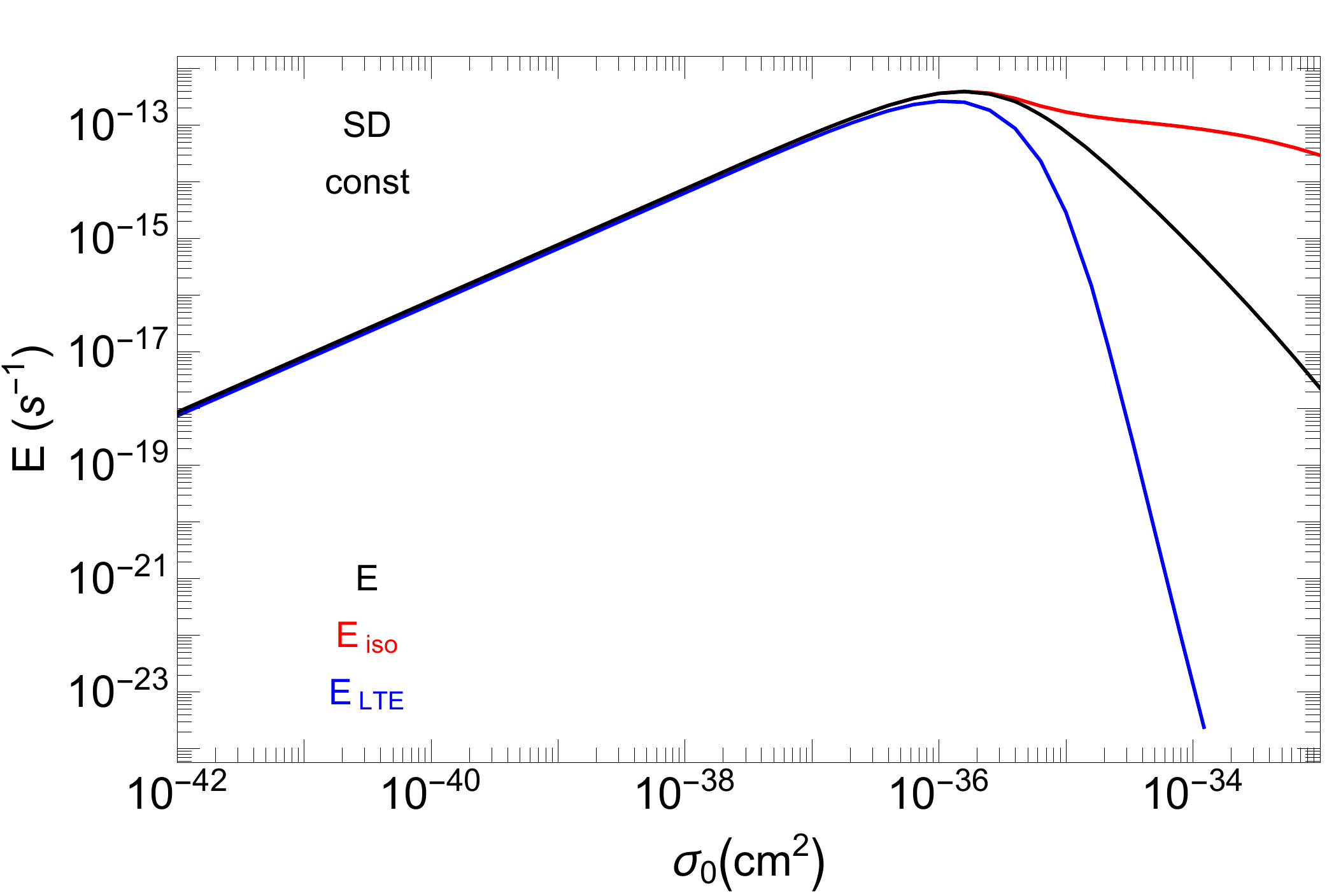}\\
    \includegraphics[width=0.45\textwidth]{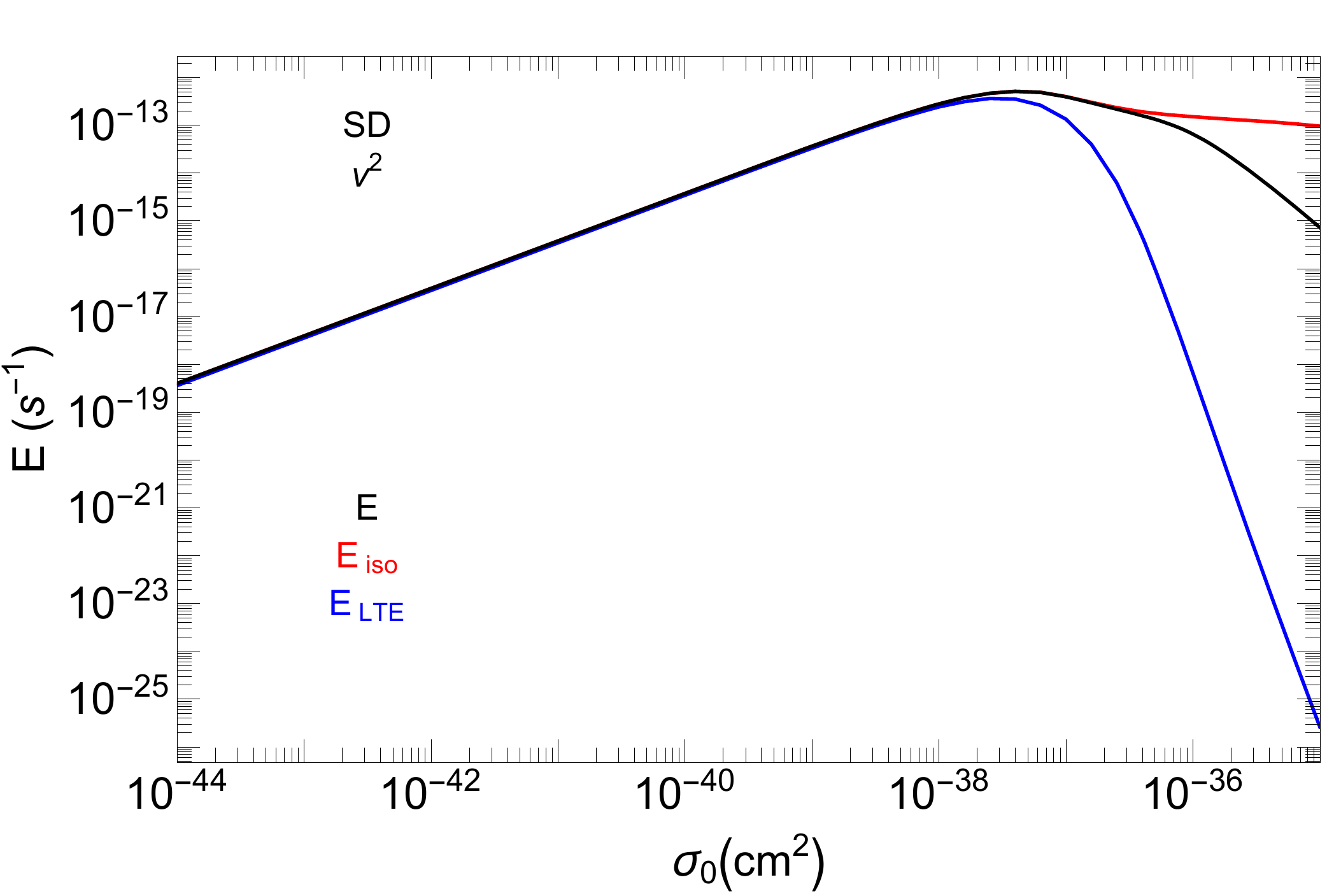}
    \includegraphics[width=0.45\textwidth]{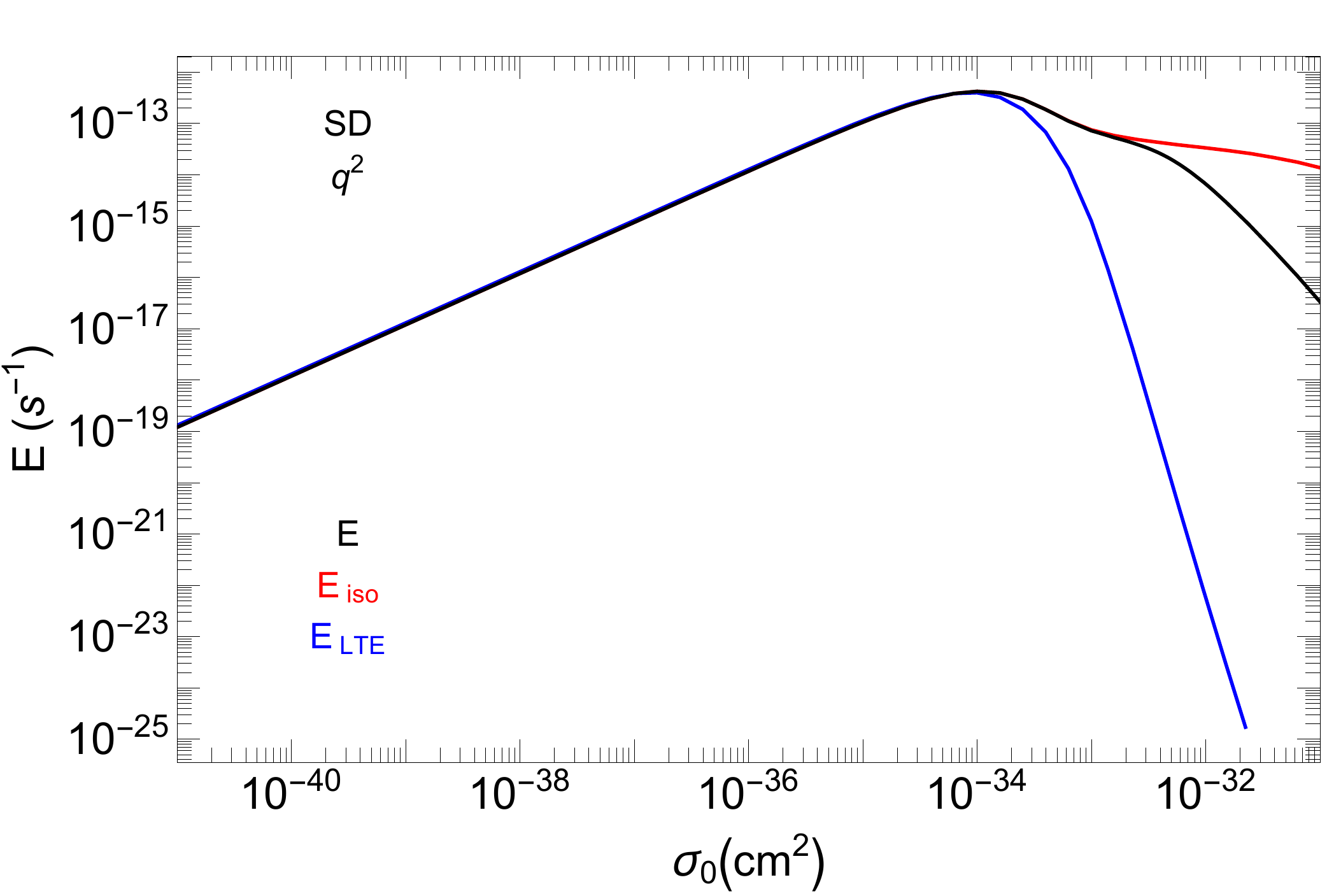}\\
    \includegraphics[width=0.45\textwidth]{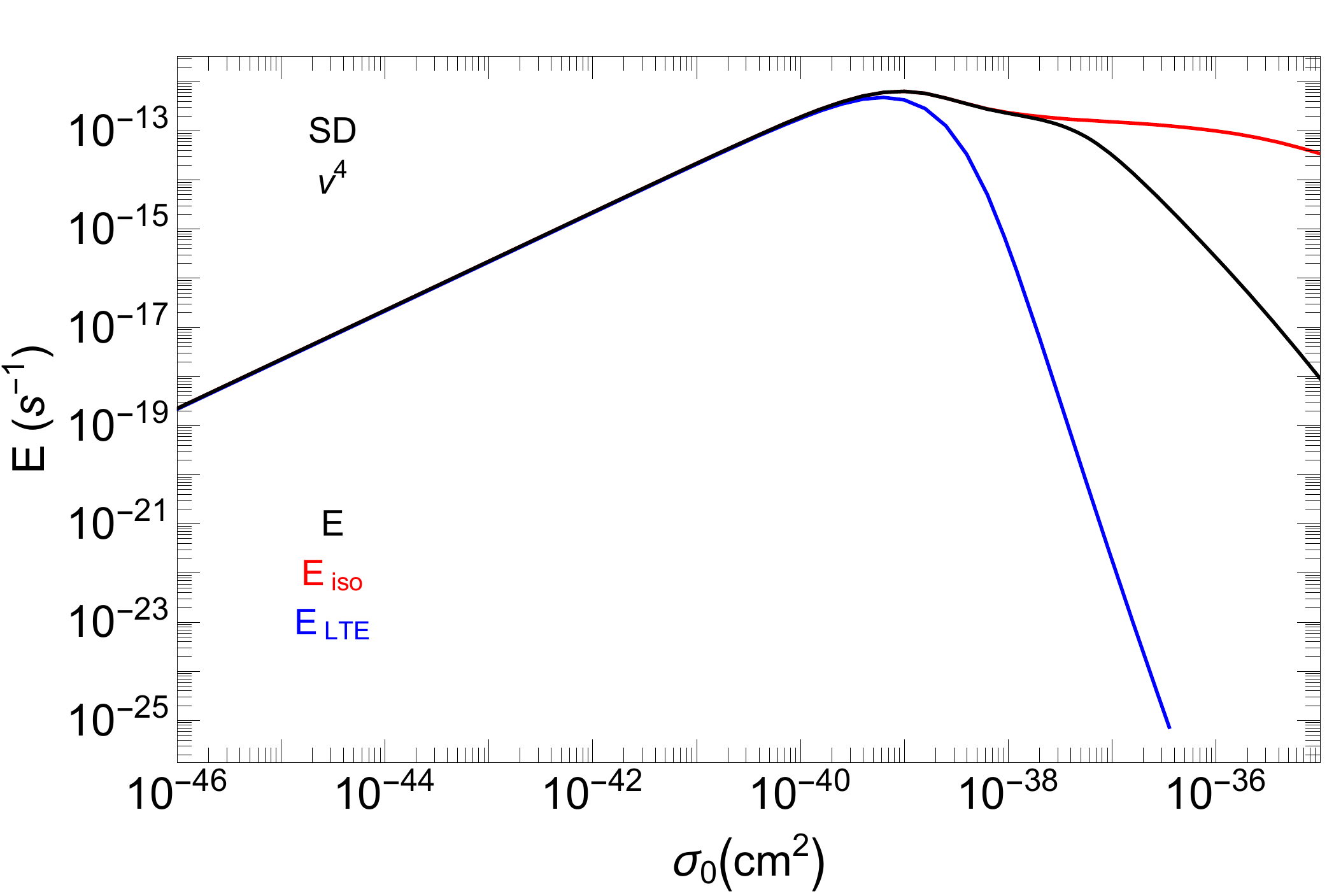}
    \includegraphics[width=0.45\textwidth]{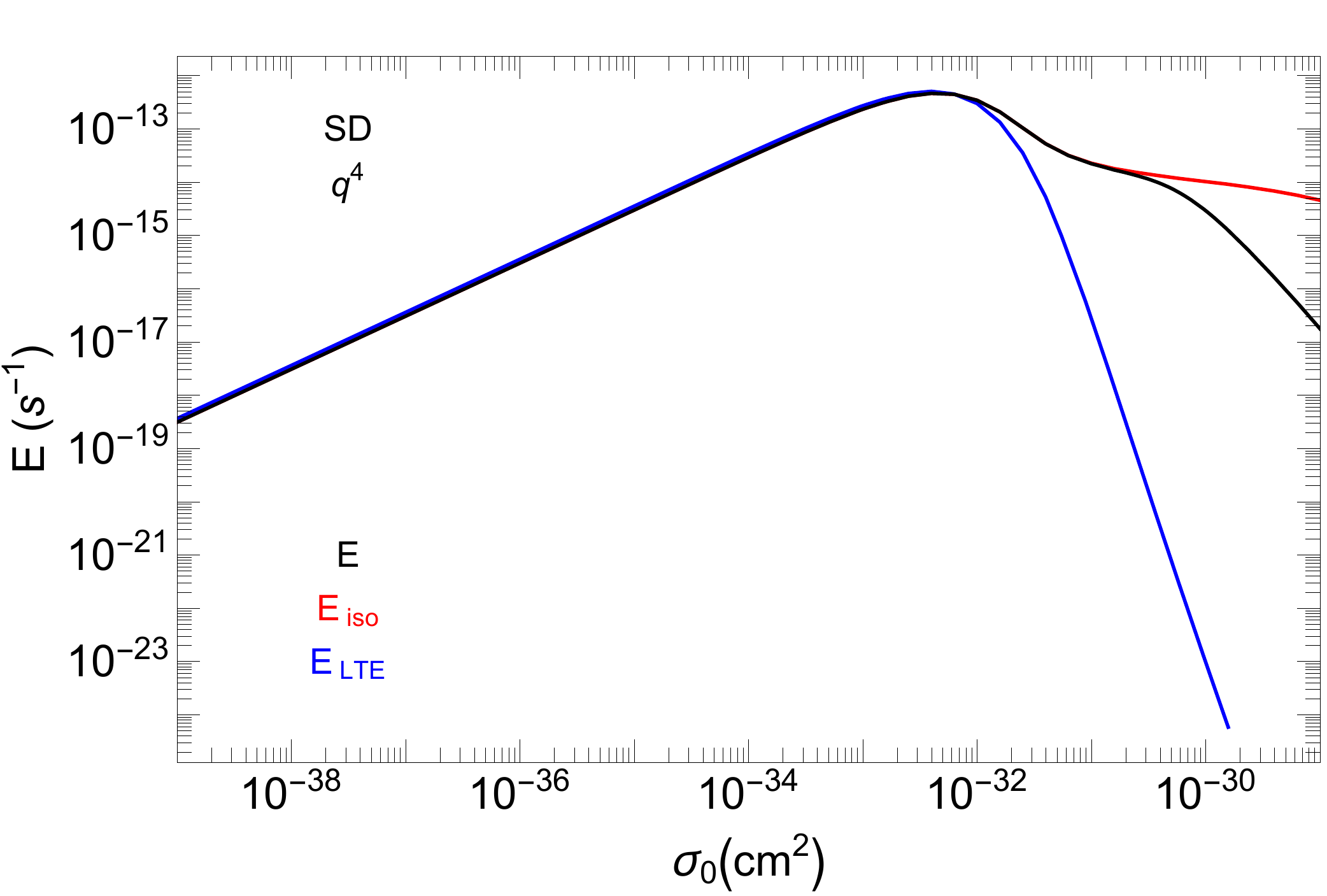}\\
    \includegraphics[width=0.45\textwidth]{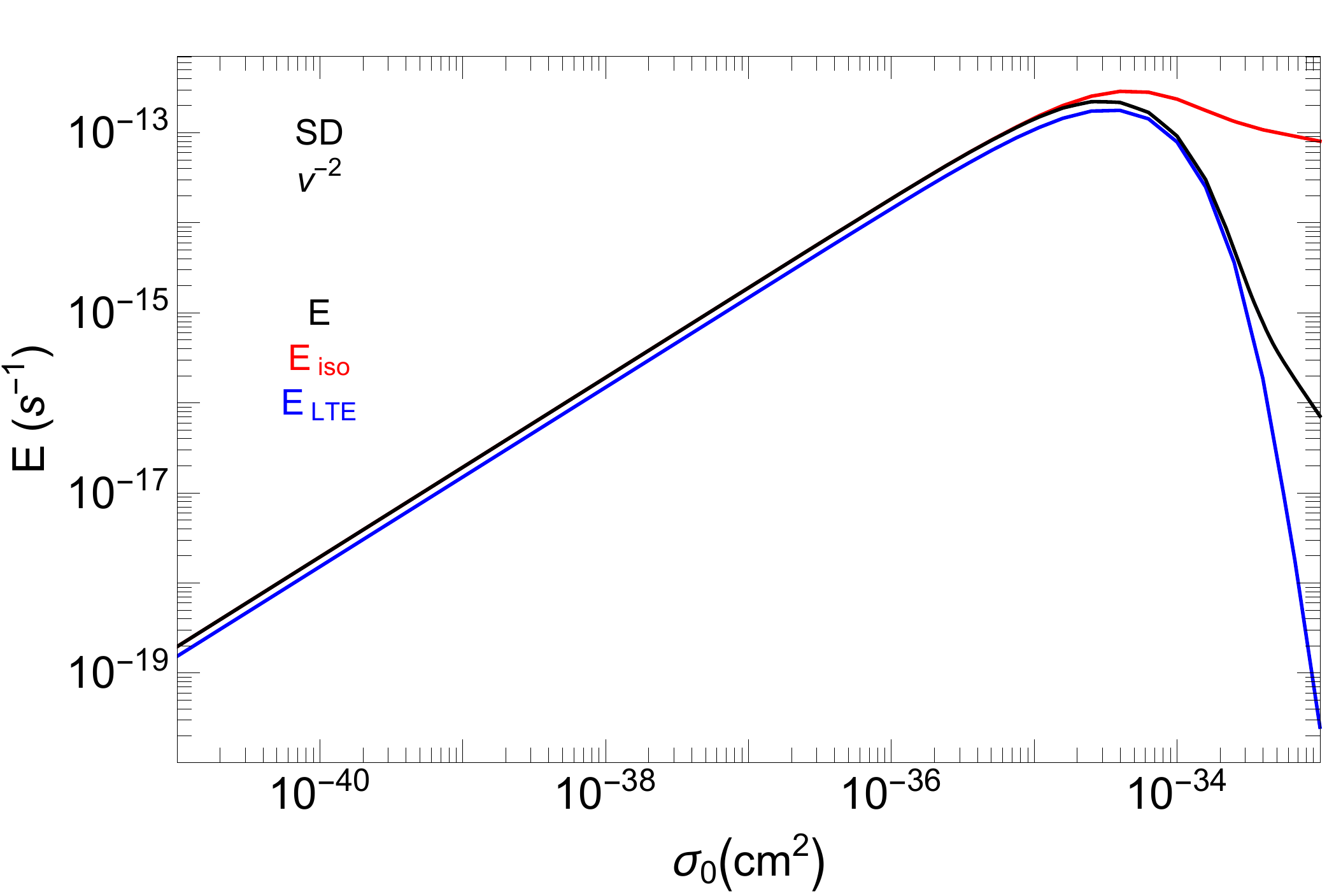}
    \includegraphics[width=0.45\textwidth]{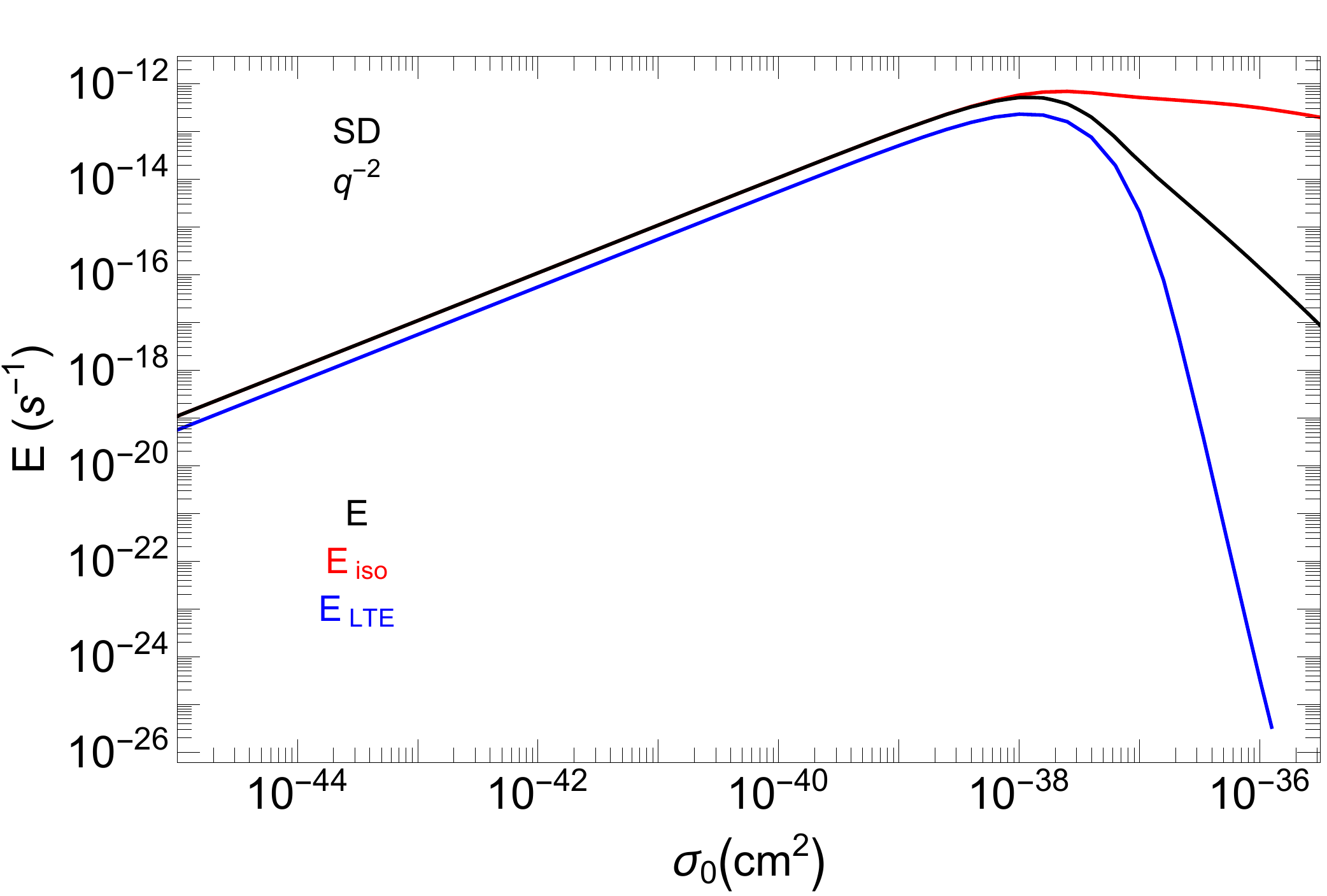}
    \caption{As per Fig.\ \protect\ref{fig:evapisolte}, but for spin-dependent interactions.}
    \label{fig:evapisolteSD}
\end{figure}

These expressions for the average cross-sections do not include the effects of nuclear form factors, which are relatively minor in the limit of small $\Lambda_i m_\chi/\mu_+$. This holds to an excellent approximation for lighter elements, and is still reasonable even for heavier elements.  Taking the example of an $m_\chi=5$\,GeV DM particle interacting via a spin-independent $q^2$ cross-section, the impacts of the form factors is only $\mathcal{O}(10\%)$.

\subsubsection{Optical depth for DM evaporation}
Turning to evaporation, the optical depth expresses the probability that a DM particle that has received a strong enough kick to evaporate does not interact again before escaping from the Sun. It is therefore necessary to use the speed distribution that the DM has \emph{after} receiving such a kick.  We therefore require the distribution of the velocities of the population of DM particles that are in the process of evaporating.  These will be the particles with velocities equal to or greater than the local escape velocity $v_e$.  In general this population will be completely dominated by those with velocities not much above $v_e$, as higher velocities will be progressively more Boltzmann suppressed.  We can therefore approximate the DM speed to be equal to $v_e$, while still assuming a Maxwell-Boltzmann distribution for the nuclei.

In fact, we can find the relative speed distribution for this case from the equivalent expression for DM capture without additional gravitational focussing (Eq.\ \ref{cap_distrib}), by treating the halo DM as having just a single speed, i.e.\ setting the width of the DM velocity distribution to zero, $v_d\rightarrow0$. In this case, when the width of the halo DM distribution is zero, the speed of DM in the frame of the Sun is simply $v_\odot$.  We can therefore set $v_\odot$ to the relevant DM velocity for evaporation, i.e.\ $v_\odot\rightarrow v_e(r)$, and immediately obtain the result for the distribution of relative speeds between nuclei and evaporating DM:
\be
f_{evap}(w_r)dw_r = \frac{w_r}{ v_e}\sqrt{\frac{m_{\chi}}{2\pi T\mu}}\left(e^{-\frac{m_{\chi}}{2T\mu}(w_r-v_e)^2}-e^{-\frac{m_{\chi}}{2T\mu}(w_r+v_e)^2}\right)dw_r.
\ee
The moments of this distribution are
\bea
\langle w_r^4    \rangle &=& v_e^4+10v_e^2 T\mu/m_{\chi} + 15 (T\mu/m_{\chi})^2\label{eq:vmomentevapn2},\\
\langle w_r^2    \rangle &=& v_e^2 + 3T\mu/m_{\chi}\label{eq:vmomentevapn1},\\
\langle w_r^{-2} \rangle &=& \frac{1}{v_e}\sqrt{\frac{\pi m_\chi}{2T\mu}}\mathrm{Erfi}\left(v_e\sqrt\frac{m_\chi}{2T\mu}\right)e^{-\frac{m_\chi v_e^2}{2T\mu}}\label{eq:vmomentevapnminus1},
\eea
and the average cross-sections can be obtained using Eqs.~(\ref{eq:sigmagen_v}) and (\ref{eq:sigmagen_q}).  Again, these expressions do not account for the effect of form factors in spin-independent interactions.
Note that all cross-sections are $r$-dependent as both $T$ and $v_e$ depend on $r$.

In Fig.\ \ref{fig:opttau} we give an example of our results for the optical depth $\tau$, while in Fig.\ \ref{fig:opteta} we plot $\eta$.  Both of these figures are for $n=0$; the results for other cases are qualitatively similar.

\section{Capture rate}
\label{sec:capture}

At this point we can derive an improved version of the traditional Gouldian capture rate \cite{Gould87b}, removing the original assumption that the Sun is optically thin to DM scattering.  For this, only the results (and assumptions) of Sec.\ \ref{sec:optdpt} are necessary.  Using the equations obtained in \ref{sec:capturecalc}, we can write the capture rate as
\be
C = \int_0^{R_\odot} dr 4\pi r^2 \eta(r) \int_0^\infty du \frac{w}{u}f_\odot^{cap}(u) \sum_i \Omega_i^-(w)\label{eq:capture},
\ee
where $f_\odot^{cap}(r)$ is the DM speed distribution in the reference frame of the Sun, normalised to the local DM number density
\be
f_{\odot}^{cap}(u) = \frac{\rho_\chi}{m_\chi} \lim_{T\rightarrow0} f_{cap}(u) =\frac{\rho_\chi}{m_\chi} \sqrt{\frac{3}{2\pi}}\frac{u}{v_\odot v_d}
\left[e^{-\frac{3 (u-v_\odot)^2}{2 v_d^2}}-
e^{-\frac{3 (u+v_\odot)^2}{2 v_d^2}}\right]\,.
\label{fvsun}
\ee

In Fig.\ \ref{fig:capturethickv} we show the resulting capture rate as a function of the cross-section and interaction type for a DM mass of 2\,GeV, comparing with both the traditional optically-thin calculation and the result obtained in the optically thick limit.  One can clearly see the smooth transition from the optically thin regime to the saturation limit.

Close inspection of Fig.\ \ref{fig:capturethickv} reveals that we have not plotted a saturation value for the $q^{-2}$ interaction. This is due to our use of the momentum-transfer cross-section for computing the thermal averages in the optical depth for $q^{-2}$ scattering, but not in the capture expression \eqref{eq:capture}. Indeed, the saturation limit relies on precise cancellation between the cross-sections in $\eta$ and $C$ in the large $\sigma$ limit, which is not possible for $q^{-2}$ interactions if the forward scattering divergence is to be removed. Instead, where the resulting capture rate would exceed the absolute geometric limit (as is the case for SD scattering), we have truncated it at this value.  Despite the slight inaccuracy introduced by this approximation and the use of the momentum-transfer cross-section, this procedure is more physically motivated than e.g. the smoothing function employed by Ref. \cite{Garani:2017jcj}, as it accounts for the saturation behaviour.

\section{Evaporation}
\label{sec:evaporation}

\subsection{Evaporation rate}
\label{sec:evaporationrate}

Using Eqs.~(\ref{eq:dmsundistrib}) and (\ref{eq:omegap}) from Appendix \ref{sec:evapcalc}, the evaporation rate is
\be
E = \int_0^{R_\odot} dr 4\pi r^2 \eta(r) \int_0^{v_e(r)} dw f_\odot^{evap}(w) \sum_i \Omega_i^+(w),
\ee
where $\Omega^{+}(w)$ is defined in Eqs.~(\ref{eq:omegap}) and (\ref{eq:Rp}):
\begin{eqnarray}
\Omega^{+}(w) &=& \int_{v_e}^\infty R^{+}(w\rightarrow v)  \mathrm{d} v, \\
R^{+}(w\rightarrow v) &=& \int_0^\infty \mathrm{d} s \int_0^\infty \mathrm{d} t \frac{32 \mu_{+}^4}{\sqrt{\pi}}k^3 n_i \frac{d\sigma_i}{d\cos\theta}(s,t,v,w) \frac{v t}{w} e^{-k^2u^2} |F_i(q_{tr})|^2\nonumber\\
&&\times\Theta(t+s-v)\Theta(w-|t-s|).
\end{eqnarray}
Due to the linearity of the evaporation rate in $f_\odot^{evap}(v)$, we can rewrite it as
\be
E = \mathfrak{f}(K) E_{\rm LTE} + \left[1-\mathfrak{f}(K)\right] E_{\rm iso},
\ee
and the two contributions can be calculated separately.  The factor $\frac{d\sigma}{d\cos\theta}$ inside $R^+$ should be expressed using Eqs.~(\ref{eq:sigmavrel}), (\ref{eq:sigmaqtr}), (\ref{vrel}) and (\ref{eq:qtravevap}).

\afterpage{\clearpage}

The resulting evaporation rates can be seen for SI interactions in Fig.\ \ref{fig:evapisolte}, and for SD interactions in Fig.\ \ref{fig:evapisolteSD}.  Here we show not only the total evaporation rate, but the individual rates $E_{\rm LTE}$ and $E_{\rm iso}$.  The evaporation rate shows the expected increase with scattering cross-section in the optically thin regime, before turning over at larger cross-sections due to the impacts of the optical depth.  In this region, dark matter can only effectively evaporate from the surface layers of the Sun, as particles up-scattered to greater than the local escape velocity deeper in the Sun nevertheless collide again and redeposit their energy in the outer layers before they can complete their escape.  It can also be seen from this figure that properly accounting for the degree of locality of energy transport in the Sun is crucial for obtaining an accurate description of evaporation in the optically thick regime.

\subsection{Effects of evaporation}
\label{sec:dmnumber}
\begin{figure}[tp]
    \centering
    \includegraphics[width=1.\textwidth]{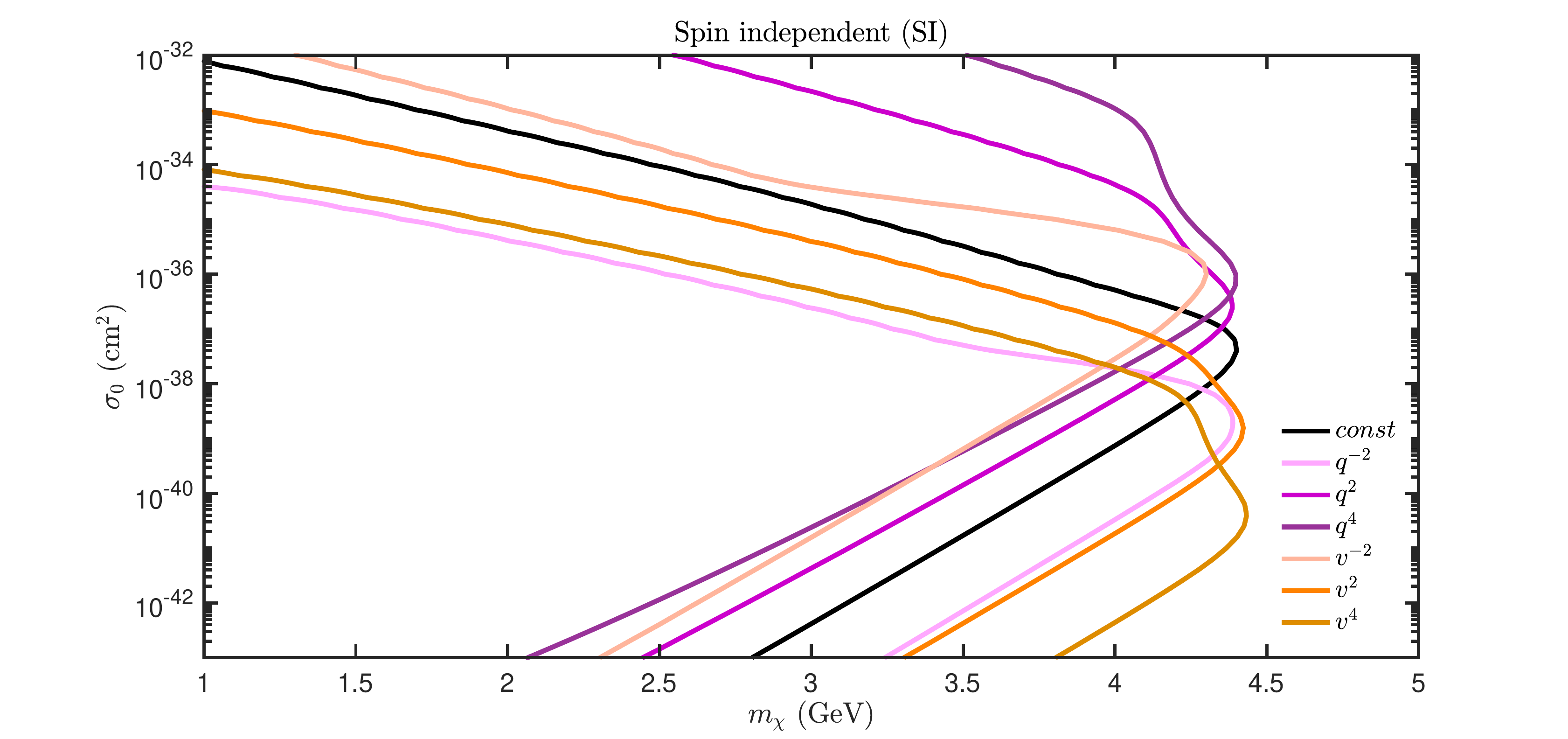}\\
    \includegraphics[width=1.\textwidth]{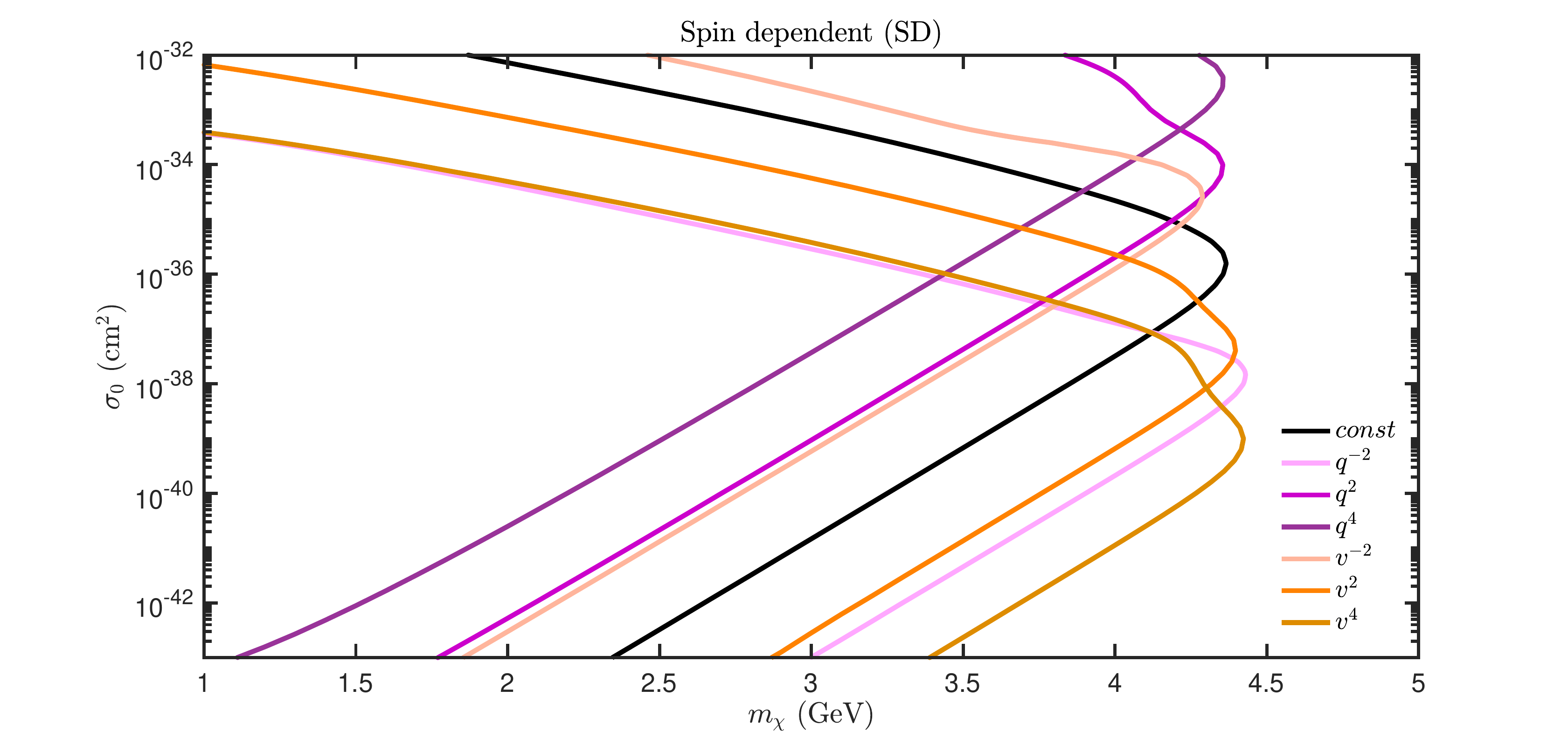}
    \caption{Regions of parameter space where the dark matter population in the Sun is depleted by more than $50\%$ due to evaporation, for different spin-independent (SI, upper panel) and spin-dependent (SD, lower panel) interactions.}
    \label{fig:dmnumb}
\end{figure}

To determine the DM population of the Sun, one must solve the equation
\be
\frac{dN}{dt} = C - E N - A N^2,
\ee
where $C$ is the capture rate, $E$ the evaporation rate and $A$ the annihilation rate. To get a rough idea of the impacts of evaporation on the DM population in the Sun for asymmetric dark matter, we can set  $A=0$ and assume that $C$ and $E$ are approximately constant for most of the life of the Sun, giving
\be
N(t_\odot) = \frac{C}{E}\left(1-e^{-E t_\odot}\right) = C t_\odot \left(\frac{1-e^{-E t_\odot}}{E t_\odot}\right).
\ee
When evaporation is negligible this simplifies to
\be
N(t_\odot) \rightarrow C t_\odot.
\ee
The factor that quantifies the importance of evaporation is thus
\be
\frac{1-e^{-E t_\odot}}{E t_\odot}.
\ee
In Fig.\ \ref{fig:dmnumb} we show for each interaction the regions of parameter space in $\sigma_0$ and $m_\chi$ where this factor is less than $0.5$, i.e.\ where evaporation depletes the solar DM population by at least 50\%.  Because the specific value of the cross-section at which evaporation passes from the optically thin to the optically thick regime varies with mass, as does the maximal evaporation rate at the transition point, we see that the effective evaporation mass for different cross-sections can be anywhere between 1 and 4\,GeV, depending on the interaction and its strength.

\afterpage{\clearpage}

This has important implications for the impact of $v_r$-dependent and $q_{tr}$-dependent scattering on helioseismology and solar neutrino rates, and by extension, their prospects for solving the Solar Abundance Problem.  In particular, these mass and interaction ranges encompass some but not all of the best-fit regions found in Ref.\ \cite{Vincent16}.  Some of these models will therefore become poorer fits, as the impacts of DM will be reduced. In some cases the effect will be only mild, at the 10--20\% level, but in others it will be quite substantial.  In other cases however, where such low masses were excluded in Ref.\ \cite{Vincent16} due to overly \textit{large} impacts of captured DM, the fits are expected to improve markedly, possibly revealing ever better solutions than seen previously.  Specifically, a substantial part of parameter space for the $v^{-2}$ and $v^{2}$ spin-dependent models can produce very large effects on solar observables while remaining compatible with direct detection experiments. A reduction in the DM population due to evaporation should bring these effects in line with helioseismological observations; full implementation of our results into solar simulations will quantify this impact.

\section{Conclusions}
\label{sec:conclusions}

In this paper we have derived the detailed expressions necessary for calculating the rate at which captured dark matter with momentum- and velocity-dependent couplings evaporates from the Sun.  For the first time, we have also included a full treatment of thermal effects and the transition to the optically thick regime in the evaporation expressions, and treated both spin-independent and spin-dependent interactions.  We have also developed an improved calculation of the capture rate for dark matter by the Sun and other stars, accounting for the effect of a finite optical depth.

Mirroring the manner in which conductive energy transport is maximised at the transition from local to non-local transport, we find that evaporation is most efficient at cross-sections intermediate between the optically thick and thin regimes.  At lower cross-sections, evaporation is reduced due to the reduced scattering rate; at higher cross-sections, would-be escapees are foiled by subsequent re-scattering in the outer atmosphere before they can effect their final departure.  Depending on the interaction type and strength, dark matter lighter than a mass of between 1 and 4\,GeV can be significantly impacted by evaporation in the Sun.  These effects will modify some (but not all) of the best-fit points identified in previous investigations of solar physics in the context of momentum- and velocity-dependent dark matter models.  They will also modify other points that were previously found to give poor fits due to an overpopulation of dark matter, potentially leading to even better fits than observed previously.

\section*{Acknowledgements} PS is supported by STFC (ST/K00414X/1 and ST/N000838/1). ACV is supported by an Imperial College Junior Research Fellowship (JRF).

\appendix
\section{Geometric and saturation limits}
\label{sec:geomlimit}
\subsection{From optical depth to the saturation limit}
\label{sec:geomlimitproof}

Here we show how, in the limit of large scattering cross-section, our treatment of the optical depth in the capture rate recovers the saturation limit.  We will demonstrate this for a constant number density $n_H$ of just one element (H). The case for many elements follows from the replacement $n_H \sigma_0\rightarrow \sum_i n_i \sigma_i$. The case of radially-varying number densities can be obtained by just replacing $n_i\rightarrow n_i(R_\odot)$, as we will see that in the high-$\sigma$ limit, the only contribution to the integral comes from the surface. Finally, for $v_r$ and $q_{tr}$-dependent cross-sections, one can replace $\sigma_0\rightarrow\langle\sigma\rangle(R_\odot)$.

For constant density, the integral in Eq.~(\ref{eq:opticdepth}) can be done analytically, giving
\be
\tau(r,z)=n_H\sigma_0\left(\sqrt{R_\odot^2-r^2(1-z^2)}-r z\right).
\ee
Let us now compute any volume integral of the form
\be
I = \int_0^{R_\odot} 4\pi r^2 f(r) n_H\sigma_0 \eta(r) dr.
\ee
From now on we will drop all indices.  In the high-$\sigma$ limit, the contribution from inside the Sun drops exponentially, as does the contribution for $z<0$. We can therefore rewrite the integral as
\be
I = \int_{R_\odot(1-\epsilon)}^{R_\odot} 2\pi r^2 dr f(r) n\sigma \int_{0}^1 e^{-n\sigma\left(\sqrt{R_\odot^2-r^2(1-z^2)}-r z\right)}dz .
\ee
Changing variables from $r$ to $\varrho\equiv1-r/R_\odot$, exchanging the order of integration and bringing all terms outside the integral that do not vary exponentially around $r=R_\odot$,
\be
I = 2\pi R_\odot^3 f(R_\odot) n\sigma\int_{0}^1 dz\int_0^\epsilon e^{-n\sigma\left(R_\odot\sqrt{1-(1-\varrho)^2(1-z^2)}-r z\right)}d\varrho .
\ee
Now we expand the exponent as a series in $\varrho$ around $\varrho=0$, giving
\be
R_\odot\sqrt{1-(1-\varrho)^2(1-z^2)}-(1-\varrho) R_\odot z = R_\odot\frac{\varrho}{z} + \mathcal{O}(\varrho^2).
\ee
Integrating in $\varrho$,
\be
I = \frac{2\pi R_\odot^3 f(R) n\sigma}{n\sigma R_\odot}\int_{0}^1 z\left(1-e^{-nR_\odot\sigma\epsilon/z}\right)dz.
\ee
It is now safe to take the limit $\sigma\rightarrow\infty$
\be
2\pi R_\odot^2 f(R_\odot)\int_{0}^1 z dz = \pi R_\odot^2 f(R_\odot),
\ee
so that the final result is
\be
\lim_{\sigma_0\rightarrow\infty}\int_0^{R_\odot} 4\pi r^2 f(r) n_H\sigma_0 \eta(r) dr  = \pi R_\odot^2 f(R_\odot)\label{eq:glim},
\ee
corresponding to an effective cross-section equal to the saturation limit $\pi R_\odot^2$.

\subsection{Capture rate and geometric/saturation limits}
\label{sec:geomlimitformulae}
Using the result of the previous section, and drawing on those of Appendix \ref{sec:calcdetails}, we now calculate the geometric and saturation limits for the capture rate. We define the geometric limit as the maximum value that the capture rate can have from pure geometrical considerations.  In contrast, the saturation limit is the limiting value of the capture rate for large cross-sections, which is necessarily less than or equal to the geometric limit. The saturation limit differs from the geometric limit by the kinematic probability that a DM particle interacts with the Sun but is not captured, and instead bounces away.

We start by calculating the saturation limit, considering $n=0$.  Using Eqs.~(\ref{CSIHv}) and (\ref{CSIv}), together with the formula for the capture rate (Eq.\ \ref{eq:capture}), we get
\bea
C &=& \int_0^{R_\odot} dr 4\pi r^2 \eta(r) \int_0^\infty du \frac{w}{u}f_\odot^{cap}(u) \Bigg\{\frac{2\mu_{+}^2}{\mu w} n_H \sigma_0 \left(v_e^2-\frac{\mu_{-}^2}{\mu_{+}^2}w^2\right) \Theta\left(v_e^2-\frac{\mu_{-}^2}{\mu_{+}^2}w^2\right)  \nn \\
&+& \sum_{i\geq \mathrm{He}} \frac{6\mu_{+}^2}{m_{\chi}^2 \Lambda^2 w} n_i \sigma_i  \Bigg(e^{-\frac{m_{\chi}^2 \Lambda^2(w^2-v_e^2)}{3\mu}}-e^{-\frac{m_{\chi}^2\Lambda^2}{3\mu_{+}^2}w^2}\Bigg) \Theta\left(v_e^2-\frac{\mu_{-}^2}{\mu_{+}^2}w^2\right) \Bigg\}. \label{eq:geom1}
\eea
The saturation limit can then be obtained using Eq.~(\ref{eq:glim}), remembering that the total cross-section is $2\sigma_i$:
\bea
C &=& \frac{1}{\sum_i 2n_i \sigma_i} \pi R_\odot^2 \int_0^\infty du \frac{w}{u}f_\odot^{cap}(u) \left\{\rule{0cm}{8mm}\frac{2\mu_{+}^2}{\mu w} n_H \sigma_0  \left(v_e^2-\frac{\mu_{-}^2}{\mu_{+}^2}w^2\right) \Theta\left(v_e^2-\frac{\mu_{-}^2}{\mu_{+}^2}w^2\right) \right. \nn \\
&+& \left.\sum_{i\geq \mathrm{He}} \frac{6\mu_{+}^2}{m_{\chi}^2 \Lambda^2 w} n_i \sigma_i  \left(e^{-\frac{m_{\chi}^2 \Lambda^2(w^2-v_e^2)}{3\mu}}-e^{-\frac{m_{\chi}^2\Lambda^2}{3\mu_{+}^2}w^2}\right) \Theta\left(v_e^2-\frac{\mu_{-}^2}{\mu_{+}^2}w^2\right) \right\} \Bigg|_{r=R_\odot}.
\eea
We can now make some approximations. First, we assume that on the surface of the Sun, the only elements relevant for capture are the lighter ones. In particular, we will neglect the contribution of iron. This is not due to low number density, as that could be partially compensated for by the $A_i^2$ cross-section enhancement, but rather due to kinematics: for heavy elements the kinematic limit expressed by the $\Theta$ function in the integral is satisfied only for the tail of $f_{cap}$, giving a negligible contribution. Second, for the light elements that contribute one has
\be
m_\chi m_{i} \Lambda^2 w^2 \sim m_\chi m_{i} \Lambda^2 v_e^2 \ll 1\label{eq:limitapprox2}.
\ee
It is therefore safe to expand the exponentials, cancelling the form factors and obtaining for the helium contribution an expression similar to the hydrogen one:
\bea
C &=& \frac{\pi R_\odot^2}{\sum_{i}n_i \sigma_i} \int_0^\infty du \frac{1}{u}f_\odot^{cap}(u) \left[\sum_{i} \frac{\mu_{+}^2}{\mu} n_i \sigma_i \left(v_e^2-\frac{\mu_{-}^2}{\mu_{+}^2}w^2\right) \Theta\left(v_e^2-\frac{\mu_{-}^2}{\mu_{+}^2}w^2\right) \right]\nn\\
 &=&  \frac{\pi R_\odot^2}{\sum_{i}n_i \sigma_i} \sum_{i} \int_0^{\sqrt{\mu}v_e/\mu_{-}} du \frac{1}{u}f_\odot^{cap}(u) \left[ \frac{\mu_{+}^2}{\mu} n_i \sigma_i \left(v_e^2-\frac{\mu_{-}^2}{\mu_{+}^2}w^2 \right)\right].\label{eq:geolimitexact}
\eea
This integral can be calculated analytically, however the result is quite long, so we now  make the further approximation
\be
\frac{\sqrt{\mu}}{\mu_{-}} v_e(R_\odot) \gg v_\odot, v_d \label{eq:limitapprox}.
\ee
This approximation is valid for $m_\chi \lesssim 6$\,GeV for H and He. For other elements it is not always true; we will see in the next section that this may cause some small loss of precision, especially for higher values of $m_\chi$ or for $q_{tr}^{2n}$ cross-sections, where heavier elements with larger momentum transfers are more important.  We now obtain
\bea
C &=&  \frac{\pi R_\odot^2}{\sum_{i}n_i \sigma_i} \sum_{i} \int_0^\infty du \frac{1}{u}f_\odot^{cap}(u) \left[ \frac{\mu_{+}^2}{\mu} n_i\sigma_i \left(v_e^2-\frac{\mu_{-}^2}{\mu_{+}^2}w^2\right) \right] \label{eq:geolimitapprox}\\
 &=& \frac{\rho_\chi \pi R_\odot^2}{\sum_i n_i \sigma_i} \sum_{i} \frac{n_i \sigma_i}{v_\odot m_\chi} \left\{\left[v_e^2 -\frac{\mu_{-}^2}{\mu} \left(v_\odot^2 + \frac{v_d^2}{3}\right)\right]\erf\left({\sqrt{\frac{3}{2}}\frac{v_\odot}{v_d}}\right) - \sqrt{\frac{6}{\pi}} \frac{\mu_{-}^2}{3\mu} v_\odot v_d e^{-\frac{3v_\odot^2}{2v_d^2}}\right\}. \nn
\eea
This formula gives the saturation limit taking into account the probability, given by kinematics, that DM interacts but bounces away. To calculate the geometric limit, we instead assume that the capture probability is one (a `sticky hard sphere'), starting from Eq.~(\ref{eq:geom1}).  After substituting
\be
P_{cap} = \frac{\mu_+^2}{\mu w^2}\left(v_e^2-\frac{\mu_{-}^2}{\mu_+^2}w^2\right) \Theta\left(v_e^2-\frac{\mu_{-}^2}{\mu_+^2}w^2\right) \rightarrow 1\label{eq:ptoone},
\ee
and going through the same steps again, we find
\bea
C &=&  \frac{\pi R_\odot^2}{\sum_i n_i \sigma_i} \sum_{i} \int_0^\infty du \frac{1}{u}f_\odot^{cap}(u) n_i \sigma_i w^2 \nn \\
  &=& \pi R_\odot^2 \int_0^\infty du \frac{1}{u}f_\odot^{cap}(u) w^2 \nn \\
  &=& \frac{\rho_\chi \pi R_\odot^2}{3 v_\odot m_\chi} \left[\left(3 v_e^2 + 3 v_\odot^2 + v_d^2\right)\erf\left({\sqrt{\frac{3}{2}}\frac{v_\odot}{v_d}}\right) + \sqrt{\frac{6}{\pi}} v_\odot v_d e^{-\frac{3v_\odot^2}{2v_d^2}}\right].\label{eq:geomupperlim}
\eea
This result matches that of Refs.\ \citep{Vincent15,Vincent16}.  Note that without setting the capture probability to 1, one necessarily finds a dependence on both the element and its number density. Note also that the upper limit given by Eq.~(\ref{eq:geomupperlim}) can never be reached, as even in the best case $\mu_{-}=0$, one has
\be
P_{cap} = \frac{v_e^2}{w^2} = \frac{v_e^2}{v_e^2+u^2} < 1.
\ee
Moreover, the best case $\mu_{-}=0$ cannot be reached simultaneously for all elements. The approximate formula Eq.~(\ref{eq:geolimitapprox}) is a very good approximation for any $m_\chi \lesssim 10\GeV$.  The geometric limit Eq.~(\ref{eq:geomupperlim}) overestimates the saturation limit by $O(10\%)$.

Eq.~(\ref{eq:geolimitapprox}) can be interpreted in the following way: the capture rate at saturation is equal to the product of the DM flux $\rho_\chi w/m_\chi$, the geometric cross-section $\pi R_\odot^2$, the probability to interact with the element $i$, $P_{int,i}= n_i \sigma_i/\sum_j n_j \sigma_j$, and the capture probability with that element $P_{cap,i}=(v_e/w)^2\mu_+^2/\mu - \mu_{-}^2/\mu$, all summed over all elements and integrated over the velocity distribution.

\subsection{Saturation limits: expressions and checks}
\label{sec:capcheck}
The saturation limits for the various cases are
\bea
C_{v,sat}^n&=&\pi R_\odot^2\,\frac{\rho_\chi v_e^2}{v_\odot m_\chi}\,\frac{1}{\sum_i n_i \sigma_i\Upsilon_n\left(v_d^2+3\frac{T\mu}{m_\chi},v_\odot^2\right)} \nn\\
&&\times \sum_{i=H,He} n_i \sigma_i \left\{\left[P_n\left(\frac{v_d^2}{v_e^2},\frac{v_\odot^2}{v_e^2}\right) - \frac{\mu_{-}^2}{\mu} Q_n\left(\frac{v_d^2}{v_e^2},\frac{v_\odot^2}{v_e^2}\right)\right]\erf\left({\sqrt{\frac{3}{2}}\frac{v_\odot}{v_d}}\right)\right. \nn\\
&+& \left. \sqrt{\frac{6}{\pi}} \left[R_n\left(\frac{v_d^2}{v_e^2},\frac{v_\odot^2}{v_e^2}\right) - \frac{\mu_{-}^2}{\mu} S_n\left(\frac{v_d^2}{v_e^2},\frac{v_\odot^2}{v_e^2}\right)\right] \frac{v_\odot v_d}{v_e^2} e^{-\frac{3v_\odot^2}{2v_d^2}}\right\}\label{eq:satlimitv}\\
C_{q,sat}^n&=&\pi R_\odot^2\,\frac{\rho_\chi v_e^2}{v_\odot m_\chi}\,\frac{1}{\sum_i n_i \sigma_i \mu_+^{-2n} \Upsilon_n\left(v_d^2+3\frac{T\mu}{m_\chi},v_\odot^2\right)} \nn\\
&&\times \sum_{i=H,He} n_i \sigma_i \mu_+^{-2n} \left\{\left[\hat{P}_n\left(\frac{v_d^2}{v_e^2},\frac{v_\odot^2}{v_e^2}\right) - \frac{T_n(\mu)}{\mu^{n+1}} \hat{Q}_n\left(\frac{v_d^2}{v_e^2},\frac{v_\odot^2}{v_e^2}\right)\right]\erf\left({\sqrt{\frac{3}{2}}\frac{v_\odot}{v_d}}\right)\right. \nn\\
&+& \left. \sqrt{\frac{6}{\pi}} \left[\hat{R}_n\left(\frac{v_d^2}{v_e^2},\frac{v_\odot^2}{v_e^2}\right) - \frac{T_n(\mu)}{\mu^{n+1}} \hat{S}_n\left(\frac{v_d^2}{v_e^2},\frac{v_\odot^2}{v_e^2}\right)\right] \frac{v_\odot v_d}{v_e^2} e^{-\frac{3v_\odot^2}{2v_d^2}}\right\}.\label{eq:satlimitq}
\eea
As the surface temperature is much smaller than the mass, one can safely neglect it in these expressions.  Here the functions $P_n,Q_n,R_n,S_n$ are
\bea
P_0(y,z) &=& 1\\
P_1(y,z) &=& 1 + z + \frac{1}{3}y\\
P_2(y,z) &=& 1+\frac{2}{3} y+2 z+\frac{1}{3} y^2+2 y z+ z^2\\
P_{-1}(y,z) &=& \frac{\int_0^\infty dt \frac{1}{1+\frac{2}{3}y t^2}\left(e^{-\left(t-\sqrt{\frac{3z}{2y}}\right)^2}-e^{-\left(t+\sqrt{\frac{3z}{2y}}\right)^2}\right)}{\int_0^\infty dt \left(e^{-\left(t-\sqrt{\frac{3z}{2y}}\right)^2}-e^{-\left(t+\sqrt{\frac{3z}{2y}}\right)^2}\right)}
\eea
\bea
Q_0(y,z) &=& \frac{1}{3}y + z\\
Q_1(y,z) &=& \frac{1}{3}y + 2 yz +\frac{1}{3} y^2 + z + z^2\\
Q_2(y,z) &=& \frac{1}{3}y \left(1+ 12 z+15 z^2\right)+\frac{1}{3} y^2 (2 +15 z)+ z (1+z)^2+\frac{5}{9} y^3\\
Q_{-1}(y,z) &=& \frac{\int_0^\infty dt \frac{\frac{2}{3}y t^2}{1+\frac{2}{3}y t^2}\left(e^{-\left(t-\sqrt{\frac{3z}{2y}}\right)^2}-e^{-\left(t+\sqrt{\frac{3z}{2y}}\right)^2}\right)}{\int_0^\infty dt \left(e^{-\left(t-\sqrt{\frac{3z}{2y}}\right)^2}-e^{-\left(t+\sqrt{\frac{3z}{2y}}\right)^2}\right)}
\eea
\bea
R_0(y,z) &=& 0\\
R_1(y,z) &=& \frac{1}{3}\\
R_2(y,z) &=& \frac{2}{3}  + \frac{5}{9} y+\frac{1}{3} z
\eea
\bea
S_0(y,z) &=& \frac{1}{3}\\
S_1(y,z) &=& \frac{1}{3}+\frac{5}{9}y+\frac{1}{3}z\\
S_2(y,z) &=& \frac{1}{3} + \frac{10}{9} y+\frac{2}{3} z+\frac{11}{9} y^2+\frac{14}{9} y z+\frac{1}{3} z^2\\
R_{-1}(y,z) &=& S_{-1}(y,z) = 0,
\eea
and the functions $\hat{P}_n,\hat{Q}_n,\hat{R}_n,\hat{S}_n,T_n$ are
\bea
\hat{P}_1(y,z) &=& 1 +\frac{2}{3} y+2 z\\
\hat{P}_2(y,z) &=& 1+y+3 z+ y^2+6  y z+3 z^2\\
\hat{Q}_1(y,z) &=& \frac{1}{3} y^2+2 y z+ z^2\\
\hat{Q}_2(y,z) &=& \frac{1}{9}\left(5 y^3+45 y^2 z+45 y z^2+9 z^3\right)\\
\hat{R}_1(y,z) &=& \frac{2}{3} \\
\hat{R}_2(y,z) &=& 1 + \frac{5}{3} y+ z\\
\hat{S}_1(y,z) &=& \frac{1}{9}\left(5 y+3 z\right)\\
\hat{S}_2(y,z) &=& \frac{1}{9}\left(11 y^2+14 y z+3 z^2\right)\\
T_1(\mu)     &=&\mu_{-}^2\left(\mu_+^2+\mu\right) = \mu_+^4-\mu^2\\
T_2(\mu)     &=& \mu_{-}^2\left(\mu_+^4+\mu\mu_+^2+\mu^2\right) = \mu_+^6-\mu^3\\
T_{-1}(\mu)     &=& \log\frac{\mu_+^2}{\mu}.
\eea
The functions $\Upsilon$ in the denominators come from the average $\sigma$, given by Eqs.~(\ref{eq:sigmagen_v}) and (\ref{eq:sigmagen_q}).  These are defined by
\bea
\langle\sigma_v\rangle_{i,cap} &=& \frac{2\sigma_{0,i} v_e^{2n}}{v_0^{2n}} \Upsilon_n\left(\frac{v_d^2+3\frac{T\mu}{m_\chi}}{v_e^2} ,\frac{v_\odot^2}{v_e^2}\right) \\
&\approx& \frac{2\sigma_{0,i} v_e^{2n}}{v_0^{2n}} \Upsilon_n\left(\frac{v_d^2}{v_e^2} ,\frac{v_\odot^2}{v_e^2}\right)\\
\langle\sigma_q\rangle_{i,cap} &=& \frac{2\sigma_i v_e^{2n}}{q_0^{2n}} \Upsilon_n\left(\frac{v_d^2+3\frac{T\mu}{m_\chi}}{v_e^2} ,\frac{v_\odot^2}{v_e^2}\right) \frac{m_{\chi}^{2n}}{2^n\mu_+^{2n}} f(n) \\
&\approx& \frac{2\sigma_i v_e^{2n}}{q_0^{2n}} \Upsilon_n\left(\frac{v_d^2}{v_e^2} ,\frac{v_\odot^2}{v_e^2}\right) \frac{m_{\chi}^{2n}}{2^n\mu_+^{2n}} f(n)\\
\Upsilon_0(y,z) &=& 1\\
\Upsilon_1(y,z) &=& 1 + y + z\\
\Upsilon_2(y,z) &=& \left(1+y+z\right)^2 +\frac{2}{3}y\left(y+2z\right)\\
\Upsilon_{-1}(y,z) &=& \frac{3}{\sqrt{2\pi 	yz}} H_{-1}\left(\sqrt{\frac{3}{2y}},\sqrt{\frac{3z}{2y}}\right),
\eea
where $H_{-1}(x,y)$ is defined in Eq.~(\ref{eq:Hm1def}). A fast way to check that the overall normalisation is fine is to check the limit $y,z \rightarrow 0$.\footnote{Note that $\lim_{x,y\rightarrow 0} \Upsilon_{-1}(y,z) = 1$, however this procedure won't work for $d\sigma \propto q^{-2}$.}

The geometric limit can be expressed using Eq.~(\ref{eq:satlimitv}) with
\bea
P_G(x,y,z) &=& 1+\frac{1}{3}y+z\\
Q_G(x,y,z) &=& 0\\
R_G(x,y,z) &=& \frac{1}{3}\\
S_G(x,y,z) &=& 0.
\eea
For positive powers of $n$, the saturation limit is
\bea
\frac{\rho_\chi \pi R_\odot^2 v_e^2}{v_\odot m_\chi} \simeq 1.05 \cdot 10^{30} \frac{1\GeV}{m_\chi} \frac{\rho_\chi}{0.4\GeV} s^{-1}.
\eea

\section{Velocity distributions}
\label{sec:reldistr}
Relative velocity distributions are calculated as follows.
We define
\be
f(A,x) = \left(\frac{A}{\pi}\right)^{3/2} e^{-A x}
\ee
For calculating the capture rate, one assumes a Maxwell-Boltzmann distribution with velocity dispersion $v_d = 270$\,km\,s$^{-1}$ for the DM halo,
\be
f^{0}_\chi(\tilde{u}) d\tilde{u} d\cos\tilde{\theta}_\chi d\tilde{\phi}_\chi = f\left(\frac{3}{2v_d^2},\tilde{u}^2\right) d^3u = \left(\frac{3}{2\pi}\right)^{3/2} \frac{\tilde{u}^2}{v_d^3} e^{-\frac{3 \tilde{u}^2}{2v_d^2}} d\tilde{u} d\cos\tilde{\theta}_\chi d\tilde{\phi}_\chi,
\ee
where $\tilde{u}$ is the DM speed in the frame where the average speed is zero (i.e.\ the Galactic rest frame).  The Sun moves at speed $|\vv{v}_\odot| = v_\odot = 220$\,km\,s$^{-1}$ relative to the frame where DM has null average speed, so the DM speed in the rest frame of the Sun is
\be
\vec{v}_\chi = \vec{\tilde{v}}_\chi + \vec{v}_\odot\label{eq:vchishift}.
\ee
The velocities of nuclei in the Sun follow a Maxwell-Boltzmann distribution with temperature $T=T_\odot(r)$,
\bea
f_{N}(v_N) dv_N d\cos\theta_N d\phi_N &=& f\left( \frac{m_N}{2T_\odot(r) } ,v_N^2\right) d^3v_N \nn\\
&=&  \left(\frac{m_N}{2\pi T_\odot(r)}\right)^{3/2} v_N^2 e^{-\frac{m_N v_N^2}{2T_\odot(r)}} dv_N d\cos\theta_N d\phi_N.
\eea
The thermal motion of the nuclei in the Sun has a negligible effect on the capture rate for a constant cross-section, so we can safely use the formulae for the capture rate with the nuclear temperature set to zero. However, in the case of a velocity-dependent cross-section, it is important to take the thermal motion into account when calculating the average cross-section, which can then be safely plugged into the formulae for the capture rate that assume zero nuclear temperature.  The joint distribution for $\tilde{u}$ and $v_N$ is the (six-dimensional) product of the individual distributions, and can be written as
\bea
f_{6D} &\equiv& f\left(\frac{3}{2v_d^2},|\vec{\tilde{u}}|^2\right) f\left( \frac{m_N}{2T_\odot(r) } ,v_N^2\right) d^3\tilde{u}  d^3v_N\nn\\
       &=&      f\left(\frac{3}{2v_d^2},|\vec{u}-\vec{v}_\odot|^2\right) f\left( \frac{m_N}{2T_\odot(r) } ,v_N^2\right) d^3u  d^3v_N \label{eq:reldistribstep1}.
\eea
The equality follows from Eq.~(\ref{eq:vchishift}) and the invariance of the phase space under translations (i.e.\ $d^3u = d^3\tilde{u}$).

At this point, to make this expression easier to manage, it helps to define a new set of coordinates rather than continuing with $\tilde{u}$ and $v_N$.  A more useful set of coordinates is the combination of $\vv{s}$, the speed (in the frame of the Sun) of the centre of mass frame of a collision between a nucleus and a halo DM particle, and the relative speed of such a collision $\vv{u}_r$.  This gives
\bea
\vv{s} &\equiv& \frac{m_N \vv{v}_N + m_\chi \vv{u}}{m_N + m_\chi},\\
\vv{u}_r &\equiv& \vv{u} - \vv{v}_N, \\
\vv{u} &=& \vv{s} + \frac{m_N}{m_\chi + m_N}\vv{u}_r,\\
\vv{v}_N &=& \vv{s} - \frac{m_\chi}{m_\chi + m_N}\vv{u}_r.
\eea
This transformation has Jacobian equal to one, so one can rewrite Eq.~(\ref{eq:reldistribstep1}) using the new variables $\vv{s}$ and $\vv{u}_r$, and then integrate over everything except $u_r =|\vv{u}_r|$ to get a relative speed distribution.  To do this, we start by rewriting the arguments of Eq.~(\ref{eq:reldistribstep1}) in terms of $\vv{s}$ and $\vv{u_r}$:
\bea
|\vv{u}-\vv{v}_\odot|^2 &=& s^2 + \left(\frac{m_N}{m_\chi + m_N}\right)^2 u_r^2 + 2 \frac{m_N}{m_\chi + m_N} s u_r \cos\theta_{sr}\nn\\
&+& v_\odot^2 -2 s v_\odot \cos\theta_{s\odot} - 2 \frac{m_N}{m_\chi + m_N} v_\odot u_r \cos\theta_{r\odot},\\
v_N^2 &=& s^2 + \left(\frac{m_\chi}{m_\chi + m_N}\right)^2 u_r^2 - 2 \frac{m_\chi}{m_\chi + m_N} s u_r \cos\theta_{sr}.
\eea
Here $\theta_{sr}$ is the angle between the vectors $\vec{s}$ and $\vec{u}_r$, $\theta_{s\odot}$ is the angle between the vectors $\vec{s}$ and $\vec{v}_\odot$, and $\theta_{r\odot}$ is the angle between vectors $\vec{u}_r$ and $\vec{v}_\odot$. We now have 3 angles, but they are not independent. Without loss of generality, we can fix the coordinate system such that $\vv{v}_\odot$ lies along the $\hat{z}$ axis.  In this case, $\theta_{s\odot}$ and $\theta_{r\odot}$ are respectively the polar angles of the vectors $\vv{s}$ and $\vv{u}_r$, and
\bea
\cos\theta_{sr} = \cos\theta_{r\odot} \cos\theta_{s\odot} + \sin\theta_{r\odot}\sin\theta_{s\odot}\cos\left(\phi_s-\phi_r\right),
\eea
where $\phi_s$ and $\phi_r$ are the azimuthal angles defining the directions of the vectors $\vv{s}$ and $\vv{u}_r$.

Integrating Eq.~(\ref{eq:reldistribstep1}) over the unnecessary variables from here is not trivial. To do this we rely on translational invariance. We can define a new coordinate
\be
\vv{\hat{s}} \equiv \vv{s} - c \vv{v}_\odot,\label{eq:VhatV}
\ee
and choose the quantity $c$ in such a way as to make the integrand independent of $\theta_{\hat{s}\odot}$. Defining
\bea
A_N &\equiv& \frac{m_N}{2T_\odot(r)},\\
A_\chi &\equiv& \frac{3}{2v_d^2},
\eea
we get
\be
c = \frac{A_\chi}{A_N+A_\chi}.
\ee
Defining
\bea
B_\chi \equiv& \hat{s}^2 + \frac{A_N^2}{(A_N+A_\chi)^2}v_\odot^2 + \frac{m_N^2}{(m_N+m_\chi)^2}u_r^2 + 2\frac{m_N}{m_N + m_\chi} u_r \hat{s} \cos\theta_{\hat{s}r}- 2\frac{A_N m_N}{(A_N+A_\chi)(m_N+m_\chi)} v_\odot u_r \cos\theta_{r\odot}\phantom{,}&\nn\\
B_N \equiv& \hat{s}^2 + \frac{A_\chi^2}{(A_N+A_\chi)^2}v_\odot^2 + \frac{m_\chi^2}{(m_N+m_\chi)^2}u_r^2 - 2\frac{m_\chi}{m_N + m_\chi} u_r \hat{s} \cos\theta_{\hat{s}r}- 2\frac{A_\chi m_\chi}{(A_N+A_\chi)(m_N+m_\chi)} v_\odot u_r \cos\theta_{r\odot},&\nn\\
\eea
we see that
\bea
|\vv{u}-\vv{v}_\odot|^2 &=& B_\chi - 2\frac{A_N}{(A_N+A_\chi)} \hat{s} v_\odot \cos\theta_{\hat{s}\odot}\\
v_N^2 &=& B_N + 2\frac{A_\chi}{(A_N+A_\chi)} \hat{s} v_\odot \cos\theta_{\hat{s}\odot}.
\eea
With this change of variables (all Jacobians so far are equal to 1), we find that the joint distribution becomes
\be
f_{6D} = f\left(A_N, v_N^2 \right) f\left(A_\chi, |\vv{u}-\vv{v}_\odot|^2 \right) d^3 \hat{s} d^3 u_r = f\left(A_N, B_N \right) f\left(A_\chi, B_\chi \right) d^3 \hat{s} d^3 u_r.
\ee
This expression now only explicitly depends on two angles, $\theta_{\hat{s}r}$ and $\theta_{r\odot}$.  We can therefore integrate easily over $\phi_s$ and $\phi_r$, as $\theta_{r\odot}$ and $\theta_{sr}$ are just the polar angles and have no dependence on either of the azimuthal angles.  The reduced distribution is then
\begin{align}
f_{4D} =& \left(2\pi\right)^2 f\left(A_N, B_N \right) f\left(A_\chi, B_\chi \right) \hat{s}^2 u_r^2 d\hat{s} du_r d\cos\theta_{\hat{s}r}d\cos\theta_{r\odot}\\
=& \left(2\pi\right)^2 f\left(A_N+A_\chi, \hat{s}^2 + 2\tfrac{A_\chi m_N-A_N m_\chi}{(A_N+A_\chi)(m_N+m_\chi)} u_r \hat{s} \cos\theta_{\hat{s}r} + \tfrac{A_\chi m_N^2+A_N m_\chi^2}{(A_N+A_\chi)(m_N+m_\chi)^2} u_r^2 \right) \nn\\
& \times f\left(\tfrac{A_N A_\chi}{A_N + A_\chi}, v_\odot^2 - 2 v_\odot u_r \cos\theta_{r\odot} \right) \hat{s}^2 u_r^2 d\hat{s} du_r d\cos\theta_{\hat{s}r}d\cos\theta_{r\odot}\label{eq:f4dstep2}.
\end{align}
If we then integrate over $\cos\theta_{\hat{s}r}$, we find
\bea
f_{3D} &=& 2\pi^2\frac{m_\chi+m_N}{A_\chi m_N-A_N m_\chi}\left\{f\left(A_N + A_\chi, \left[\hat{s} - u_r\tfrac{A_\chi m_N - A_N m_\chi}{(A_N+A_\chi)(m_N+m_\chi)}\right]^2 \right)\right.\nn\\
       && \left.-f\left(A_N + A_\chi, \left[\hat{s} + u_r\tfrac{A_\chi m_N - A_N m_\chi}{(A_N+A_\chi)(m_N+m_\chi)}\right]^2 \right) \right\}\nn\\
       && \times f\left(\frac{A_N A_\chi}{A_N + A_\chi}, v_\odot^2 - 2 v_\odot u_r \cos\theta_{r\odot} + u_r^2 \right)  \hat{s} u_r d\hat{s} du_r d\cos\theta_{r\odot}.
\eea
Integrating over $\hat{s}$, this becomes,
\be
f_{2D} = 2\pi f\left(\frac{A_N A_\chi}{A_N + A_\chi}, v_\odot^2 - 2 v_\odot u_r \cos\theta_{r\odot} + u_r^2\right) u_r^2 du_r d\cos\theta_{r\odot},
\ee
and finally over $\cos\theta_{r\odot}$, we obtain
\bea
f_{cap}(u_r)du_r &=& \frac{u_r}{v_\odot} \sqrt{\frac{3}{2\pi(v_d^2+3T\mu/m_{\chi})}} \left(e^{-\frac{3(u_r-v_\odot)^2}{2(3T\mu/m_{\chi}+v_d^2)}}-e^{-\frac{3(u_r+v_\odot)^2}{2(3T\mu/m_{\chi}+v_d^2)}}\right)du_r.
\label{cap_distrib_appendix}
\eea
This expression gives the distribution of relative velocities between nuclei in the Sun, and DM particles in a halo an infinite distance away from the Sun.  When DM falls into the gravitational potential well of the Sun, it gets accelerated. The actual relative DM-nucleus speed in collisions taking place at some distance $r$ from the centre of the Sun is therefore not $u_r$, but the squared sum of the relative speed at infinity and the escape velocity (cf.\ Eq.\ \ref{eq:wtou}), i.e.
\be
w_r^2(r) = u_r^2+v_e^2(r).
\ee
The final expression for the distribution of actual relative speeds is therefore
\be
f_{cap}(w_r)dw_r = \frac{w_r}{v_\odot} \sqrt{\frac{3}{2\pi(v_d^2+3T\mu/m_{\chi})}} \left\{\exp\left[-\tfrac{3\left(\sqrt{w_r^2-v_e^2}-v_\odot\right)^2}{2(3T\mu/m_{\chi}+v_d^2)}\right]-\exp\left[-\tfrac{3\left(\sqrt{w_r^2-v_e^2}+v_\odot\right)^2}{2(3T\mu/m_{\chi}+v_d^2)}\right]\right\}dw_r.
\ee
We see that the final result has a rather intuitive form: it is the DM speed distribution in the frame of the Sun, but with a velocity dispersion instead given by the sum in quadrature of the halo dispersion $v_d$ and the nuclear velocity dispersion $\sqrt{3T/m_N}$.

\section{Capture and evaporation rate calculation}
\label{sec:calcdetails}

\subsection{Definitions of variables}
We define the following kinematic variables in the DM-nucleus collision:
\begin{itemize}
\item $\vv{w}, w$ DM speed before the collision, and its modulus
\item $\vv{v}_N, v_N$ Nucleon speed before the collision, and modulus
\item $\vv{v}, v$ DM speed after the collision, and modulus
\item $\vv{s}, s$ Center of mass speed, and modulus
\item $\vv{t}, t$ DM speed before collision, and modulus, in the center of mass reference system
\item $\vv{t'}, t'$ DM speed after collision, and modulus, in the center of mass reference system
\item $\theta_{xy}$ angle between two vectors $\vv{x}$ and $\vv{y}$.
\item $\mu = m_{\chi}/m_{N_i}$
\item $\mu_{+}=\frac{\mu+1}{2}$
\item $\mu_{-}=\frac{\mu-1}{2}$
\item $u$ speed of DM far away from the Sun
\end{itemize}
Here are some important relations between these quantities:
\begin{eqnarray}
|\vv{w}-\vv{v}_N| &=& t(1+\mu) \label{vrel}\\
|\mu\vv{w}+\vv{v}_N| &=& s(1+\mu)\\
\cos \theta_{st'} &=& \frac{s^2+t^2-v^2}{2st}\\
\cos \theta_{wv_N} &=& -\frac{w^2+v_N^2-t^2(1+\mu)^2}{2v_N w}\\
t' &=& t\\
\cos \theta_{st} &=& \frac{s^2+t^2-w^2}{2st}\\
v_N^2 &=& 2\mu \mu_{+} t^2 + 2 \mu_{+} s^2 - \mu w^2\\
w^2 &=& u^2+v_e^2(r) \label{eq:wtou}
\end{eqnarray}

We would like to express $v_r, q_{tr}$ as a function of the variables listed above. For the relative velocity we can just use Eq.~(\ref{vrel}). For $q_{tr}$, we need some geometry: the angle between $\vv{t}$ and $-\vv{t'}$ is
\begin{equation}
\cos (\theta_{st'} \pm \theta_{st}) = \cos \theta_{st'} \cos \theta_{st} + \sin \theta_{st'} \sin\theta_{st}\sin \phi,
\end{equation}
where $\phi$ is an azimuthal angle over which we have averaged. Thus,
\begin{eqnarray}
q_{tr} &=& m_{\chi} 2t \sin \left(\frac{|\theta_{st'} \pm \theta_{st}|}{2}\right),\\
q_{tr}^2 &=& 2 m_{\chi}^2 t^2 \left\{\rule{0cm}{7mm}1-\frac{(s^2+t^2-v^2)(s^2+t^2-w^2)}{4s^2t^2} \nonumber\right. \\
 &&\left.+\sqrt{\left[1-\frac{(s^2+t^2-v^2)^2}{4s^2t^2}\right]\left[1-\frac{(s^2+t^2-w^2)^2}{4s^2t^2}\right]}\sin\phi\right\},\\
\langle q_{tr}^2 \rangle &=& 2 m_{\chi}^2 t^2 \left[1-\frac{(s^2+t^2-v^2)(s^2+t^2-w^2)}{4s^2t^2}\right].\label{eq:qtravevap}
\end{eqnarray}

\subsection{Capture}
\label{sec:capturecalc}
We follow the approach of \cite{Gould87a, Gould87b}. To calculate the capture rate, we have to evaluate $\Omega^{-}(w)$.

\begin{eqnarray}
\Omega^{-}(w) &=& \int_0^{v_e} R^{-}(w\rightarrow v)  \mathrm{d} v\\
R^{-}(w\rightarrow v) &=& \int_0^\infty \mathrm{d} s \int_0^\infty \mathrm{d} t \frac{32 \mu_{+}^4}{\sqrt{\pi}}k^3 n_i \frac{d\sigma_i}{d\cos\theta}(s,t,v,w) \frac{v t}{w} e^{-k^2v_N^2} \nonumber \\
&&\times \Theta(t+s-w)\Theta(v-|t-s|),
\end{eqnarray}
where we have defined
\begin{equation}
k^2 = \frac{m_i}{2T},
\end{equation}
and the DM-proton cross-section is given by Eqs.~(\ref{eq:sigmavrel}) and (\ref{eq:sigmaqtr}). $\Theta(x)$ is the standard Heaviside step function.
The DM-nucleus cross-section can be obtained from Eq.~(\ref{eq:nucleoncrosssection}). For the calculation of the average cross-section, it is important to retain the dependence on the temperature of the nuclei.  In contrast, for the kinematics associated with capture, the effect of the thermal motion of the nuclei is either negligible (constant cross section, $n=0$), or mild (other cases) \citep{Garani:2017jcj}), so we decide to set their temperature to zero, as this allows us to simplify the 5-dimensional integral to a 1-dimensional or 2-dimensional one. In this limit,
\begin{equation}
\lim_{T \to 0} \frac{8 \mu_{+}^2}{\sqrt{\pi}}k^3  t \mu e^{-k^2v_N^2}\Theta(t+s-w) \rightarrow \delta\left(s-\frac{w \mu}{2\mu_+}\right)\delta\left(t-\frac{w}{2\mu_+}\right),
\end{equation}
and
\bea
v_r &=& |\vv{w}-\vv{u}| = t(1+\mu) = w; \\ s(1+\mu) &=& \mu w \\
q_{tr}^2 &=& 2 m_{\chi}^2 t^2 \left[1-\frac{(s^2+t^2-v^2)(s^2+t^2-w^2)}{4s^2t^2}\right] = m_{\chi}^2 \frac{w^2-v^2}{\mu}.
\eea
Thus,
\begin{eqnarray}
R^{-}(w\rightarrow v) &=& \frac{4 \mu_{+}^2}{\mu} n_i \frac{v}{w} \int_0^\infty \mathrm{d} t \frac{d\sigma_i}{d\cos\theta}\left(\frac{w \mu}{2\mu_+},t,v,w\right) \delta\left(t-\frac{w}{2\mu_+}\right) \Theta\left(v-w\frac{|\mu_-|}{\mu_+}\right)\nn\\
 &=& \frac{4\mu_{+}^2}{\mu} n_i \frac{v}{w} \frac{d\sigma_i}{d\cos\theta}\left(\frac{w \mu}{2\mu_+},\frac{w}{2\mu_+},v,w\right) \Theta\left(v-w\frac{|\mu_-|}{\mu_+}\right)\\
\Omega^{-}(w) &=& \int_0^{v_e} \mathrm{d} v \frac{4 \mu_{+}^2}{\mu} n_i \frac{v}{w} \frac{d\sigma_i}{d\cos\theta}\left(\frac{w \mu}{2\mu_+},\frac{w}{2\mu_+},v,w\right) \Theta\left(v-w\frac{|\mu_-|}{\mu_+}\right)\\
&=& \frac{4\mu_{+}^2}{\mu w} n_i \int_{w\frac{|\mu_-|}{\mu_+}}^{v_e} \mathrm{d} v v \frac{d\sigma_i}{d\cos\theta}\left(\frac{w \mu}{2\mu_+},\frac{w}{2\mu_+},v,w\right).
\end{eqnarray}

We now evaluate the above integral for generalised form factor DM. If the differential cross-section depends on the relative velocity, then for hydrogen
\begin{eqnarray}
\Omega^{-}(w) &=& \frac{4 \mu_{+}^2}{\mu w} n_i \sigma_0 \int_{w\frac{|\mu_-|}{\mu_+}}^{v_e} \mathrm{d} v v \left(\frac{w}{v_0}\right)^{2n} \nn\\
&=& \frac{2 \mu_{+}^2}{\mu w} n_i \sigma_0 \left(\frac{w}{v_0}\right)^{2n} \int_{w\frac{|\mu_-|}{\mu_+}}^{v_e} \mathrm{d} v v\nn\\
&=& \frac{2\mu_{+}^2}{\mu w} n_i \sigma_0 \left(\frac{w}{v_0}\right)^{2n} \left(v_e^2-\frac{\mu_{-}^2}{\mu_{+}^2}w^2\right) \Theta\left(v_e^2-\frac{\mu_{-}^2}{\mu_{+}^2}w^2\right) \label{CSIHv}
\end{eqnarray}
and for other elements
\begin{eqnarray}
\Omega^{-}(w) &=& \frac{4\mu_{+}^2}{\mu w} n_i \sigma_i \left(\frac{w}{v_0}\right)^{2n} \int_{w\frac{|\mu_-|}{\mu_+}}^{v_e} \mathrm{d} v v |F_i(q_{tr})|^2\\
&=& \frac{6\mu_{+}^2}{m_{\chi}^2 \Lambda^2 w} n_i \sigma_i \left(\frac{w}{v_0}\right)^{2n} \left(e^{-\frac{m_{\chi}^2 \Lambda^2(w^2-v_e^2)}{3\mu}}-e^{-\frac{m_{\chi}^2\Lambda^2}{3\mu_{+}^2}w^2}\right) \Theta\left(v_e^2-\frac{\mu_{-}^2}{\mu_{+}^2}w^2\right). \label{CSIv}
\end{eqnarray}
Note that these differ from the case of a constant cross-section only by a constant factor $({w}/{v_0})^{2n}$.

If the differential cross-section instead depends on the momentum transferred, for hydrogen we have
\begin{eqnarray}
\Omega^{-}(w) &=& \frac{4\mu_{+}^2}{\mu w} n_i \sigma_0 \int_{w\frac{|\mu_-|}{\mu_+}}^{v_e} \mathrm{d} v v \left(\frac{m_{\chi}^2}{q_0^2}\right)^{n} \left(\frac{w^2-v^2}{\mu}\right)^n \\
&=& \frac{2n_i \sigma_0 \mu_{+}^2 m_\chi^{2n}}{(n+1)\mu^{n+1} w q_0^{2n}}\left[\left(\frac{w^2\mu}{\mu_+^2}\right)^{n+1}-u^{2(n+1)}\right] \\
&=& \frac{2\mu_{+}^2}{\mu w} n_i \sigma_0 \left\{
  \begin{array}{lr}
    \frac{m_{\chi}^2}{q_0^2} \frac{1}{2\mu}\left(\frac{w^4 \mu^2}{\mu_+^4}-u^4\right) &: n =1\\
    \frac{m_{\chi}^4}{q_0^4} \frac{1}{3\mu^2}\left(\frac{w^6 \mu^3}{\mu_+^6}-u^6\right) &: n=2\\
    \frac{q_0^2}{m_{\chi}^2} \mu\log{\frac{\mu w^2}{\mu_+^2 u^2}} &: n=-1,\\
    \end{array}\right.
\end{eqnarray}
and for the other elements,
\begin{eqnarray}
\Omega^{-}(w) &=& \frac{4 \mu_{+}^2}{\mu w} n_i \sigma_i \left(\frac{m_{\chi}^2}{q_0^2}\right)^{n} \int_{w\frac{|\mu_-|}{\mu_+}}^{v_e} \mathrm{d} v v |F_i(q_{tr})|^2\left(\frac{w^2-v^2}{\mu}\right)^n\\
&=& \frac{2\mu_{+}^2}{\mu w} n_i \sigma_i \left\{
  \begin{array}{lr}
    \frac{m_{\chi}^2}{q_0^2} \frac{u^4}{\mu \gamma_u^2}\left[(1+\gamma_u)e^{-\gamma_u}-(1+\gamma_w)e^{-\gamma_w}\right] & : n = 1\\
    \frac{m_{\chi}^4}{q_0^4} \frac{2u^6}{\mu^2\gamma_u^3}\left[(1+\gamma_u+\gamma_u^2/2)e^{-\gamma_u} -(1+\gamma_w+\gamma_w^2/2)e^{-\gamma_w}\right] & : n=2\\
    \frac{q_0^2}{m_{\chi}^2} \mu\left[G(-\gamma_w)-G(-\gamma_u)\right] & : n=-1\\
    \end{array}\right.\nn,
\end{eqnarray}
where
\begin{equation}
G(x) \equiv - \int_{-x}^\infty \mathrm{d} y \frac{e^{-y}}{y},
\end{equation}
and
\bea
\gamma_u &\equiv& \frac{\Lambda^2 m_{\chi}^2 u^2}{3\mu}\\
\gamma_w &\equiv& \frac{\Lambda^2 m_{\chi}^2 w^2}{3\mu_+^2}.
\eea
We have suppressed the $\Theta$ functions in these formulae to keep them short(er).  They agree with the ones of Ref.\ \citep{Vincent15}.

\subsection{Evaporation}
\label{sec:evapcalc}
For evaporation, the temperature of the nuclei has a dominant effect on the kinematics, so it is not possible to take the zero temperature limit, and we must retain the full expressions.
\begin{eqnarray}
\Omega^{+}(w) &=& \int_{v_e}^\infty R^{+}(w\rightarrow v)  \mathrm{d} v \label{eq:omegap}\\
R^{+}(w\rightarrow v) &=& \int_0^\infty \mathrm{d} s \int_0^\infty \mathrm{d} t \frac{{32} \mu_{+}^4}{\sqrt{\pi}}k^3 n_i \frac{d\sigma_i}{d\cos\theta}(s,t,v,w) \frac{v t}{w} e^{-k^2v_N^2}\nonumber\\
&&\times\Theta(t+s-v)\Theta(w-|t-s|)\label{eq:Rp}.
\end{eqnarray}
Because we cannot use the $T_N\rightarrow0$ limit, it is not possible to evaluate these integrals analytically (unlike in the case of capture). To speed up the computation, one can still calculate $\Omega^{+}(w)$ on a grid for the variables $v_e, m_\chi$ and $t_\chi$, and for each element and differential cross-section. After that, the evaporation rate can be computed for different solar models in the same way, and potentially with the same computation speed as for the capture rate. This is how we intend to apply the results of this paper to solar simulations in the future.


\bibliographystyle{JHEP_pat}

\label{Bibliography}

\bibliography{Bibliography} 

\providecommand{\href}[2]{#2}\begingroup\raggedright\begin{thebibliography}{10}

\bibitem{LUXRun2}
D.~S. {Akerib}, S.~{Alsum}, {\em et.~al.}, {\it {Results from a search for dark
  matter in the complete LUX exposure}},
  \href{http://arxiv.org/abs/1608.07648}{{\tt arXiv:1608.07648}}.

\bibitem{CDMSLite}
{SuperCDMS Collaboration}, R.~{Agnese}, {\em et.~al.}, {\it {WIMP-Search
  Results from the Second CDMSlite Run}},  {\em \prl} (2015) 071301,
  [\href{http://arxiv.org/abs/1509.02448}{{\tt arXiv:1509.02448}}].

\bibitem{PICO15}
C.~{Amole}, M.~{Ardid}, {\em et.~al.}, {\it {Dark Matter Search Results from
  the PICO-2L C$_{3}$F$_{8}$ Bubble Chamber}},  {\em \prl} {\bf 114} (2015)
  231302, [\href{http://arxiv.org/abs/1503.00008}{{\tt arXiv:1503.00008}}].

\bibitem{PICO60}
C.~{Amole}, M.~{Ardid}, {\em et.~al.}, {\it {Dark Matter Search Results from
  the PICO-60 CF$\_3$I Bubble Chamber}},
  \href{http://arxiv.org/abs/1510.07754}{{\tt arXiv:1510.07754}}.

\bibitem{PandaX2016}
{PandaX-II Collaboration}, {:}, {\em et.~al.}, {\it {Dark Matter Results from
  First 98.7-day Data of PandaX-II Experiment}},
  \href{http://arxiv.org/abs/1607.07400}{{\tt arXiv:1607.07400}}.

\bibitem{Aprile:2012zx}
XENON1T: E.~Aprile, {\it {The XENON1T Dark Matter Search Experiment}},  {\em
  Springer Proc. Phys.} {\bf 148} (2013) 93--96,
  [\href{http://arxiv.org/abs/1206.6288}{{\tt arXiv:1206.6288}}].

\bibitem{Zurek14}
K.~M. {Zurek}, {\it {Asymmetric Dark Matter: Theories, signatures, and
  constraints}},  {\em \physrep} {\bf 537} (2014) 91--121,
  [\href{http://arxiv.org/abs/1308.0338}{{\tt arXiv:1308.0338}}].

\bibitem{Gould87a}
A.~{Gould}, {\it {Weakly interacting massive particle distribution in and
  evaporation from the sun}},  {\em \apj} {\bf 321} (1987) 560--570.

\bibitem{Taoso10}
M.~{Taoso}, F.~{Iocco}, G.~{Meynet}, G.~{Bertone}, and P.~{Eggenberger}, {\it
  {Effect of low mass dark matter particles on the Sun}},  {\em \prd} {\bf 82}
  (2010) 083509, [\href{http://arxiv.org/abs/1005.5711}{{\tt
  arXiv:1005.5711}}].

\bibitem{FrandsenSarkar}
M.~T. {Frandsen} and S.~{Sarkar}, {\it {Asymmetric Dark Matter and the Sun}},
  {\em \prl} {\bf 105} (2010) 011301,
  [\href{http://arxiv.org/abs/1003.4505}{{\tt arXiv:1003.4505}}].

\bibitem{Widmark:2017yvd}
A.~Widmark, {\it {Thermalization time scales for WIMP capture by the Sun in
  effective theories}},  \href{http://arxiv.org/abs/1703.06878}{{\tt
  arXiv:1703.06878}}.

\bibitem{Bergstrom98b}
L.~{Bergstr{\"o}m}, J.~{Edsj{\"o}}, and P.~{Gondolo}, {\it {Indirect detection
  of dark matter in km-size neutrino telescopes}},  {\em \prd} {\bf 58} (1998)
  103519, [\href{http://arxiv.org/abs/hep-ph/9806293}{{\tt hep-ph/9806293}}].

\bibitem{Barger02}
V.~{Barger}, F.~{Halzen}, D.~{Hooper}, and C.~{Kao}, {\it {Indirect search for
  neutralino dark matter with high energy neutrinos}},  {\em \prd} {\bf 65}
  (2002) 075022, [\href{http://arxiv.org/abs/hep-ph/0105182}{{\tt
  hep-ph/0105182}}].

\bibitem{Blennow08}
M.~{Blennow}, J.~{Edsj{\"o}}, and T.~{Ohlsson}, {\it {Neutrinos from WIMP
  annihilations obtained using a full three-flavor Monte Carlo approach}},
  {\em \jcap} {\bf 1} (2008) 21, [\href{http://arxiv.org/abs/0709.3898}{{\tt
  arXiv:0709.3898}}].

\bibitem{Wikstrom09}
G.~{Wikstr{\"o}m} and J.~{Edsj{\"o}}, {\it {Limits on the WIMP-nucleon
  scattering cross-section from neutrino telescopes}},  {\em \jcap} {\bf 4}
  (2009) 9, [\href{http://arxiv.org/abs/0903.2986}{{\tt arXiv:0903.2986}}].

\bibitem{IC22Methods}
P.~{Scott}, C.~{Savage}, J.~{Edsj{\"o}}, and {the IceCube Collaboration:
  R.~Abbasi et al.}, {\it {Use of event-level neutrino telescope data in global
  fits for theories of new physics}},  {\em \jcap} {\bf 11} (2012) 57,
  [\href{http://arxiv.org/abs/1207.0810}{{\tt arXiv:1207.0810}}].

\bibitem{Silverwood12}
H.~{Silverwood}, P.~{Scott}, {\em et.~al.}, {\it {Sensitivity of
  IceCube-DeepCore to neutralino dark matter in the MSSM-25}},  {\em \jcap}
  {\bf 3} (2013) 27, [\href{http://arxiv.org/abs/1210.0844}{{\tt
  arXiv:1210.0844}}].

\bibitem{SuperK15}
K.~{Choi}, K.~{Abe}, {\em et.~al.}, {\it {Search for Neutrinos from
  Annihilation of Captured Low-Mass Dark Matter Particles in the Sun by
  Super-Kamiokande}},  {\em \prl} {\bf 114} (2015) 141301.

\bibitem{IC79}
IceCube Collaboration: M.~G. {Aartsen}, R.~{Abbasi}, {\em et.~al.}, {\it
  {Search for Dark Matter Annihilations in the Sun with the 79-String IceCube
  Detector}},  {\em \prl} {\bf 110} (2013) 131302,
  [\href{http://arxiv.org/abs/1212.4097}{{\tt arXiv:1212.4097}}].

\bibitem{IC79_SUSY}
IceCube Collaboration: M.~G. {Aartsen} {\em et.~al.}, {\it {Improved limits on
  dark matter annihilation in the Sun with the 79-string IceCube detector and
  implications for supersymmetry}},  {\em \jcap} {\bf 04} (2016) 022,
  [\href{http://arxiv.org/abs/1601.00653}{{\tt arXiv:1601.00653}}].

\bibitem{Spergel85}
D.~N. {Spergel} and W.~H. {Press}, {\it {Effect of hypothetical, weakly
  interacting, massive particles on energy transport in the solar interior}},
  {\em \apj} {\bf 294} (1985) 663--673.

\bibitem{Nauenberg87}
M.~{Nauenberg}, {\it {Energy transport and evaporation of weakly interacting
  particles in the Sun}},  {\em \prd} {\bf 36} (1987) 1080--1087.

\bibitem{GouldRaffelt90a}
A.~{Gould} and G.~{Raffelt}, {\it {Thermal conduction by massive particles}},
  {\em \apj} {\bf 352} (1990) 654--668.

\bibitem{GouldRaffelt90b}
A.~{Gould} and G.~{Raffelt}, {\it {Cosmion energy transfer in stars - The
  Knudsen limit}},  {\em \apj} {\bf 352} (1990) 669--680.

\bibitem{Vincent13}
A.~C. {Vincent} and P.~{Scott}, {\it {Thermal conduction by dark matter with
  velocity and momentum-dependent cross-sections}},  {\em \jcap} {\bf 4} (2014)
  19, [\href{http://arxiv.org/abs/1311.2074}{{\tt arXiv:1311.2074}}].

\bibitem{Vincent14}
A.~C. {Vincent}, P.~{Scott}, and A.~{Serenelli}, {\it {Possible Indication of
  Momentum-Dependent Asymmetric Dark Matter in the Sun}},  {\em \prl} {\bf 114}
  (2015) 081302, [\href{http://arxiv.org/abs/1411.6626}{{\tt
  arXiv:1411.6626}}].

\bibitem{Vincent15}
A.~C. {Vincent}, A.~{Serenelli}, and P.~{Scott}, {\it {Generalised form factor
  dark matter in the Sun}},  {\em \jcap} {\bf 8} (2015) 40,
  [\href{http://arxiv.org/abs/1504.04378}{{\tt arXiv:1504.04378}}].

\bibitem{Vincent16}
A.~C. {Vincent}, P.~{Scott}, and A.~{Serenelli}, {\it {Updated constraints on
  velocity and momentum-dependent asymmetric dark matter}},  {\em \jcap} {\bf
  11} (2016) 007, [\href{http://arxiv.org/abs/1605.06502}{{\tt
  arXiv:1605.06502}}].

\bibitem{Geytenbeek16}
B.~{Geytenbeek}, S.~{Rao}, {\em et.~al.}, {\it {Effect of electromagnetic
  dipole dark matter on energy transport in the solar interior}},  {\em
  submitted to \jcap} (2016) [\href{http://arxiv.org/abs/1610.06737}{{\tt
  arXiv:1610.06737}}].

\bibitem{AspIV}
M.~{Asplund}, N.~{Grevesse}, A.~J. {Sauval}, C.~{Allende Prieto}, and
  D.~{Kiselman}, {\it {Line formation in solar granulation. IV. [O I], O I and
  OH lines and the photospheric O abundance}},  {\em \aap} {\bf 417} (2004)
  751--768, [\href{http://arxiv.org/abs/astro-ph/0312290}{{\tt
  astro-ph/0312290}}].

\bibitem{Bahcall05}
J.~N. {Bahcall}, S.~{Basu}, M.~{Pinsonneault}, and A.~M. {Serenelli}, {\it
  {Helioseismological Implications of Recent Solar Abundance Determinations}},
  {\em \apj} {\bf 618} (2005) 1049--1056,
  [\href{http://arxiv.org/abs/astro-ph/0407060}{{\tt astro-ph/0407060}}].

\bibitem{Serenelli09}
A.~Serenelli, S.~Basu, J.~W. Ferguson, and M.~Asplund, {\it {New Solar
  Composition: The Problem With Solar Models Revisited}},  {\em \apjl} {\bf
  705} (2009) L123--L127, [\href{http://arxiv.org/abs/0909.2668}{{\tt
  arXiv:0909.2668}}].

\bibitem{Scott09Ni}
P.~{Scott}, M.~{Asplund}, N.~{Grevesse}, and A.~J. {Sauval}, {\it {On the Solar
  Nickel and Oxygen Abundances}},  {\em \apjl} {\bf 691} (2009) L119--L122,
  [\href{http://arxiv.org/abs/0811.0815}{{\tt arXiv:0811.0815}}].

\bibitem{AGSS}
M.~{Asplund}, N.~{Grevesse}, A.~J. {Sauval}, and P.~{Scott}, {\it {The chemical
  composition of the Sun}},  {\em \araa} {\bf 47} (2009) 481--522,
  [\href{http://arxiv.org/abs/0909.0948}{{\tt arXiv:0909.0948}}].

\bibitem{Serenelli11}
A.~M. {Serenelli}, W.~C. {Haxton}, and C.~{Pe{\~n}a-Garay}, {\it {Solar Models
  with Accretion. I. Application to the Solar Abundance Problem}},  {\em \apj}
  {\bf 743} (2011) 24, [\href{http://arxiv.org/abs/1104.1639}{{\tt
  arXiv:1104.1639}}].

\bibitem{Villante14}
F.~L. {Villante}, A.~M. {Serenelli}, F.~{Delahaye}, and M.~H. {Pinsonneault},
  {\it {The Chemical Composition of the Sun from Helioseismic and Solar
  Neutrino Data}},  {\em \apj} {\bf 787} (2014) 13,
  [\href{http://arxiv.org/abs/1312.3885}{{\tt arXiv:1312.3885}}].

\bibitem{Serenelli16}
A.~{Serenelli}, P.~{Scott}, {\em et.~al.}, {\it {Implications of solar wind
  measurements for solar models and composition}},  {\em \mnras} {\bf 463}
  (2016) 2--9, [\href{http://arxiv.org/abs/1604.05318}{{\tt
  arXiv:1604.05318}}].

\bibitem{CRESST_momdep}
G.~{Angloher}, A.~{Bento}, {\em et.~al.}, {\it {Limits on momentum-dependent
  asymmetric dark matter with CRESST-II}},  {\em ArXiv e-prints} (2016)
  [\href{http://arxiv.org/abs/1601.04447}{{\tt arXiv:1601.04447}}].

\bibitem{Garani:2017jcj}
R.~Garani and S.~Palomares-Ruiz, {\it {Dark matter in the Sun: scattering off
  electrons vs nucleons}},  \href{http://arxiv.org/abs/1702.02768}{{\tt
  arXiv:1702.02768}}.

\bibitem{Busoni13}
G.~{Busoni}, A.~{De Simone}, and W.-C. {Huang}, {\it {On the minimum dark
  matter mass testable by neutrinos from the Sun}},  {\em \jcap} {\bf 7} (2013)
  010, [\href{http://arxiv.org/abs/1305.1817}{{\tt arXiv:1305.1817}}].

\bibitem{Liang16}
Z.-L. {Liang}, Y.-L. {Wu}, Z.-Q. {Yang}, and Y.-F. {Zhou}, {\it {On the
  evaporation of solar dark matter: spin-independent effective operators}},
  {\em \jcap} {\bf 9} (2016) 018, [\href{http://arxiv.org/abs/1606.02157}{{\tt
  arXiv:1606.02157}}].

\bibitem{Lopes14}
I.~{Lopes}, P.~{Panci}, and J.~{Silk}, {\it {Helioseismology with Long-range
  Dark Matter-Baryon Interactions}},  {\em \apj} {\bf 795} (2014) 162,
  [\href{http://arxiv.org/abs/1402.0682}{{\tt arXiv:1402.0682}}].

\bibitem{Bramante:2017xlb}
J.~Bramante, A.~Delgado, and A.~Martin, {\it {Multiscatter stellar capture of
  dark matter}},  \href{http://arxiv.org/abs/1703.04043}{{\tt
  arXiv:1703.04043}}.

\bibitem{Helm:1956zz}
R.~H. Helm, {\it {Inelastic and Elastic Scattering of 187-Mev Electrons from
  Selected Even-Even Nuclei}},  {\em Phys. Rev.} {\bf 104} (1956) 1466--1475.

\bibitem{Spergel:1984re}
D.~N. Spergel and W.~H. Press, {\it {Effect of hypothetical, weakly
  interacting, massive particles on energy transport in the solar interior}},
  {\em Astrophys. J.} {\bf 294} (1985) 663--673.

\bibitem{Gould:1989tu}
A.~{Gould}, {\it {Evaporation of WIMPs with arbitrary cross sections}},  {\em
  \apj} {\bf 356} (1990) 302--309.

\bibitem{Gould87b}
A.~{Gould}, {\it {Resonant enhancements in weakly interacting massive particle
  capture by the earth}},  {\em \apj} {\bf 321} (1987) 571--585.

\end{thebibliography}\endgroup

\end{document}